\documentclass[12pt]{article}
\usepackage{latexsym}
\usepackage{bbm}
\usepackage{theorem}
\usepackage{amsopn}
\usepackage{amsfonts}
\usepackage{amssymb}
\usepackage{amsmath}
\usepackage[english]{babel}
\renewcommand{\baselinestretch}{1.2}
\normalsize
\newtheorem{Theorem}{Theorem}[section]
\newtheorem{Proposition}[Theorem]{Proposition}
\newtheorem{Lemma}[Theorem]{Lemma}

\newtheorem{Definition}[Theorem]{Definition}

\theorembodyfont{\rmfamily}

\DeclareMathAlphabet{\Ma}{U}{msa}{m}{n}
\DeclareMathAlphabet{\Mb}{U}{msb}{m}{n}
\DeclareMathAlphabet{\Meuf}{U}{euf}{m}{n}

\def\got#1{\Meuf{#1}}

\DeclareSymbolFont{ASMa}{U}{msa}{m}{n}
\DeclareSymbolFont{ASMb}{U}{msb}{m}{n}
\DeclareMathSymbol{\hrist}{\mathord}{ASMa}{"16}
\DeclareMathSymbol{\varkappa}{\mathalpha}{ASMb}{"7B}
\DeclareMathSymbol{\CrPr}{\mathord}{ASMb}{"6F}

\newfont{\EinsFont}{cmr7 scaled 1070}
\def\EINS{{\mathchoice{
 \mbox{\unitlength1cm\begin{picture}(.25,.2)\put(0,0){$1$}%
 \put(0.105,0){{\mbox{\fontfamily{cmr}\upshape\small l}}}\end{picture}}}{%
 \mbox{\unitlength1cm\begin{picture}(.25,.2)\put(0,0){$1$}%
 \put(0.105,0){{\mbox{\fontfamily{cmr}\upshape\small l}}}\end{picture}}}{%
 \mbox{\unitlength1cm\begin{picture}(.18,.15)\put(0,0){$\scriptstyle 1$}%
 \put(0.07,0){{\mbox{\fontfamily{cmr}\upshape\EinsFont l}}}\end{picture}}}{%
 \mbox{\unitlength1cm\begin{picture}(.18,.15)\put(0,0){$\scriptstyle 1$}%
 \put(0.07,0){{\mbox{\fontfamily{cmr}\upshape\EinsFont l}}}\end{picture}}}}}
\def\un{\EINS}
\def\restriction{{\mathchoice{
 \mbox{\unitlength1cm\begin{picture}(.2,.4)%
  \bezier{5}(.07,.3)(.1,.27)(.13,.24)%
  \put(.07,.35){\line(0,-1){.5}}\end{picture}}}{
 \mbox{\unitlength1cm\begin{picture}(.2,.4)%
  \bezier{5}(.07,.3)(.1,.27)(.13,.24)%
  \put(.07,.35){\line(0,-1){.5}}\end{picture}}}{
  \hrist}{\hrist}}}

  \def\al #1.{{\mathcal{#1}}}
  \def\ot #1.{{\got{#1}}}

  \def\ccr #1,#2.{\overline{\Delta(#1,\,#2)}}

  \def\b #1.{{\bf #1}}
  \def\cross#1.{\mathrel{\mathop{\times}\limits_{#1}}}
  \def\tensor#1.{\mathrel{\mathord{\otimes}_{#1}}}
  \def\B{\Theta}
  \def\C{\Mb{C}}
  
  \def\N{\Mb{N}}
  
  \def\R{\Mb{R}}
  
  \def\T{\Mb{T}}
  \def\Z{\Mb{Z}}
  \def\Cn{{\bf V}}

  \def\wt{\widetilde}
\def\ilim{\mathop{{\rm lim}}\limits_{\longrightarrow}}
\def\cross #1.{\mathrel{\raise 3pt\hbox{$\mathop\times\limits_{#1}$}}}
\def\crossr #1.{\mathrel{\raise 1pt\hbox{$\mathop\rtimes\limits_{#1}$}}}
\def\maprightu #1;{\mathrel{\smash{\mathop{\longrightarrow}\limits^{#1}}}}
\def\maprightb #1;#2.{\mathrel{\smash{\mathop{\longrightarrow}\limits_{#2}^{#1}}}}
\def\s #1.{_{\smash{\lower2pt\hbox{\mathsurround=0pt $\scriptstyle #1$}}\mathsurround=5pt}}
\def\cl #1.{{\cal #1}} \def\al #1.{{\cal #1}}
\def\ker{{\rm Ker}\,}
\def\aut{{\rm Aut}\,}
\def\gau{{\rm Gau}\,}
\def\gaud{{\rm Gau}_d\,}
\def\gaue{{\rm Gau}^e\,}
\def\gaued{{\rm Gau}_d^e\,}

\def\Lgau{{\got{gau}}\,}

\def\gauc{{\rm Gau}\,}

\def\f #1,#2.{\mathsurround=0pt \hbox{${#1\over #2}$}\mathsurround=5pt}

\def\ol#1.{\overline{#1}}

\def\CAR #1.{{{\rm CAR}(#1)}}

\def\fl #1,#2.{{\mathord{\rm Fl}^{#1}_{#2}}}

\def\XP#1!{\renewcommand{\baselinestretch}{.7}\marginpar{{\footnotesize
$\leftarrow$#1}\hfil}\renewcommand{\baselinestretch}{1.5}}
\def\f #1,#2.{\mathsurround=0pt \hbox{${#1\over #2}$}\mathsurround=5pt}
   
\def\set #1,#2.{\left\{\,#1\;\bigm|\;#2\,\right\}}

\newcommand{\Aut}{\mathop{{\rm Aut}}\nolimits}

\newcommand{\into}{\hookrightarrow}
\renewcommand{\:}{\colon}

\newcommand{\supp}{\mathop{{\rm supp}}\nolimits}

\renewcommand{\hat}{\widehat}
\renewcommand{\tilde}{\widetilde}

\newcommand{\la}{\langle}
\newcommand{\ra}{\rangle}
\newcommand{\1}{\mathbf{1}}

\def\slim{\mathop{\hbox{\rm s-lim}}}
\def\bn{{\bf n}}

\begin{document}


\title{\bf QCD on an infinite lattice}

\author{
  {\sc Hendrik Grundling}                                            \\[1mm]
 {\footnotesize Department of Mathematics,}                         \\
 {\footnotesize University of New South Wales,}                      \\
 {\footnotesize Sydney, NSW 2052, Australia.}                             \\
 {\footnotesize hendrik@maths.unsw.edu.au}                  \\
 {\footnotesize FAX: +61-2-93857123} \\
\and
 {\sc  Gerd Rudolph }                                            \\[1mm]
 {\footnotesize Institut f\"ur Theoretische Physik,}                         \\
 {\footnotesize Universit\"at Leipzig,}                      \\
 {\footnotesize Postfach 100 920, D-4109 Leipzig.}                             \\
 {\footnotesize rudolph@rz.uni-leipzig.de}                  \\
 {\footnotesize FAX: +49-341-9732548}}

\maketitle


\begin{abstract}
We construct a mathematically well--defined framework for the kinematics of Hamiltonian
QCD on an infinite lattice in $\R^3$, and it is done in a C*-algebraic context.
This is based on the finite lattice model for Hamiltonian QCD developed by
Kijowski, Rudolph e.a.  (cf.~\cite{KR, KR1}). To extend this model to
an infinite lattice,  we need to take an infinite tensor product
of nonunital C*-algebras, which is a nonstandard situation. We use a recent construction
for such situations, developed in \cite{GrNe}. Once the field C*-algebra is constructed
for the fermions and gauge bosons,
we define local and global gauge transformations, and identify the Gauss law constraint.
The full field algebra is  the crossed product of the previous one with the
local gauge transformations.
The rest of the paper is concerned with
enforcing the Gauss law constraint to obtain the C*-algebra of quantum observables.
For this, we use the method of enforcing quantum constraints developed by Grundling and Hurst
(cf.~\cite{GrSrv}). In particular, the natural inductive limit structure
of the field algebra is a central component of the analysis, and the constraint system
defined by the Gauss law constraint is a system of local constraints in the sense of
\cite{GL}. Using the techniques developed in that area, we solve the full constraint system
by first solving the finite (local) systems and then combining the results appropriately.
We do not consider dynamics.
\end{abstract}

\newpage

\vspace{0.5cm}

\tableofcontents

\newpage


\setcounter{equation}{0}
\section{Introduction}


QCD is an important component of the standard model, and
the explicit construction of a field C*-algebra for it is still an unsolved problem in
mathematical physics. The construction of a field algebra
is a kinematics problem and it precedes the hard problem of dynamics, which
involves interactions, so it seems more tractable.
There is a deep body of theory developed for the locality properties of the field algebras of
quantum field theories in space-time (cf.~\cite{Haag92} for a survey), and of course
any explicitly constructed field algebra of this system must be consistent with that.
There is also extensive work on the Hamiltonian model of a Fermion in a
nonabelian {\it classical} gauge potential in $\R^3$, cf.~\cite{CR, La1, Mic1},
and it leads to interesting proposals for the field algebra of the fully quantized model
\cite{Han}.

Thus far, the only explicit rigorous constructions of appropriate field algebras for QCD have been for lattice approximations
of Hamiltonian QCD in $\R^3$ cf.~\cite{KR, KR1}.
Unfortunately due to a technical problem explained below, these models have been
confined to finite lattices. This is the main problem which we want to address here,
i.e. we want to construct the field C*-algebra for QCD
 on an infinite lattice in $\R^3$.
Using this field algebra, we then want to define gauge transformations
and solve the Gauss law constraint, hence identifying the physical observables.

More specifically;- for the model of QCD on a finite lattice developed by Kijowski, Rudolph e.a.
{\cite{KR, KR1}}, one finds that the field algebra is isomorphic to the
algebra of compact operators $\cl K.(\cl H.)$ on a separable infinite dimensional Hilbert space $\cl H.$.
As this has (up to unitary equivalence) only one irreducible representation, one obtains
a generalized von Neumann uniqueness theorem for the system.
For an infinite lattice, when passing to infinitely many
degrees of freedom, one has to expect inequivalent representations.
Explicitly, for the gauge part of the algebra, one needs to take an infinite tensor product of the algebras
associated to the links of the lattice (these are also isomorphic to $\cl K.(\cl H.)$).
This means that the standard theory for infinite tensor products does not apply.
However, there is a little-known definition for an infinite tensor product of nonunital algebras developed by
Blackadar cf.~\cite{Bla2}, which however has some drawbacks.
Recently this approach was further developed by Grundling and Neeb in \cite{GrNe}, where an infinite tensor product of
nonunital $C^*$-algebras was constructed
which has good representation properties. This is what we use for our construction of the field algebra of our model,
and as expected, this new field algebra has many inequivalent representations.

Once we have the field algebra of our model, we can define (local and global) gauge transformations,
extend the field algebra to include the implementers of these,
and identify the Gauss law constraint.
 Enforcement of quantum constraints is not a simple matter,
in fact compared with Quantum Electrodynamics (cf.~\cite{KRT, KRS}),  the analysis of the Gauss law is much more complicated.
This is due to the fact that in QCD the Gauss law constraint is neither built from gauge invariant operators nor is it linear
in the gauge connection fields.
Here we use the general method of enforcing quantum constraints developed by Grundling and Hurst
(the T--procedure, cf.~\cite{GrSrv}). It is crucial for this, that the constraint system
defined by the Gauss law constraint is a system of local constraints in the sense of
Grundling and  Lled\'o~\cite{GL}. This allows us to  solve the full constraint system
by first solving the finite (local) systems and then combining the results appropriately.
This method of constraint enforcement needs no gauge fixing (i.e. the
selection of one representative in each
gauge orbit) hence the Gribov problem does not occur.

Finally, we will discuss some of the types of physical observables
which occurred in  the historical papers of Kogut and Susskind \cite{K,KS}
and show how to fit these into our framework. Some of these observables, e.g.  Casimirs built from the colour electric fields,
are unbounded, so they are not in any $C^*$-algebra. However for a finite lattice, the
colour electric fields are closely related to our field C*-algebra, and we show this link
concretely in Subsection~\ref{PM}. More abstractly, they generate part of the field
C*-algebra in the sense of  Woronowicz, cf. Example 3 in Sect 3 of~\cite{unb}.

In this paper, we do not consider boundary effects, and we postpone colour charge
analysis to a separate project.
Boundary effects were analyzed for finite lattice systems by Kijowski and Rudolph in \cite{KR}, where
it was shown that from the local Gauss equation one can  extract a gauge invariant, additive law for operators with
eigenvalues in ${\mathbb Z}_3$. As in QED, this
implies a gauge invariant conservation law:- the
global ${\mathbb Z}_3$-valued colour charge is equal to a ${\mathbb Z}_3$-valued gauge invariant quantity obtained from the
color electric flux ``at infinity". The discussion of the boundary data yielding this flux is a subtle task, see \cite{KR1}.

Our paper is organized as follows. We start Sect.~\ref{FieldAlgebra} with a statement of
our initial assumptions,
and in Sect.~\ref{PM} we discuss the underlying models on which our analysis is based.
In Sect.~\ref{FAlg}    we construct the Fermion algebra for the full lattice. In Sect.~\ref{GFA}
we define for each link the field algebra for the gauge connection, recall the method developed in
\cite{GrNe}, and then use it to construct an infinite tensor product of the link algebras.
We then take the tensor algebra of this gauge field algebra with the fermion algebra,
and consider a natural inductive structure of it in  Sect.~\ref{FullFA}.
We call it the  kinematic field algebra.
In Sect.~\ref{LocalGT} we define the action of the local gauge transformations on the
kinematic algebra, and in Sect.~\ref{GlobalGT} we do this for global gauge transformations.
This requires us to choose a gauge invariant approximate identity
in the link algebras, and we analyze this issue.
In Sect~\ref{FFAlg} we construct the full field algebra as a (discrete) crossed product of
the kinematic algebra with the local gauge transformations. This contains all the relevant information of
the system, and in Sect.~\ref{GaussLaw} we define the local Gauss law constraint.
The rest of the paper is dedicated to the enforcement of this constraint.
We first review  in Sect.~\ref{GL-Tproc} the T-procedure of enforcing
constraints (cf.~\cite{GrSrv}), and show how the current constraint system fits into it.
In Sect.~\ref{FinLat-Tproc} we solve the constraint system for a finite lattice
in terms of the T-procedure. These results are used in Sect.~\ref{LocC-syst}
to solve the constraint system for the local algebras in the inductive limit of the
full field algebra. Finally, in Sect.~\ref{GL-Tprfull} we show that the full system of constraints
is a system of local quantum constraints in the sense of~\cite{GL}.
Using techniques from~\cite{GL} we then solve the constraint system fully for the local observables,
but for global observables the constraining remains unresolved.
In Sect.~\ref{ObsPh} we consider how standard observables from physics fit into our algebra.
There is an appendix to state a result on
constraint subsystems which we need.


\setcounter{equation}{0}
\section{The Kinematic Field Algebra}
\label{FieldAlgebra}

%
%

We consider a model for QCD in the Hamiltonian framework on an infinite regular
cubic lattice in $\Z^3.$
For basic notions concerning lattice gauge
theories including fermions, we refer to \cite{Seiler} and
references therein.

We first fix notation.
For the lattice, define a triple $\Lambda:=(\Lambda^0, \Lambda^1, \Lambda^2)$ as follows:
\begin{itemize}
\item{}
 $\Lambda^0:=\{(n,m,r)\in\R^3\,\mid\,n,\,m,\,r\in\Z\}\cap X$ where 
$X$ is an open connected set in $\R^3.$ Thus $\Lambda^0$ is a unit cubic lattice
(possibly infinite) contained in $X,$ and its elements are called sites.
\item{} Let $\wt\Lambda^1$ be the set of all directed edges (or links) between nearest neighbours, i.e.\\
$\wt\Lambda^1:= \{(x,y)\in\Lambda^0\times\Lambda^0\,\mid\,y=x\pm \b e._i\;\;\hbox{for some $i$}\}\;\;$
where the $\b e._i\in\R^3$ are the standard unit basis vectors.  Define a map $\eta:\wt\Lambda^1\to\cl P.(\Lambda^0)\equiv$ power set of
$\Lambda^0,$ by $\eta((x,y)):=\{x,y\},$ i.e. it is the map which ``forgets'' the orientation of links, then
$\Lambda^1$ will denote a choice of orientation of $\wt\Lambda^1,$ i.e. it is a section of $\eta,$ i.e. for each
$\{x,y\}\in\eta(\wt\Lambda^1)$ it contains either $(x,y)$ or $(y,x)$ but not both. Thus the pair $(\Lambda^0,\Lambda^1)$
is a directed graph, and we assume that it is connected.
\item{} Let $\wt\Lambda^2$ be the set of all directed faces (or plaquettes) of the unit cubes comprising the lattice i.e.\\
$\wt\Lambda^2:= \{(\ell_1,\ell_2,\ell_3,\ell_4)\in\big(\wt\Lambda^1\big)^4\,\mid\,Q_2\ell_i=Q_1\ell_{i+1}\;
\hbox{for}\;i=1,2,3,\;\hbox{and}\; Q_2\ell_4=Q_1\ell_1\}$\\
where $Q_i:\Lambda^0\times\Lambda^0\to\Lambda^0$ is the projection onto the $i^{\hbox{th}}$  component.
Note that for a plaquette $p=(\ell_1,\ell_2,\ell_3,\ell_4)\in\wt\Lambda^2,$ it has an orientation given by the order of the edges,
and there is only one other orientation in $\wt\Lambda^2$ for the same face, i.e. the reverse ordering
$\overline{p}=(\overline\ell_4,\overline\ell_3,\overline\ell_2,\overline\ell_1)$ where
$\overline\ell={(y,x)}$ if $\ell={(x,y)}\in \wt\Lambda^1$.
In analogy to the last point, we let
$\Lambda^2$ be a choice of orientation in $\wt\Lambda^2$.
\item{} Sometimes we need to identify the elements of $\Lambda^i$ with subsets of $\R^3$ and we will make the natural
identifications, e.g. a link $\ell=(x,y)\in\Lambda^1$ is the undirected closed line segment from $x$ to $y$.
\end{itemize}
Below, in Subsection~\ref{FAlg} onwards,
the set $X$ will play no role,  so that one may just consider the lattice $\Lambda^0=\Z^3$.
If one wants to analyze surface effects, $X$ will become  important.

For a classical matter field on $\R^3$  with a classical gauge
connection field acting on it, we describe its usual approximation on the lattice $\Lambda^0$.
First, let $G$ be a connected, compact Lie group (the gauge structure group),
and let ${\big(\Cn,\,(\cdot,\cdot)_\Cn\big)}$ be a finite dimensional complex
Hilbert space  (the space of internal degrees of freedom of the matter field)
on which $G$ acts smoothly as unitaries, so we take $G\subset U(\Cn)$.

The classical matter fields on $\R^3$  are the elements of $C^\infty(\R^3,\Cn).$
Its lattice approximation is
given by restricting $C^\infty(\R^3,\Cn)$ to the lattice $\Lambda^0$. This produces $\prod\limits_{x\in\Lambda^0}\Cn$
as the classical configuration space for the matter field on the lattice.

Given a connection on the trivial principal bundle $P:=\R^3\times G$, then parallel transport in $P$ along a curve in $\R^3$
is given by left multiplication by an element of $G.$ It is natural to model the connection on the lattice by
associating to each link $\ell\in\Lambda^1$ an element of $G$ (the parallel transporter along the link).
So the lattice approximation
of a classical connection is a map $\Phi:\Lambda^1\to G.$
Thus our classical configuration space for the connections is $\prod\limits_{\ell\in\Lambda^1}G,$
hence its phase space is $\prod\limits_{\ell\in\Lambda^1}T^*G\cong \prod\limits_{\ell\in\Lambda^1}(G\times\mathfrak{g}^* ).$

 The gauge group $C^\infty(\R^3,G)$ acts on  $C^\infty(\R^3,\Cn)$ by pointwise multiplication,
hence this restricts on $\Lambda^0$ to the group $\prod\limits_{x\in \Lambda^0}G=G^{\Lambda^0}=\big\{\gamma:\Lambda^0\to G\big\}$.
This group is the lattice approximation of the local gauge group.
Its pointwise action on the elements of $\prod\limits_{x\in\Lambda^0}\Cn$ produces the transformation law for the matter field.
Moreover, a gauge transformation $\gamma \in\prod\limits_{x\in \Lambda^0}G=\big\{\gamma:\Lambda^0\to G\big\}$
acts on a connection $\Phi:\Lambda^1\to G$ by
\begin{equation}
\label{GTr-classGP}
\big(\gamma\cdot\Phi)(\ell)=
\gamma(x_\ell)\,\Phi(\ell)\,\gamma(y_\ell)^{-1}
\qquad\hbox{for all}\qquad \ell=(x_\ell,y_\ell)\in\Lambda^1.
\end{equation}
This follows from the transformation law of the parallel transporter under
vertical automorphisms of the bundle $P$.

To summarize, the classical configuration space is $\big(\prod\limits_{x\in\Lambda^0}\Cn\big)\times
\big(\prod\limits_{\ell\in\Lambda^1}G\big)$, and the local gauge group $\prod\limits_{x\in \Lambda^0}G$ acts on it  by
\begin{equation}
\label{ClassGTr}
\big(\prod_{x\in\Lambda^0}v_x\big)\times\big(\prod_{\ell\in\Lambda^1}g_\ell\big)\mapsto
\big(\prod_{x\in\Lambda^0}\gamma(x)\cdot v_x\big)\times\big(\prod_{\ell\in\Lambda^1}\gamma(x_\ell)\,g_\ell\,\gamma(y_\ell)^{-1}\big)
\end{equation}
where  $ \ell=(x_\ell,y_\ell)$ and $\gamma \in\prod\limits_{x\in \Lambda^0}G$.
Note that the  orientation of links in $\Lambda^1$ was used in the action
because it treats the $x_\ell$ and $y_\ell$ differently.

This is the basic classical kinematical model for which we want to
obtain the quantum counterpart.


\subsection{The finite lattice model.}
\label{PM}

In this section we want to state the model for finite lattice approximation
of Hamiltonian QCD in $\R^3$ developed by Kijowski, Rudolph e.a.~\cite{KR, KR1}.
Briefly, given a finite lattice $\Lambda^0$, it quantizes the classical model on  $\Lambda^0$ above, by replacing
for each lattice site $x\in\Lambda^0$, the classical matter configuration space $\Cn$ with
the algebra for a fermionic particle on $\Cn$
(the quarks), and for each link $\ell\in\Lambda^1$ we replace the classical connection configuration space
$G$ by an algebra which describes
a bosonic particle on $G$ (the gluons). We state the model first concretely, in terms of operators
on a Hilbert space, and then we give an appropriate C*-algebra which can reproduce it.

Equip the space of classical matter fields $\prod\limits_{x\in\Lambda^0}\Cn=\{f:\Lambda^0\to\Cn\}$ with the natural pointwise inner product
$\la f, h \ra=\sum\limits_{x\in\Lambda^0}\big({f(x)},\, h(x)\big)_{\Cn}$, and take for the quantized
matter fields the CAR-algebra ${\mathfrak F}_{\Lambda} :=  {\rm CAR}\big(\prod\limits_{x\in\Lambda^0}\Cn)$.
That is, for each classical matter field $f\in\prod\limits_{x\in\Lambda^0}\Cn$,
we associate a fermionic field $a(f)\in{\mathfrak F}_{\Lambda}$, and these
satisfy the usual CAR--relations:
\[
\{a(f), a(h)^*\}  = \la f, h \ra \1
\quad \hbox{ and } \quad  \{a(f),a(h)\} = 0
\quad \mbox{ for } \quad f, h\in \prod\limits_{x\in\Lambda^0}\Cn
\]
where  $\{A,B\} := AB + BA$ and ${\mathfrak F}_{\Lambda}$ is generated by the set of all $a(f)$.
As $\Lambda^0$ is finite,
${\mathfrak F}_{\Lambda}$ is a full matrix algebra, hence up to unitary equivalence
it has only one irreducible representation.

In physics notation,  the quark at $x$ is given by
$a(\delta_x{\bf v}_i)=\psi_i(x)$ where $\{{\bf v}_i\mid i=1,\ldots n\}$ is an orthonormal basis for $\Cn$
and $\delta_x:\Lambda^0\to\R$ is the characteristic function of $\{x\}$.
More elaborate indices may be included, e.g.
if $\Cn = \b W.\otimes\C^k$ where $\b W.$ has the non--gauge degrees of freedom, and $\C^k$ has the gauge degrees of freedom.
In particular, there is a smooth irreducible unitary action of the structure group $G$ on $\C^k$ (if  $G=SU(3)$ we take $k=3$)
which produces a smooth unitary action of $G$ on $\Cn$. If $\{w_1,\ldots,w_m\}$ is an orthonormal basis of $\b W.$
and  $\{e_1,\ldots,e_k\}$ is an orthonormal basis of $\C^k$, then w.r.t. the orthonormal basis $\{w_j\otimes e_b\mid j=1,\ldots,m,\,
b =1,\ldots,k\}$ of $\Cn$, we obtain the usual physics indices
\[
a(w_j\otimes e_b\cdot\delta_x)=:\psi_{jb}(x)
\]
for the quark field generators. Clearly ${\mathfrak F}_{\Lambda}$ is generated as a C*-algebra
by the set of components $\{\psi_{jb}(x)\mid j=1,\ldots,m,\,
b =1,\ldots,k,\,x\in\Lambda^0\}.$

Next, to quantize the classical gauge connection fields $\prod\limits_{\ell\in\Lambda^1}G,$
consider first for a single link $\ell$ how to produce a  bosonic particle on $G$.
In the case that the configuration space is $\R=G$, then in the Schr\"odinger representation on $L^2(\R)$ we have the usual
position and momentum operators $Q$ and $P$. As these are unbounded operators, we consider (equivalently) the bounded
operators
\[
(U_t\varphi)(s):=\varphi(s-t)\quad\hbox{and}\quad\big(T_f\varphi)(s):=f(s)\varphi(s)\quad\hbox{for}\quad
\varphi\in L^2(\R),
\]
$s,\,t\in\R$ and $f\in L^\infty(\R)$ from which $Q$ and $P$ can be reconstructed.
Note in particular that $P$ is the generator of translations, which suggest that for
a  bosonic particle on $G$ its momentum should be a generator of (left) translations w.r.t. $G$.
Thus we consider a generalized Schr\"odinger representation on $L^2(G)$
given by
\begin{equation}
\label{GSRp}
(U_g\varphi)(h):=\varphi(g^{-1}h)\quad\hbox{and}\quad\big(T_f\varphi)(h):=f(h)\varphi(h)\quad\hbox{for}\quad
\varphi\in L^2(G),
\end{equation}
$g,\,h\in G$ and $f\in L^\infty(G)$, and it is irreducible in the sense that the commutant
of ${U_G\cup T_{L^\infty(G)}}$ consists of the scalars.
This is the natural concrete quantization for a quantum particle with phase space $T^*G$
which we will assume below.
Note that there is a natural ground state unit vector $\psi_0\in L^2(G)$ given by the constant
function $\psi_0(h)=1$ for all $h\in G$ (assuming that the Haar measure of $G$ is normalized).
 Then $U_g\psi_0=\psi_0$, and $\psi_0$
is cyclic w.r.t. the *-algebra generated by ${U_G\cup T_{L^\infty(G)}}$ (by irreducibility).

 The generalized canonical commutation relations are obtained from the
intertwining relation $U_gT_fU^*_g=T\s\lambda_g(f).$ where
\begin{equation}
\label{leftact}
\lambda:G\to\aut C(G)\, , \quad
 \lambda_g(f)(h) := f(g^{-1} h) \quad\hbox{for}\quad g,\,h\in G
\end{equation}
is the usual left translation. In particular, given $X\in\mathfrak{g}$, define
its associated momentum operator
\begin{eqnarray*}
P_X:C^\infty(G)\to C^\infty(G)\quad&\hbox{by}&\quad
P_X\varphi:=i{d\over dt}U(e^{tX})\varphi\Big|_{t=0}\,.\\[1mm]
\hbox{Then}\qquad
\big[P_X,\,T_f\big]\varphi
&=&i{d\over dt}U(e^{tX})T_fU(e^{-tX})\varphi\Big|_{t=0}
=i{d\over dt}\,T\s\lambda_{\exp(tX)}(f).\varphi\Big|_{t=0}\\[1mm]
&=& iT\s X^R(f). \varphi\qquad\hbox{for}\quad
f,\,\varphi\in C^\infty(G),
\end{eqnarray*}
where $X^R\in\ot X.(G)$ is the associated right-invariant vector field.
(Note that whilst the differential is defined w.r.t. the $L^2\hbox{--topology,}$
as all functions are smooth and of compact support, we may use the pointwise differential).
As $P_X=dU(X)$, it defines  a representation of the Lie algebra $\mathfrak{g}$,
and clearly $P_X\psi_0=0$.

To identify the quantum connection $\Phi(\ell)$ at link $\ell$ in this context,
use the irreducible action of the structure group $G$ on $\C^k$ to define the function $\Phi_{ij}(\ell)\in C(G)$ by
\begin{equation}
\label{Def-QuConn}
\Phi_{ij}(\ell)(g):=(e_i,ge_j),\quad g\in G,
\end{equation}
 using  the orthonormal basis $\{ e_i\mid
i =1,\ldots,k\}$ of $\C^k$. Then  the matrix components of the quantum connection are
taken to be the operators $T_{\Phi_{ij}(\ell)}$,
which we will see transform correctly w.r.t. gauge transformations.
As the $\Phi_{ij}(\ell)$ are matrix elements of elements of $G$, there are obvious
relations between them which reflect the structure of $G$.
Note that the algebra generated by the functions $\Phi_{ij}(\ell)$ (w.r.t. pointwise operations)
separates the points in $G$, hence by the Weierstrass theorem, it is a dense subalgebra of $C(G)$.
So the C*-algebra generated by the operators $\{ T_{\Phi_{ij}(\ell)}\,\mid\, i,\,j=1,\ldots,k\}$
is $T\s C(G).$.

To define gauge momentum operators, we first assign to each link an element of $\mathfrak{g}$,
i.e. we choose a map $\Psi:\Lambda^1\to \mathfrak{g}.$ Given such a $\Psi$, it is then natural to take
for the associated quantum gauge momentum at $\ell$, the operator
$P\s\Psi(\ell).:C^\infty(G)\to C^\infty(G)$. To obtain the generalized canonical commutation relations
for these, recall that $U_gT_fU^*_g=T\s\lambda_g(f).$ and
\[
{\lambda_g\big(\Phi_{ij}(\ell)\big)(h)} = (e_i,g^{-1} he_j)
=\sum_m(e_i,g^{-1}e_m)(e_m,he_j)\quad  
\hbox{for}\quad g,\,h\in G.
\]
If $g=\exp(t\Psi(\ell))$ for $\Psi(\ell)\in{\mathfrak g}$, $t\in\R$, then
for $\varphi\in C^\infty(G)$ we have
\[
\Big(\big[P\s\Psi(\ell).,\,T\s\Phi_{ij}(\ell).\big]\varphi\Big)(h)=i{d\over dt}\sum_m(e_i,g^{-1}e_m)(e_m,he_j)\,\varphi(h)\big|_{t=0}
=\sum_m\Psi(\ell)_{im}\Phi_{mj}(\ell)(h)\varphi(h)
\]
where $\Psi(\ell)_{im}:=(e_i,\Psi(\ell)e_m).$ Thus the generalized canonical commutation relations are
\begin{equation}
\label{genCCR}
\big[P\s\Psi(\ell).,\,T\s\Phi_{ij}(\ell).\big]=\sum_mT\s\Psi(\ell)_{im}\Phi_{mj}(\ell).\quad\hbox{on}\quad C^\infty(G).
\end{equation}
To obtain the G--electrical fields at $\ell$, we choose an appropriate basis
$\{Y_r\mid r=1,\ldots,{\rm dim}( \mathfrak{g})\}\subset \mathfrak{g}$
and substituting for $\Psi$ the constant map $\Psi(\ell)=Y_r$ we set
$E_r(\ell):=P_{Y_r}$. In the case that $G=SU(3)$, these are the colour electrical fields,
and one takes  the basis $\{Y_r\}$ to be the traceless selfadjoint Gell--Mann matrices
satisfying $Y_rY_s=\delta_{rs}$. We then define
\[
E_{ij}(\ell):=\sum_r(Y_r)_{ij}E_r(\ell)=\sum_r(Y_r)_{ij}P\s{Y_r}.
\]
and for these we obtain  from $(\ref{genCCR})$
the commutation formulii in \cite{KR, KR1} for the colour electrical field.
As the set of $E_{ij}(\ell)$ span all of $P_{\mathfrak{g}}$, it is clear that the unitary group they generate is
all of $U_G\subset M(C^*(G))$. From  Example 3 in Sect 3 of~\cite{unb} and \cite{WN}, we
also see that they generate $C^*(G)$ in the sense of  Woronowicz,
which is a component of our field C*-algebra for a link, $C(G)\rtimes_\lambda G$, below.

The full collection of operators which comprise the dynamical
variables of the model are as follows. The representation Hilbert space is
$\cl H.=\cl H._F\otimes\mathop{\bigotimes}\limits_{\ell\in\Lambda^1}L^2(G)$ where
${\pi_F:{\mathfrak F}_{\Lambda}}\to\cl B.(\cl H._F)$ is any
irreducible representation of ${\mathfrak F}_{\Lambda}$.
As $\Lambda^1$ is finite, $\cl H.$ is well--defined. Then
$\pi_F\otimes\un:{\mathfrak F}_{\Lambda}\to\cl B.(\cl H.)$ will be the action of ${\mathfrak F}_{\Lambda}$
on $\cl H.$. The quantum connection is given by the set of operators
\[
\{\hat{T}_{\Phi_{ij}(\ell)}^{(\ell)}\,\mid\,\ell\in \Lambda^1,\;i,j=1,\ldots, k\}\quad
 \hbox{where}\quad \hat{T}_f^{(\ell)}:=\un\otimes\big(\un\otimes\cdots\otimes\un\otimes T^{(\ell)}_f\otimes\un\big)
\]
and $T^{(\ell)}_f$ is the multiplication operator on the $\ell^{\rm th}$ factor, hence $\hat{T}_{\Phi_{ij}(\ell)}^{(\ell)}$
acts as the identity on all the other factors of $\cl H.$. Likewise, for the gauge momenta we take
\[
\hat{P}^{(\ell)}_{\Psi(\ell)}:=\un\otimes\big(\un\otimes\cdots\otimes\un\otimes P^{(\ell)}_{\Psi(\ell)}\otimes\un\big),\quad
\ell\in \Lambda^1
\]
where $P^{(\ell)}_X$ is the $P_X$ operator on the subspace $C^\infty(G)\subset L^2(G)$ of the $\ell^{\rm th}$ factor.
These are obviously generators of $\hat{U}_g^{(\ell)}:=\un\otimes\big(\un\otimes\cdots\otimes\un\otimes{U}_g^{(\ell)}\otimes\un\big)$
when $g=\exp(t\Psi(\ell))$
where $U^{(\ell)}_g$ is the $U_g$ operator on the $\ell^{\rm th}$ factor.
Thus the quantum G--electrical field $\hat{E}_r$ is a map from  $\Lambda^1$ to operators on
the dense domain $\cl H._F\otimes\mathop{\bigotimes}\limits_{\ell\in\Lambda^1}C^\infty(G)$, given by
$\hat{E}_r(\ell):=\hat{P}^{(\ell)}_{Y_r}$.

Next, we want to define gauge transformations. Recall from Equation~(\ref{ClassGTr}) that
the local gauge group is $\gauc \Lambda=\prod\limits_{x\in \Lambda^0}G=\{\gamma:\Lambda^0\to G\}$, and it acts on the
classical configuration space  $\big(\prod\limits_{x\in\Lambda^0}\Cn\big)\times
\big(\prod\limits_{\ell\in\Lambda^1}G\big)$  by
\[
\big(\prod_{x\in\Lambda^0}v_x\big)\times\big(\prod_{\ell\in\Lambda^1}g_\ell\big)\mapsto
\big(\prod_{x\in\Lambda^0}\gamma(x)\cdot v_x\big)\times\big(\prod_{\ell\in\Lambda^1}\gamma(x_\ell)\,g_\ell\,\gamma(y_\ell)^{-1}\big)
\]
where  $ \ell=(x_\ell,y_\ell)$ and $\gamma \in\prod\limits_{x\in \Lambda^0}G$.
Thus for the Fermion algebra we define an action
$\alpha^1:\gauc \Lambda \to\aut{\mathfrak F}_{\Lambda}$ by
\[
\alpha_\gamma^1(a(f)):=a(\gamma\cdot f)\qquad \hbox{where}\qquad (\gamma\cdot f)(x):={\gamma(x)}f(x)\quad \hbox{for all}\quad x\in\Lambda^0,
\;f\in\prod\limits_{x\in\Lambda^0}\Cn
\]
since $f\mapsto \gamma\cdot f$ defines a unitary on $\prod\limits_{x\in\Lambda^0}\Cn$
where  $\gamma\in\gauc \Lambda$. As ${\mathfrak F}_{\Lambda}$ has up to unitary equivalence
only one irreducible representation, it follows that ${\pi_F:{\mathfrak F}_{\Lambda}}\to\cl B.(\cl H._F)$ is
equivalent to the Fock representation, hence it is covariant w.r.t. $\alpha^1$, i.e. there is a
(continuous) unitary representation $U^F:\gauc \Lambda\to \cl U.(\cl H._F)$ such that
$\pi_F(\alpha^1_\gamma(A))=U^F_\gamma\pi_F(A)U^F_{\gamma^{-1}}$ for $A\in {\mathfrak F}_{\Lambda}$.

On the other hand,
if the classical configuration space  $G$ corresponds to
a link  $\ell=(x_\ell,y_\ell)$, then the gauge transformation is $\gamma\cdot g=
\gamma(x_\ell)\,g\,\gamma(y_\ell)^{-1}$ for all $g\in G$. Using this, we can
define a unitary $W_\gamma:L^2(G)\to L^2(G)$ by
\[
(W_\gamma\varphi)(h):=\varphi(\gamma^{-1}\cdot h)=\varphi(\gamma(x_\ell)^{-1}\,h\,\gamma(y_\ell))
\]
 using the fact that $G$
is unimodular, where the inverse was introduced to ensure that $\gamma\to W_\gamma$ is a homomorphism.
Note that $W_\gamma\psi_0=\psi_0$.
So for the
quantum observables ${U_G\cup T_{L^\infty(G)}}$, the gauge transformation becomes
\begin{equation}
\label{GTfell}
T_f\mapsto W_\gamma T_f W_\gamma^{-1}=T_{W_\gamma f}\quad\hbox{and}\quad
U_g\mapsto W_\gamma U_g W_\gamma^{-1}=U_{\gamma(x_\ell)g\gamma(x_\ell)^{-1}}
\end{equation}
for $f\in L^\infty(G)\subset L^2(G)$ and $g\in G$. Moreover each $W_\gamma$ preserves
the space $C^\infty(G)$, hence Equation~(\ref{GTfell}) also implies that
\[
 W_\gamma P_X W_\gamma^{-1}=P\s{\gamma(x_\ell)X\gamma(x_\ell)^{-1}}.\quad\hbox{for}\quad
 X\in\mathfrak{g}.
\]
Thus for the full system we define on $\cl H.=\cl H._F\otimes\mathop{\bigotimes}\limits_{\ell\in\Lambda^1}L^2(G)$
the unitaries
\[
\hat{W}_\gamma:=U^F_\gamma\otimes\big(\bigotimes_{\ell\in\Lambda^1}W^{(\ell)}_\gamma\big),\quad
\gamma\in\gauc \Lambda
\]
where $W^{(\ell)}_\gamma$ is the $W_\gamma$ operator on the $\ell^{\rm th}$ factor, then the gauge transformation
produced by $\gamma$ on the system of operators is given by ${\rm Ad}(\hat{W}_\gamma)$.

In particular, recalling $W_\gamma T\s\Phi_{ij}(\ell). W_\gamma^{-1}=T\s{W_\gamma \Phi_{ij}(\ell)}.$ we see that
\begin{eqnarray}
\big(W_\gamma \Phi_{ij}(\ell)\big)(g)&=&\Phi_{ij}(\ell)(\gamma(x_\ell)^{-1}\,g\,\gamma(y_\ell))
=\big(e_i,\gamma(x_\ell)^{-1}\,g\,\gamma(y_\ell)e_j\big)\nonumber\\[1mm]
\label{PhiTfs}
&=&\sum_{n,m}[\gamma(x_\ell)^{-1}]_{in}\,\Phi_{nm}(\ell)(g)\,[\gamma(y_\ell)]_{mj}
\end{eqnarray}
where $[\gamma(x_\ell)]_{in}=(e_i,\gamma(x_\ell)e_n)$ are the usual matrix elements, so
it is clear that the indices of the quantum connection $T_{\Phi_{ij}(\ell)}$ transform correctly
for the gauge transformation $\gamma^{-1}$.
This is consistent with the transformation of the fermions. To see this, recall
 that the map $f\mapsto a(f)$ is conjugate linear.
So if we make the identification $a(\delta_xe_i)=\psi_i(x)$ with the heuristic field
(assuming $\Cn=\C^k$,
otherwise $\Cn=\C^k\times{\bf W}$ and there are more indices), then we  obtain that under $\gamma(x)\in G$ it transforms by
\[
\psi_i(x)\to a(\delta_x \,\gamma(x)\cdot e_i)=\sum_j a(\delta_x {[\gamma(x)]_{ji}}e_j)=\sum_j \overline{[\gamma(x)]_{ji}}\, a(\delta_x e_j)
=\sum_j {[\gamma(x)^{-1}]_{ij}}\, \psi_j(x)
\]
where $[g]_{ji}=(e_j,ge_i)$ are the usual matrix elements, and we used the fact that $G$ is a subgroup of the unitary group.
 This is consistent with the transformation
of the connection in equation~(\ref{PhiTfs}) because it implies that
\begin{eqnarray*}
\sum_j\Phi_{ij}(\ell)\psi_j(y_\ell)&\mapsto&\sum_{n,m,j,k}[\gamma(x_\ell)^{-1}]_{in}\,
\Phi_{nm}(\ell)\,[\gamma(y_\ell)]_{mj}{[\gamma(y_\ell)^{-1}]_{jk}}\, \psi_k(y_\ell)\\[1mm]
&=&\sum_{n,m}[\gamma(x_\ell)^{-1}]_{in}\,
\Phi_{nm}(\ell)\, \psi_m(y_\ell)
\end{eqnarray*}
i.e. it transforms exactly like $\psi(x_\ell)$.
(We used the obvious short-hand of indicating the operator $T\s\Phi_{ij}(\ell).$ simply as $\Phi_{ij}(\ell)$.)

Finally, we wish to construct the appropriate C*-algebra for the field algebra of this model.
For the fermion part, we already have the C*-algebra ${\mathfrak F}_{\Lambda} =  {\rm CAR}\big(\prod\limits_{x\in\Lambda^0}\Cn)$,
so we only need to consider the appropriate C*-algebras for the link operators
${U_G,\; T_{L^\infty(G)}},\; P_{\mathfrak{g}}$ associated with each link $\ell$.

Fix a link $\ell$, hence a specific copy of $G$ in the configuration space.
Above in (\ref{leftact}) we had the distinguished action
 $\lambda:G\to\aut C(G)$ by
\[
\lambda_g(f)(h) := f(g^{-1} h) \, \, , \, \,
f \in C(G),\; g,h\in G.
\]
The generalized Schr\"odinger representation $(T,U)$ above in~(\ref{GSRp}) is a covariant representation
for the action $\lambda:G\to\aut C(G)$ so it is natural to take for our field algebra
 the crossed product C*-algebra $C(G)\rtimes_\lambda G $ whose representations are exactly the
 covariant representations of the $C^*$-dynamical system defined by $\lambda$.
The algebra $C(G)\rtimes_\lambda G $ is also called the generalised Weyl algebra, and it
 is well--known that
$C(G)\rtimes_\lambda G\cong\cl K.\big(L^2(G)\big)$ cf.~\cite{Rief} and
Theorem~II.10.4.3 in~\cite{Bla1}.
In fact $\pi_0\big(C(G)\rtimes_\lambda G\big)=\cl K.\big(L^2(G)\big)$
where $\pi_0:C(G)\rtimes_\lambda G\to L^2(G)$ is the generalized Schr\"odinger representation.
Since the algebra of compacts $\cl K.\big(L^2(G)\big)$
has only one irreducible representation up to unitary equivalence, it follows that
the generalized Schr\"odinger representation is the unique irreducible
covariant representation of $\lambda$ (up to equivalence). Moreover, as $\psi_0$ is cyclic for
$\cl K.\big(L^2(G)\big)$,  the
generalized Schr\"odinger representation is unitary equivalent to the GNS--representation
of the vector state $\omega_0$ given by $\omega_0(A):={(\psi_0,\pi_0(A)\psi_0)}$ for $A\in
C(G)\rtimes_\lambda G$.

Note that the operators $U_g$ and $T_f$
in equation~(\ref{GSRp}) are not compact, so they are not in $\cl K.\big(L^2(G)\big)=\pi_0\big(C(G)\rtimes_\lambda G\big)$,
but are in fact in its multiplier algebra. This is not a problem, as
a state or representation on $C(G)\rtimes_\lambda G$ has a unique extension to its multiplier algebra,
so will be fully determined on these elements.
If one chose $C^*(U_G\cup T_{L^\infty(G)})$ as the field algebra instead of
$C(G)\rtimes_\lambda G$, then this will contain many inappropriate representations, e.g.
 covariant representations
for $\lambda:G\to\aut C(G)$ where the implementing unitaries are discontinuous w.r.t. $G$.
Thus, our choice for the field algebra of a link remains as ${C(G)\rtimes_\lambda G}\cong\cl K.\big(L^2(G)\big)$.
Clearly, as the momentum operators $P_X$ are unbounded, they cannot be in any C*-algebra, but
they are obtained from $U_G$
in the generalized Schr\"odinger representation.

We combine these C*-algebras into the kinematic field algebra, which is
\[
{\mathfrak A}_{\Lambda} :=
{\mathfrak F}_{\Lambda} \otimes \bigotimes_{\ell\in\Lambda^1}\big(C(G)\rtimes_\lambda G\big)
\]
which is is well--defined as $\Lambda^1$ is finite, and the cross--norms are unique as all
algebras in the entries are nuclear. (If $\Lambda^1$ is infinite, the tensor product
${\mathop{\bigotimes}\limits_{\ell\in\Lambda^1}\big(C(G)\rtimes_\lambda G\big)}$ is undefined, as
$C(G)\rtimes_\lambda G$ is nonunital).
Moreover, since $C(G)\rtimes_\lambda G\cong\cl K.\big(L^2(G)\big)$ and
$\cl K.(\cl H._1)\otimes\cl K.(\cl H._2)\cong\cl K.(\cl H._1\otimes\cl H._2),$ it follows
that
\[
\mathop{\bigotimes}\limits_{\ell\in\Lambda^1}\big(C(G)\rtimes_\lambda G\big)\cong\cl K.\big(
\mathop{\otimes}\limits_{\ell\in\Lambda^1}L^2(G)\big)\cong\cl K.(\cl L.)
\]
 as $\Lambda^1$ is finite, where $\cl L.$ is a generic infinite dimensional separable Hilbert space.
So
\[
{\mathfrak A}_{\Lambda} ={\mathfrak F}_{\Lambda} \otimes\mathop{\bigotimes}\limits_{\ell\in\Lambda^1}\big(C(G)\rtimes_\lambda G\big)\cong
{\mathfrak F}_{\Lambda} \otimes\cl K.\big(\mathop{\otimes}\limits_{\ell\in\Lambda^1}L^2(G)\big)\cong\cl K.(\cl L.)
\]
as ${\mathfrak F}_{\Lambda}$ is a full matrix algebra. This shows that for a finite lattice
there will be only one irreducible representation, up to unitary equivalence. Also, ${\mathfrak A}_{\Lambda}$ is
simple, so all representations are faithful.

The algebra ${\mathfrak A}_{\Lambda}$
is faithfully and irreducibly represented on $\cl H.=\cl H._F\otimes\mathop{\bigotimes}\limits_{\ell\in\Lambda^1}L^2(G)$
by $\pi=\pi_F\otimes\big(\mathop{\bigotimes}\limits_{\ell\in\Lambda^1}\pi_{\ell}\big)$
where $\pi_\ell:C(G)\rtimes_\lambda G\to L^2(G)$ is the generalized Schr\"odinger representation for the
$\ell^{\rm th}$ entry. Then $\pi\big({\mathfrak A}_{\Lambda}\big)$ contains in its multiplier algebra the
operators $\hat{T}^{(\ell)}_{\Phi_{ij}(\ell)},\;\hat{U}^{(\ell)}_g$ for all $\ell\in \Lambda^1$.

To complete the picture, we also need to define the action of the local gauge group on ${\mathfrak A}_{\Lambda}$.
Recall that in $\pi$ it is given by $\gamma\to{\rm Ad}(\hat{W}_\gamma)$, and this clearly preserves
$\pi\big({\mathfrak A}_{\Lambda}\big)=\cl K.(\cl H.)$ and defines a strongly continuous action $\alpha$
of $\gauc \Lambda$ on $\pi\big({\mathfrak A}_{\Lambda}\big)$ (hence on ${\mathfrak A}_{\Lambda} $)
 as $\gamma\to\hat{W}_\gamma$ is strong operator continuous.
By construction ${(\pi,\hat{W}_\gamma)}$ is a covariant representation for the
C*-dynamical system given by  $\alpha:\gauc \Lambda \to\aut{\mathfrak A}_{\Lambda}$.
As $\gauc \Lambda=\prod\limits_{x\in \Lambda^0}G$ is locally compact, we can construct the crossed product
${\mathfrak A}_{\Lambda}\rtimes_\alpha \gauc \Lambda$ which has as representation space all covariant
representations of $\alpha:\gauc \Lambda \to\aut{\mathfrak A}_{\Lambda}$. As it is convenient to have an
identity in the algebra, our full field algebra for the system will be taken to be:
\[
\al F._e:=({\mathfrak A}_{\Lambda}\oplus\C)\rtimes_\alpha (\gauc \Lambda)
\]
where ${\mathfrak A}_{\Lambda}\oplus\C$ denotes ${\mathfrak A}_{\Lambda}$ with an identity adjoined.
Our first aim in the sections below is to extend these constructions to an infinite lattice.

\smallskip
\noindent {\bf Remark:}\\
\smallskip
The model written above goes back to the model constructed in the classical paper of
Kogut~\cite{K} (which elaborates the earlier one of Kogut and Susskind~\cite{KS}).
The fermions (quarks) at lattice points are treated in exactly the same way as above, when
we rewrite it in physics notation as indicated. Regarding the bosons (gluons)
on the links,
note that the generalized canonical commutation relations~(\ref{genCCR}) appear in Kogut~\cite{K}, if one
identifies $P\s\Psi(\ell).$ with a colour electric field on $\ell$ as above, and  $T\s\Phi_{ij}(\ell).$ with the
gluonic gauge field on $\ell$. Thus for a fixed link $\ell$, these operators produce a covariant representation
for the action $\lambda:G\to\aut C(G)$, hence it is a direct sum of copies of the generalized Schr\"odinger representation
above, since $C(G)\rtimes_\lambda G\cong\cl K.\big(L^2(G)\big)$.
However Kogut~\cite{K} assumes the same ground state $\psi_0$ as we do above (the defining relation is
that $P_X\psi_0=0$ for all $X$, hence $\psi_0$ must be a constant function w.r.t. translations in $G$).
As his representation space is constructed from observables applied to the vector  $\psi_0$,
his representation is the GNS-representation for $\omega_0$. This is unitarily equivalent to the
 generalized Schr\"odinger representation.
 We conclude that for a finite lattice, the operator theory in the  generalized Schr\"odinger representation
given above, is the appropriate mathematical framework for the lattice QCD model of Kogut~\cite{K}.

One may ask whether the complicated structure of the classical
configuration space of a gauge theory has an impact at all at the quantum level .
 There are hints that the stratified structure of this
space, see e.g. \cite{RSV, RSV1}, does show up on quantum level, see \cite{Huebsch,RS} for a case study.

\subsection{The Fermion algebra.}
\label{FAlg}

It is unproblematic to define the Fermion field on an infinite lattice $\Lambda=(\Lambda^0, \Lambda^1, \Lambda^2)$:
\begin{Definition}
\label{CAR1}
\sl Assume the quantum matter field algebra on $\Lambda$ is:
\begin{equation}
\label{fermifieldalgebra}
{\mathfrak F}_{\Lambda} :=  \CAR \ell^2(\Lambda^0,\Cn). =C^*\big(\mathop{\bigcup}_{x\in\Lambda^0}{\mathfrak F}_x\big)
\end{equation}
where ${\mathfrak F}_x:=\CAR V_x.$ and $V_x:=\{f\in\ell^2(\Lambda^0,\Cn)\,\mid\, f(y)=0\;\;\hbox{if}\;\; y\not=x\}\cong \Cn.$
We interpret ${\mathfrak F}_x\cong\CAR \Cn.$ as the field algebra for a fermion at $x.$
We denote the generating elements of $\CAR \ell^2(\Lambda^0,\Cn).$ by $a(f),$ $f\in\ell^2(\Lambda^0,\Cn),$ and these
satisfy the usual CAR--relations:
\begin{equation}
  \label{eq:car}
\{a(f), a(g)^*\}  = \la f, g \ra \1
\quad \hbox{ and } \quad  \{a(f),a(g)\} = 0
\quad \mbox{ for } \quad f, g \in \ell^2(\Lambda^0,\Cn)
\end{equation}
where  $\{A,B\} := AB + BA$.
\end{Definition}
Note that the odd parts of ${\mathfrak F}_x$ and ${\mathfrak F}_y$
w.r.t. the fields $a(f)$ anticommute if $x\not=y.$ Moreover, as $\Lambda^0$ is infinite,
${\mathfrak F}_{\Lambda}$ has inequivalent irreducible representations.

This  defines the quantum matter fields on the lattice sites, and as above, the correspondence with the physics notation is
$a(\delta_x{\bf v}_i)=\psi_i(x)$ where $\{{\bf v}_i\mid i=1,\ldots n\}$ is an orthonormal basis for $\Cn$
and $\delta_x:\Lambda^0\to\R$ is the characteristic function of $\{x\}$.


\subsection{The gauge field algebra.}
\label{GFA}


Following the discussion in  Subsection \ref{PM},  for every link $\ell\in\Lambda^1$
we will assume a generalised Weyl algebra  $C(G)\rtimes_\lambda G $
where $G$ is  our compact gauge group.  Since for the classical
connection field, the phase space is $\prod\limits_{\ell\in\Lambda^1}T^*G$,
it seems that for the quantum system we must
take  a tensor product $\mathop{\bigotimes}\limits_{\ell\in\Lambda^1}\big(C(G)\rtimes_\lambda G\big).$
In the case of a finite lattice as we saw, this is fine, and the C*-tensor norms are unique.
In the case of an infinite lattice, the situation is considerably different,
and we will expect  inequivalent representations when passing to infinitely many
degrees of freedom (as in quantum field theory).

First, note that since
 ${C(G)\rtimes_\lambda G}\cong\cl K.\big(L^2(G)\big)$ is nonunital, the standard theory for
infinite tensor products breaks down, i.e. an infinite tensor product of these is undefined.
The problem of infinite tensor products for nonunital C*-algebras is still relatively undeveloped,
in fact Takesaki states in~\cite{Tak3} on p84 that ``the
infinite tensor product of non-unital C*-algebras is not defined.''
There is however a little-known definition for an infinite tensor product of nonunital algebras developed by
Blackadar cf.~\cite{Bla2}, but this uses a choice of reference projections in the sequence,
and representations of the resultant C*-algebra, depends on the choice of projections.
Recently in \cite{GrNe}, extending Blackadar's construction,  an infinite tensor product of $\cl K.(\cl H.)$ was constructed
which has good representation properties w.r.t. a natural Weyl algebra in its multiplier algebra.
This is very close to the situation which we have here, so we will choose this
method of construction for the full bosonic field algebra. We describe the construction.
Further details, and proofs of the rest of the claims in this subsection can be found in \cite{GrNe}.

Observe first, that the representation theory of $\cl K.(\cl H.)$ (resp.
$\mathop{\bigotimes}\limits_{n=1}^k\cl K.(\cl H.)$)   is precisely
the regular representation theory of the Weyl algebra ${\rm CCR}(\R^2)\subset M(\cl K.(\cl H.)$
(resp. $\mathop{\bigotimes}\limits_{n=1}^k{\rm CCR}(\R^2)={\rm CCR}(\R^{2k})$ using minimal tensor norm)
hence we would expect that  the representation theory of
 ``$\mathop{\bigotimes}\limits_{n=1}^\infty\cl K.(\cl H.)$''
 (if this object is given a proper meaning) should be the
regular representations of the Weyl algebra $\mathop{\bigotimes}\limits_{n=1}^\infty{\rm CCR}(\R^2)$,
where the latter tensor product is well-defined as ${\rm CCR}(\R^2)$  is unital.
This is precisely what we have for the construction in \cite{GrNe}.

We start with Blackadar's construction~\cite{Bla2}.
Let $\cl L._n:=\cl K.(\cl H.)$, and choose a sequence of ``reference projections,'' i.e.
for each $n\in\N$, choose a nonzero projection
$P_n\in{\cal L}_n$. Define $C^*$-embeddings
$$\Psi_{\ell k}:{\cal L}^{(k)}\to
{\cal L}^{(\ell)}\qquad\hbox{by}\qquad
\Psi_{\ell k}(A_1\otimes\cdots\otimes A_k):=
A_1\otimes\cdots\otimes A_k\otimes P_{k+1}\otimes\cdots\otimes P_\ell, $$
where $k<\ell$ and ${\cal L}^{(k)}:=\mathop{\bigotimes}\limits_{n=1}^k{\cal L}_n$.
 Then the inductive limit makes sense, so we define
$$ {\cal L} := \bigotimes_{n=1}^\infty{\cal L}_n:=\ilim\big\{
{\cal L}^{(n)},\,\Psi_{\ell k}\big\}$$
and write $\Psi_k \: {\cal L}^{(k)} \to {\cal L}$ for the corresponding embeddings,
satisfying $\Psi_k \circ \Psi_{kj} = \Psi_j$ for $j \leq k$.
Since each ${\cal L}_n$ is simple, so are the finite tensor products
${\cal L}^{(k)}$ (\cite{WO93}, Prop.~T.6.25),
and as inductive limits of simple $C^*$-algebras
are simple (\cite{KR83}, Prop.~11.4.2), so is $\cl L.$.
It is also clear that $\cl L.$ is
separable, and it is nuclear as it is an inductive limit of nuclear algebras.

Since
$\Psi_{k+n,k}(L_k)=L_k\otimes P_{k+1}\otimes\cdots\otimes P_{k+n},$
where $L_k\in{\cal L}^{(k)}$, this means that we can consider ${\cal L}$
to be built up out of elementary tensors of the form
\begin{equation}
\label{elemTens}
\Psi_k(L_1 \otimes \cdots \otimes L_k)=
L_1\otimes L_2\otimes\cdots\otimes L_k\otimes P_{k+1}\otimes P_{k+2}\otimes\cdots\;,
\quad\hbox{where}\quad L_i\in{\cal L}_i\qquad
\end{equation}
i.e. eventually they are of the form $\cdots\otimes P_{k}\otimes P_{k+1}\otimes\cdots$.
We will use this picture below, and generally will not indicate the maps $\Psi_k\,.$
By componentwise multiplication, we can also identify elementary tensors
$\un\otimes\cdots\otimes\un\otimes P_{k}\otimes P_{k+1}\otimes\cdots$ in the multiplier algebra
$M(\cl L.).$
The representations $\pi$ of $\cl L.$ are well-behaved w.r.t. the reference sequence $\{P_k\}_{k=1}^\infty$
in the sense that
\[
\slim_{k\to\infty}\pi(\un\otimes\cdots\otimes\un\otimes P_{k}\otimes P_{k+1}\otimes\cdots)=\un\,,
\]
and this restricts the corresponding regular representations on
$\mathop{\bigotimes}\limits_{n=1}^\infty{\rm CCR}(\R^2)\subset M(\cl L.).$
Thus, if we do not want our representations to depend on the choice
of the reference sequence of projections,
we will need to go beyond a single Blackadar product $\cl L..$

As we saw, for every sequence of projections $P_k\in{\cal L}_k$ we obtained
a Blackadar product $\cl L..$ We now want to examine a collection of them, where our choices of
$P_k\in{\cal L}_k$ will ``fill out'' the full Hilbert space $\cl H..$

There is a (countable) approximate
identity $(E_n)_{n \in \N}$ in ${\cal K}(\cl H.)$ consisting of a strictly increasing
sequence of projections $E_n$ with $\dim (E_n\cl H.) <\infty$.
For each $k$, choose such an approximate identity $(E_n^{(k)})_{n \in \N}\subset{\cal L}_k= {\cal K}(\cl H.)$,
then for each sequence ${\bf n}=(n_1,\,n_2,\ldots)\in\N^\infty := \N^\N$,
we have a sequence of projections $\big(E_{n_1}^{(1)},\,E_{n_2}^{(2)},\ldots\big)$
from which we can construct an infinite tensor product as above, and we will
denote it by ${\cal L}[{\bf n}]$.
For the elementary tensors, we streamline the notation to:
\[
A_1\otimes\cdots\otimes A_k\otimes E[{\bf n}]_{k+1}:=
A_1\otimes\cdots\otimes A_k\otimes E_{n_{k+1}}^{(k+1)}\otimes E_{n_{k+2}}^{(k+2)}
\otimes\cdots\in {\cal L}[{\bf n}],
\]
  where $A_i\in{\cal L}_i$, and their closed span
is the simple $C^*$-algebra ${\cal L}[{\bf n}]$.

Next we define componentwise multiplication between different
C*-algebras ${\cal L}[{\bf n}]$ and ${\cal L}[{\bf m}]$.
For componentwise multiplication, the sequences give:
$$\big(E_{n_1}^{(1)},\,E_{n_2}^{(2)},\ldots\big)\cdot\big(E_{m_1}^{(1)},\,E_{m_2}^{(2)},\ldots\big)
=\big(E_{p_1}^{(1)},\,E_{p_2}^{(2)},\ldots\big)$$
where $p_j:={\rm min}(n_j,\,m_j)$, i.e. multiplication reduces the entries, and hence
the sequence ${\big(E_1^{(1)},\,E_1^{(2)},\,E_1^{(3)}\ldots\big)}$ is invariant
under such multiplication. So we define an embedding
 ${\cal L}[{\bf n}]\subseteq M({\cal L}[{\bf 1}])$ for all ${\bf n}$,
where ${\bf 1}:=(1,\,1,\ldots)$ by
\begin{eqnarray*}
\big(A_1\otimes\cdots\otimes A_k\otimes E[{\bf n}]_{k+1}\big)\cdot
\big(B_1\otimes\cdots\otimes B_n\otimes E[{\bf 1}]_{n+1}\big)\qquad\qquad\qquad\qquad\qquad\qquad  \\
\qquad\qquad:= \begin{cases}
A_1B_1\otimes\cdots\otimes A_nB_n\otimes A_{n+1}E_1^{(n+1)}\cdots\otimes A_kE_1^{(k)}\otimes E[{\bf 1}]_{k+1}
& \text{if $n\leq k$} \\
A_1B_1\otimes\cdots\otimes A_kB_k\otimes E_{n_{k+1}}^{(k+1)} B_{k+1}\cdots
\otimes E_{n_n}^{(n)} B_{n}\otimes E[{\bf 1}]_{n+1}
& \text{if $n\geq k$}
\end{cases}
\end{eqnarray*}
for the left action, and similar for the right action on ${\cal L}[{\bf 1}]\,.$
Since multiplication by elements of ${\cal L}[{\bf 1}]$ can separate the
elements of  ${\cal L}[{\bf n}]$, the embeddings are faithful.
Using these embeddings ${\cal L}[{\bf n}]\subseteq M({\cal L}[{\bf 1}])$ we see that
\begin{equation}
\label{4.3}
{\cal L}[{\bf n}]\cdot{\cal L}[{\bf m}]
\subseteq{\cal L}[{\bf p}],
\end{equation}
where $p_j:={\rm min}(n_j,\,m_j)$, and in fact
\begin{equation}
\label{4.4}
{\cal L}[{\bf n}]\subset M({\cal L}[{\bf p}])\supset
{\cal L}[{\bf m}].
\end{equation}
Since
${\cal L}[{\bf n}]\subseteq M({\cal L}[{\bf 1}])$ for all ${\bf n}$,
we can define the
$C^*$-algebra in $M({\cal L}[{\bf 1}])$ generated by all ${\cal L}[{\bf n}]$,
and denote it by ${\cal L}[E]$. By (\ref{4.3}), this is just the closed span
of all ${\cal L}[{\bf n}]$ and hence the closure of
the dense *-subalgebra $\cl L._0\subset{\cal L}[E]$, where
$$ \cl L._0:= \sum_{{\bf n}\in\N^\infty} \cl L.[{\bf n}]_0 \quad\hbox{and}\quad
\cl L.[{\bf n}]_0:=\bigcup_{k \in \N} {\cal L}^{(k)} \otimes E[\bn]_{k+1}. $$
Note that if two sequences ${\bf n}$ and ${\bf m}$ differ only  in a finite number of entries,
then ${\cal L}[{\bf n}]={\cal L}[{\bf m}]$, and hence we actually have
that the correct index set for the algebras ${\cal L}[{\bf n}]$ is
not the sequences $\N^\infty$, but the set of equivalence classes
$\N^\infty\big/\mathord{\sim}$ where ${\bf n}\sim{\bf m}$ if they
differ only in finitely many entries.
We have a partial ordering
of equivalence classes defined by $[\b n.]\geq [\b m.]$ if
for any representatives $\b n.$ and $\b m.$ resp., we have that
there is an $N$ (depending on the representatives)
such that $n_k\geq m_k$ for all $k>N\,.$
In particular, we note that
products reduce sequences, i.e., we have $\cl L.[\b n.]\cdot\cl L.[\b p.]\subseteq
 \cl L.[\b q.]$ for $q_i={\rm min}(n_i,p_i),$ so $[\b n.]\geq[\b q.]\leq[\b p.]$.

Let $\phi:\N^\infty\big/\mathord{\sim}
\to\N^\infty$ be a section of the factor map.
Then
 ${\cal L}[E]$ is the $C^*$-algebra generated in $M({\cal L}[{\bf 1}])$ by
${\big\{{\cal L}[\phi(\gamma)]\,\big|\,\gamma\in\N^\infty\big/\mathord{\sim}
\big\}}$, and it is the closure of the span of the elementary tensors
in this generating set.

{}From the reducing property of products, we already know that ${\cal L}[E]$
has the ideal ${\cal L}[{\bf 1}]$ (we will see that it is proper), hence
that it is not simple. However, it has in fact infinitely many proper ideals
and each of the generating algebras $\cl L.[\b n.]$ is contained in such an ideal:
\begin{Proposition}
For the $C^*$-algebra ${\cal L}[E]$, we have the following:
\begin{itemize}
\item[(i)] ${\cal L}[E]$ is nonseparable,
\item[(ii)] Define $\cl I.[\b n._1,\ldots,\b n._k]$ to be the closed span of
$$\big\{ {\cal L}[{\bf q}]_0 \,\mid\, [\b q.]\leq[\b n._\ell]\;\hbox{for some}\;\;\ell=1,\ldots,k\big\}\,.$$
Let $[\b p.]>[\b n._\ell]$ strictly for all $\ell\in\{1,\ldots,\,k\}\,,$ then
$\cl L.[\b p.]\cap\cl I.[\b n._1,\ldots,\b n._k]=\{0\}\,.$
\item[(iii)] $\cl I.[\b n._1,\ldots,\b n._k]$ is a proper closed
two sided ideal of  ${\cal L}[E]\,.$
\item[(iv)] Define $\cl L.[\b n._1,\ldots,\b n._k]:= C^*\left(
 \cl L.[\b n._1]\cup\cdots\cup\cl L.[\b n._k]\right)\,.$ \\
 Then
 $\cl L.[\b n._1,\ldots,\b n._k]\subset\cl I.[\b n._1,\ldots,\b n._k]$
and
$$C^*\left(\cl L.[\b n._1,\ldots,\b n._k]
 \cdot\cl L.[\b n._{k+1}]\right)
\subseteq \cl L.[\b q._1,\ldots,\b q._k],\quad\hbox{where}\quad
 (\b q._{j})_\ell={{\rm min}\big((\b n._j)_\ell,\,
 (\b n._{k+1})_\ell\big)}\,.$$ 
 \end{itemize}
 \end{Proposition}
\noindent The main attraction of the C*-algebra ${\cal L}[E]$, is that its representation theory
 is exactly the regular representations of $\mathop{\bigotimes}\limits_{n=1}^\infty{\rm CCR}(\R^2)$,
 which naively is what one would require for the representation theory of
 ``$\mathop{\bigotimes}\limits_{n=1}^\infty\cl K.(\cl H.)$''.
 One of the main costs of using it, is that the finite tensor products $\mathop{\bigotimes}\limits_{n=1}^N\cl K.(\cl H.)$
 are not contained in ${\cal L}[E]$, but are contained in its multiplier algebra  $M({\cal L}[E]).$
 This is not a serious problem because a representation (resp. state) on ${\cal L}[E]$ extends uniquely
 to $M({\cal L}[E])$ on the same representation space (resp. as a state), and hence to subalgebras of
  $M({\cal L}[E])$.

One could interpret the sequences of projections as specifying the ``type'' of infinite lattice in which
we embed our finite systems. As these sequences restrict the representations, they have physical content,
so in the next  main section we will try to obtain sequences which are natural from the
physical point of view (e.g. being gauge invariant). To conclude:
\begin{Definition}
 \label{FieldAlg}
 The field algebra for the quantum connection fields on a lattice is ${\cal L}[E]$,
where the components $\cl L._\ell=\cl K.(\cl H.)\cong C(G)\rtimes_\lambda G$
are labelled by links $\ell\in\Lambda^1.$
  \end{Definition}

\subsection{The kinematic field algebra.}
\label{FullFA}

From the matter and gauge field algebras, it is now natural  to take:
 \begin{Definition}
 \label{KinFieldAlg}
The kinematic field algebra is ${\mathfrak A}_{\Lambda} :=
{\mathfrak F}_{\Lambda} \otimes {\cal L}[E]$.    
It has a unique tensor norm as ${\mathfrak F}_{\Lambda}$ is nuclear.
  \end{Definition}
Note that ${\mathfrak A}_{\Lambda}$ is not unital since  ${\cal L}[E]$ is not unital, and it is not simple since
 ${\cal L}[E]$ is not simple. As mentioned, we will restrict our choice of
 approximate identities $(E_n^{(k)})_{n \in \N}\subset{\cal L}_k= {\cal K}(\cl H.)$ below
 when we have defined gauge transformations. In fact  ${\mathfrak A}_{\Lambda}$ is not yet the full field algebra,
 since information of important physical transformations is still absent. Below we will extend it
 to a crossed product of the gauge transformations, to obtain the full field algebra.

We next consider a natural inductive limit structure for this field algebra.
Let $\cl S.$ be a directed set of open, bounded convex subsets of $\R^3$ such that
$\bigcup\limits_{S\in\cl S.}S=\R^3,$ where the partial ordering is set inclusion.
 Let $\Lambda_S^i=\{x\in\Lambda^i\,\mid\,x\subset S\}$ (using the natural identification of elements of
 $\Lambda^i$ with subsets of $\R^3$),
and note that $S_1\subseteq S_2$ implies $\Lambda_{S_1}^i\subseteq\Lambda_{S_2}^i$
and $\bigcup\limits_{S\in\cl S.}\Lambda_S^i=\Lambda^i.$
Define
${\mathfrak F}_{S}:=C^*\big(\mathop{\cup}\limits_{x\in\Lambda_S^0}{\mathfrak F}_x\big)\subset{\mathfrak F}_{\Lambda}$
and then ${\mathfrak F}_{\Lambda}=\ilim{\mathfrak F}_{S}$ is an inductive limit w.r.t. the partial ordering
in $\cl S..$

To identify the analogous inductive limit for ${\cal L}[E],$ enumerate the links $\{\ell_1,\ell_2,\ldots\}=\Lambda^1$
and recall that ${\cal L}[E]$ has the dense *-subalgebra
$$ \cl L._0:= \sum_{{\bf n}\in\N^\infty} \cl L.[{\bf n}]_0 \quad\hbox{and}\quad
\cl L.[{\bf n}]_0:=\bigcup_{k \in \N} {\cal L}^{(k)} \otimes E[\bn]_{k+1}\quad\hbox{where}\quad
{\cal L}^{(k)}=\cl L._{1}\otimes\cdots\otimes\cl L._{k}. $$
This suggests that for an $S\in\cl S.$ we should take those elementary tensors in each
$\cl L.[{\bf n}]_0$ which can only differ from $ E[\bn]_{1}=E_{n_{1}}^{(1)}\otimes E_{n_{2}}^{(2)}
\otimes\cdots$ in entries
corresponding to links in $\Lambda_S^1.$
Denote the set of these elementary tensors by $\cl E._S[\bn],$ and define
 \[
 {\cal L}_S[E]:=C^*\big(\bigcup_{{\bf n}\in\N^\infty}\cl E._S[\bn]\big)\subset {\cal L}[E],
 \]
 then again we have the inductive limit structure ${\cal L}[E]=\ilim{\cal L}_S[E]$ w.r.t. set inclusion,
 since $\cl E._{S_1}[\bn]\subseteq \cl E._{S_2}[\bn]$ if $S_1\subseteq S_2,$
 and $\cl L.[{\bf n}]_0=\bigcup\limits_{S\in\cl S.}\cl E._{S}[\bn]$.

\begin{Proposition}
\label{indLimitA}
Given as above, a directed set $\cl S.$  of open, bounded convex subsets of $\R^3$ such that
$\bigcup\limits_{S\in\cl S.}S=\R^3,$ partially ordered by inclusion, then
 \[  {\mathfrak A}_{\Lambda} =\ilim{\mathfrak A}_S=\ilim\big({\mathfrak F}_{S}\otimes{\cal L}_{S}[E]\big)\]
 where ${\mathfrak F}_{S}:=C^*\big(\mathop{\cup}\limits_{x\in\Lambda_S^0}{\mathfrak F}_x\big)$ and
 ${\cal L}_S[E]:=C^*\big(\mathop{\cup}\limits_{{\bf n}\in\N^\infty}\cl E._S[\bn]\big)$.
\end{Proposition}
 \noindent {\bf Proof:}
  Now the  field algebra ${\mathfrak A}_{\Lambda} =
{\mathfrak F}_{\Lambda} \otimes {\cal L}[E]=\big(\ilim{\mathfrak F}_{S}\big)\otimes\big(\ilim{\cal L}_{S'}[E]\big)$
and we want to show that this is isomorphic to $\ilim\big({\mathfrak F}_{S}\otimes{\cal L}_{S}[E]\big)$.
Note first that for a fixed $S\in\cl S.$ that ${\rm Span}\{F\otimes L\,\mid\,F\in{\mathfrak F}_{S},\,L\in{\cal L}_{S}[E]\}
\subset{\mathfrak A}_{\Lambda}$ is the algebraic tensor product of ${\mathfrak F}_{S}$ with ${\cal L}_{S}[E]$,
and that the restriction of the C*-norm of ${\mathfrak A}_{\Lambda}$ to this is still a cross--norm
(as it is one on the full algebra). Thus the closure of this space in ${\mathfrak A}_{\Lambda}$
is precisely ${\mathfrak F}_{S}\otimes{\cal L}_{S}[E]=:{\mathfrak A}_S$ as this algebra has a unique tensor norm.
By construction we have that ${\mathfrak A}_{S_1}\subseteq{\mathfrak A}_{S_2}$ if $S_1\subseteq S_2,$ and
the *-algebra $\bigcup\limits_{S\in\cl S.}{\mathfrak A}_S$ contains all of
$\big(\mathop{\cup}\limits_{x\in\Lambda^0}{\mathfrak F}_x\big)\otimes\mathop{\cup}\limits_{{\bf n}\in\N^\infty}\cl L.[{\bf n}]_0$,
hence it is dense. Thus ${\mathfrak A}_{\Lambda} =\ilim{\mathfrak A}_S=\ilim\big({\mathfrak F}_{S}\otimes{\cal L}_{S}[E]\big)$
as required.
 \hfill $\rule{2mm}{2mm}$
\vspace{3mm}\break
 Recall though that the algebras ${\cal L}_{S}[E]$ are not the local algebras $\mathop{\bigotimes}\limits_{\ell_k\in\Lambda^1_S}\cl L._k
\subset M({\cal L}[E])$, since the elementary tensors  $A_1\otimes\cdots\otimes A_k\otimes E[{\bf n}]_{k+1}\in \cl E._S[\bn]$
generating the  ${\cal L}_{S}[E]$  contain the extra parts
$E[{\bf n}]_{k+1}$. As remarked above, this
 is not a serious problem because a representation (resp. state) on ${\cal L}[E]$ extends uniquely
 to $M({\cal L}[E])$  on the same representation space (resp. as a state), and hence to subalgebras of
  $M({\cal L}[E])$. Thus ${\cal L}[E]$ determines states and representations on all the local algebras
   $\mathop{\bigotimes}\limits_{\ell_k\in\Lambda^1_S}\cl L._k$.

\section{Gauge transformations and the local Gauss law}
\label{Gauge--Gauss}

We next consider the gauge transformations. Classically,
for the trivial principal bundle $P=\R^3\times G$ we have
$\gau P=C^\infty(\R^3,G)$.
However, $\R^3$ is not compact, and in this case it is customary to assume
that local gauge transformations are of compact support
(cf.~\cite{Is}). The global gauge transformations are taken to be the constant maps $\gamma:\R^3\to G$
(for nontrivial $P$ global gauge transformations need not exist).
These maps restrict in the obvious way to the lattice $\Lambda^0\subset\R^3$.

\subsection{Local gauge transformations.}
\label{LocalGT}

As the local gauge transformations are of compact support, they restrict
 on the lattice $\Lambda^0$ to
the group of maps $\gamma:\Lambda^0\to G$ of finite support, i.e.
\[
\gauc\Lambda := G^{(\Lambda^0)}=\big\{\gamma:\Lambda^0\to G\,\mid\,\big|{\rm supp}(\gamma)\big|<\infty\big\},\qquad
{\rm supp}(\gamma):=\{x\in\Lambda^0\,\mid\,\gamma(x)\not=e\}.
\]
This is an inductive limit indexed by the finite subsets $S\subset\Lambda^0$,
of the subgroups ${\rm Gau}_S \Lambda:=\{\gamma:\Lambda^0\to G\,\mid\,{\rm supp}(\gamma)\subseteq S\}
\cong\prod\limits_{x\in S}G$, and we give it the inductive limit
topology. As the groups $\prod\limits_{x\in S}G$ are compact, $\gauc \Lambda$ is amenable,
hence any continuous automorphic action of it on a C*-algebra will have an invariant state.
Moreover, as $G$ is connected, so is any finite product $\prod\limits_{x\in S}G$, and as every element of
$\gauc \Lambda$ is in one of these, $\gauc \Lambda$ is connected (a more general result is in Prop.~4.4 of \cite{Gl03}).
By choosing a strictly increasing chain of finite subsets $S\subset\Lambda^0$ with union $\Lambda^0$, we conclude from \cite{Gl03} that
the  inductive limit $\gauc \Lambda$ is an infinite dimensional Lie group, with (infinite dimensional) Lie algebra
\[
\Lgau\Lambda = {\mathfrak g}^{(\Lambda^0)}=\big\{\nu:\Lambda^0\to {\mathfrak g}\,\mid\,\big|{\rm supp}(\nu)\big|<\infty\big\}
={\rm Span}\{Y\cdot\delta_x\,\mid\,Y\in{\mathfrak g},\,x\in\Lambda^0\}
\]
where $\delta_x:\Lambda^0\to\R$ is $\delta_x(y)=1$ if $y=x$ and zero otherwise.

Next, we consider the action of the gauge group on the lattice.
Recall that the action of $\gamma\in\gauc \Lambda$ on classical configuration space $\big(\prod\limits_{x\in\Lambda^0}\Cn\big)\times
\big(\prod\limits_{\ell\in\Lambda^1}G\big)$ is by
\[
\big(\prod_{x\in\Lambda^0}v_x\big)\times\big(\prod_{\ell\in\Lambda^1}g_\ell\big)\mapsto
\big(\prod_{x\in\Lambda^0}\gamma(x)\cdot v_x\big)\times\big(\prod_{\ell\in\Lambda^1}\gamma(x_\ell)\,g_\ell\,\gamma(y_\ell)^{-1}\big)
\qquad\hbox{where}\qquad \ell=(x_\ell,y_\ell).
\]
For the quantum case, we define
following the discussion in  Subsection \ref{PM},
an analogous action $\alpha:\gauc \Lambda \to\aut{\mathfrak A}_{\Lambda}$ as follows.
Using the tensor structure ${\mathfrak A}_{\Lambda} =
{\mathfrak F}_{\Lambda} \otimes {\cal L}[E]$, we define $\alpha$ as a product action:
\[
\alpha_\gamma:=\alpha_\gamma^1\otimes\alpha_\gamma^2\qquad
\hbox{where}\qquad
\alpha^1:\gauc \Lambda \to\aut{\mathfrak F}_{\Lambda}\qquad\hbox{and}\qquad\alpha^2:\gauc \Lambda \to\aut{\cal L}[E]
\]
for $\gamma\in\gauc \Lambda$. The first component of the action
is given by:
\[
\alpha_\gamma^1(a(f)):=a(\gamma\cdot f)\qquad \hbox{where}\qquad (\gamma\cdot f)(x):={\gamma(x)}f(x)\quad \hbox{for all}\quad x\in\Lambda^0,
\;f\in\ell^2(\Lambda^0,\Cn)
\]
since $f\mapsto \gamma\cdot f$ defines a unitary on $\ell^2(\Lambda^0,\Cn)$.
For the second component action $\alpha^2$, we first show how to define it on an individual
tensor factor ${\cal L}_k=C(G)\rtimes_\lambda G$ of ${\cal L}[E]$.
Fix a pair $x,\,y\in\Lambda^0$ and guided by (\ref{GTfell}) define:
\[
\tau:\gauc \Lambda \to\aut C(G)\qquad\hbox{by}\qquad(\tau_\gamma f)(g):=f\big(\gamma(x)^{-1}g\,\gamma(y)\big)
\]
which corresponds to the classical action on $G$. Since $\tau_\gamma\circ\lambda_h=\lambda\s
\gamma(x)h\,\gamma(x)^{-1}.\circ\tau_\gamma,$ recalling that $C(G)\rtimes_\lambda G$ is generated by
$\psi \in L^1(G,C(G)),$ 
we extend $\tau_\gamma$ to  an automorphism on $C(G)\rtimes_\lambda G$
by setting $(\theta_\gamma(\psi))(g):=\tau_\gamma\big(\psi\big(\gamma(x)^{-1}g\,\gamma(x)\big)\big)$.
Since the product and adjoint in
$L^1(G,C(G))\subset C(G)\rtimes_\lambda G$ are given by
\begin{eqnarray*}
\big(\psi_1\times\psi_2)(g)&:=&
\int\psi_1(s)\,\lambda_s(\psi_2(s^{-1}g))\,ds\\
\psi^*(g)&:=&\lambda_g\big(\psi(g^{-1})^*\big)
\end{eqnarray*}
it is clear by straightforward verification that $\theta$ is an automorphic action.
In fact, as it only uses the evaluations of $\gamma$ at two points, it is a compact action
\[
\theta:G\times G\to{\rm Aut}\left(C(G)\rtimes_\lambda G\right)\,.
\]
This can be simplified by recalling that the crossed product $C(G)\rtimes_\lambda G$
is just the closure of the space spanned by $L^1(G)\cdot C(G)$, using the canonical containments
$L^1(G)\subset C^*(G)\subset M\big(C(G)\rtimes_\lambda G\big)\supset C(G)$ (cf. Thm~2.6.1
in \cite{Wil}). Thus, if we consider $\varphi\cdot f\in L^1(G)\cdot C(G)$ for
$\varphi\in L^1(G),$ $f\in C(G)$  then
\[
\theta_\gamma(\varphi\cdot f)=\sigma_\gamma(\varphi)\cdot\tau_\gamma(f)\quad\hbox{where}\quad
\sigma_\gamma(\varphi)(g):=\varphi\big(\gamma(x)^{-1}g\,\gamma(x)\big)\quad\hbox{and}\quad
\]
$(\tau_\gamma f)(g):=f\big(\gamma(x)^{-1}g\,\gamma(y)\big)$ is as above.
This is consistent with the gauge transformation on a link obtained in (\ref{GTfell}).
Thus $d\theta(\nu)=d\sigma(\nu)+d\tau(\nu)$ for $\nu\in \Lgau\Lambda$ on the span of
$(L^1(G)\cap C^\infty(G))\cdot C^\infty(G).$
This will be useful below.

Next, to define  $\alpha^2$,  we combine these actions for the full algebra ${\cal L}[E]$.
Recall that we enumerated the links
$\Lambda^1=\{\ell_n=(x_n,y_n)\,\mid\,n\in\N\}$, and that ${\cal L}[E]$ is generated by the elements
\[
A_1\otimes\cdots\otimes A_k\otimes E[{\bf n}]_{k+1}\in{\cal L}^{(k)} \otimes E[\bn]_{k+1}\qquad \hbox{where}\qquad A_i\in\cl L._{i}=C(G)\rtimes_\lambda G
\]
(note that ${\cal L}^{(j)} \otimes E[\bn]_{j+1}\subset{\cal L}^{(k)} \otimes E[\bn]_{k+1}$ if $j<k$, simply by putting some
$A_i=E_{n_i}^{(i)}$).
For a given $\gamma\in\gauc \Lambda$ there is always an $m$ large enough so that\\
 ${\rm supp}(\gamma)\subset\{x_n,y_n\,\mid\,n=1,\ldots,m\}.$ Thus
 \[
 \alpha_\gamma^2\big(A_1\otimes\cdots\otimes A_k\otimes E[{\bf n}]_{k+1}\big):=
 \theta^1_\gamma(A_1)\otimes\cdots\otimes \theta^k_\gamma(A_k)\otimes E[{\bf n}]_{k+1}\quad\hbox{for all}\quad k\geq m
 \]
where $\theta^j_\gamma(A_j)$ is $\theta_\gamma(A_j)$ where the pair $(x,y)$ is replaced by $(x_j,y_j)=\ell_j$ in the definition above.
Explicitly, if we let $A_j=\varphi\cdot f\in L^1(G)\cdot C(G)$, then
 \begin{equation}
\label{thetaform}
\theta^j_\gamma(A_j)(g)=\sigma^j_\gamma(\varphi)\cdot\tau^j_\gamma(f)\quad\hbox{where}\quad
\sigma^j_\gamma(\varphi)(g):=\varphi\big(\gamma(x_j)^{-1}g\,\gamma(x_j)\big)\quad\hbox{and}\quad
\end{equation}
$(\tau^j_\gamma f)(g):=f\big(\gamma(x_j)^{-1}g\,\gamma(y_j)\big)$.
This completely
defines $\alpha^2:\gauc \Lambda \to\aut{\cal L}[E]$ and hence $\alpha_\gamma:=\alpha_\gamma^1\otimes\alpha_\gamma^2$.
Note that $\alpha$ is continuous w.r.t. the inductive limit topology of $\gauc \Lambda$.

\smallskip
\noindent {\bf Remarks:}
\begin{enumerate}
\item
Note that the  orientation of links in $\Lambda^1$ was used in the definition of $\alpha^2$,
because the definition of $\theta$ based on a pair $(x,y)$ treated the $x$ and $y$ differently.
\item
The use of compact support for the gauge transformations was crucial. If one did not assume this,
then it may not be possible to define $\alpha^2_\gamma$ because $\gamma$ may not map elementary
tensors of the type $A_1\otimes\cdots\otimes A_k\otimes E[{\bf n}]_{k+1}$
to  one of the type $B_1\otimes\cdots\otimes B_j\otimes E[{\bf m}]_{j+1}$
as it may not preserve the approximate identities which they are based on.
This means that global gauge transformations cannot be defined, unless one chooses
approximate identities
$ (E_n^{(k)})_{n \in \N}$ which are invariant with respect
to the gauge action. This is what we will do in the next subsection.
\end{enumerate}

\subsection{Global gauge transformations.}
\label{GlobalGT}

As mentioned in the last remark, the action $\alpha^2:\gauc \Lambda \to\aut{\cal L}[E]$
cannot in general be extended to the constant maps, unless the $ (E_n^{(k)})_{n \in \N}$
are chosen to be  gauge  invariant. We examine this issue. Recall that for $\gamma\in\gauc \Lambda$,
$\alpha_\gamma^2$ is given by  $\theta^k_\gamma$ in the $k^{\rm th}$ factor for all $k$, so we consider the invariance of
$ (E_n^{(k)})_{n \in \N}$ w.r.t. $\theta^k:\gauc \Lambda\to\aut\al L._k$. Explicitly this action is
$\theta^k_\gamma(L)=\theta\s(\gamma(x_k),\gamma(y_k)).(L)$ where
\[
\theta:G\times G\to{\rm Aut}\left(C(G)\rtimes_\lambda G\right)
\]
is given as follows. Let  $\varphi\in L^1(G),$ $f\in C(G)$,  then for
$L=\varphi\cdot f\in L^1(G)\cdot C(G)\subset C(G)\rtimes_\lambda G$ we have
\[
\theta_{(h,s)}(L)=\theta_{(h,s)}(\varphi\cdot f)=\sigma\s{(h,s)}.(\varphi)\cdot\tau\s{(h,s)}.(f)\quad\hbox{where}\quad
\sigma\s{(h,s)}.(\varphi)(g):=\varphi\big(h^{-1}g\,h\big)
\]
and $(\tau\s{(h,s)}. f)(g):=f\big(h^{-1}g\,s\big)$.
\begin{Lemma}
\label{Lem.2.12B}
\begin{itemize}
\item[(i)] Let $\pi_0:C(G)\rtimes_\lambda G\to\cl B.\big(L^2(G)\big)$  be the irreducible representation
given by  $\pi_0(\varphi\cdot f)=\pi_1(\varphi)\pi_2( f)$ for $\varphi\in L^1(G)$ and  $f\in C(G)$ where
 \[
(\pi_1(\varphi)\psi)(g):=\int\varphi(h)\psi(h^{-1}g)\,{\rm d}h \qquad\hbox{and}\qquad
 (\pi_2(f)\psi)(g):=f(g)\psi(g)
 \]
 for all $\psi\in L^2(G)$ (Schr\"odinger representation). Then  $\pi_0$ is a covariant representation for $\theta$ with
 unitary implementers $W\s{(h,s)}.\in U(L^2(G))$, $h,\,s\in G$, given by
 $\left(W\s{(h,s)}.\psi\right)(g):=\psi(h^{-1}gs)$. Constant vectors, i.e. $\psi(g)=c\in\C$ for all $g$
 are in $L^2(G)$ and invariant w.r.t. $W.$
\item[(ii)]
There is an approximate identity of commuting projections $ (E_n)_{n \in \N}$ for $C(G)\rtimes_\lambda G$
which is invariant w.r.t. $\theta:G\times G\to{\rm Aut}\left(C(G)\rtimes_\lambda G\right)$.
\end{itemize}
\end{Lemma}
 \noindent {\bf Proof:}
(i) We have that $\pi_0\big(C(G)\rtimes_\lambda G)=\al K.(L^2(G))$ (cf. Theorem~II.10.4.3 in~\cite{Bla1}),
 hence that $\pi_0$ is irreducible. Direct verification also shows  that $W:G\times G\to U(L^2(G))$
 is a continuous unitary representation. We verify implementation of $\theta$:
 \begin{eqnarray*}
&&\!\!\!\!\!\!\!\!
 \left(W\s{(h,s)}.\pi_1(\varphi)W\s{(h,s)}.^{-1}\psi\right)(g)=\left(\pi_1(\varphi)W\s{(h,s)}.^{-1}\psi\right)(h^{-1}gs)\\[1mm]
&& =\int\varphi(t)\left(W\s{(h,s)}.^{-1}\psi\right)(t^{-1}h^{-1}gs)\,{\rm d}t=\int\varphi(t)\left(\psi\right)(ht^{-1}h^{-1}g)\,{\rm d}t
 \\[1mm]
 &&=\int\varphi(h^{-1}th)\left(\psi\right)(t^{-1}g)\,{\rm d}t=\left(\pi_1(\sigma\s{(h,s)}.(\varphi))\psi\right)(g)\\[1mm]
 &&\!\!\!\!\!\!\!\!
  \left(W\s{(h,s)}.\pi_2(f)W\s{(h,s)}.^{-1}\psi\right)(g)=\left(\pi_2(f)W\s{(h,s)}.^{-1}\psi\right)(h^{-1}gs)\\[1mm]
 &&= f(h^{-1}gs)\left(W\s{(h,s)}.^{-1}\psi\right)(h^{-1}gs)=\left(\pi_2(\tau\s{(h,s)}. f)\psi\right)(g)
  \end{eqnarray*}
  which produces $W\s{(h,s)}.\pi(L)W\s{(h,s)}.^{-1}=\pi\big(\theta_{(h,s)}(L)\big)$ as required.

  (ii)
  Since $G\times G$ is compact, the representation $W:G\times G\to U(L^2(G))$ is a direct orthogonal sum of
  finite dimensional irreducible representations of $G\times G$. The projections onto these finite dimensional subspaces
  are therefore in $\cl K.\big(L^2(G)\big)=\pi_0\big(C(G)\rtimes_\lambda G)$, and as these projections commute with $W$ they are
 invariant w.r.t. $\theta:G\times G\to{\rm Aut}\left(C(G)\rtimes_\lambda G\right)$. Moreover, they form a commuting set
 with total sum the identity, hence by taking larger and larger sums of them we obtain the desired
 approximate identity.
 \hfill $\rule{2mm}{2mm}$
\vspace{3mm}\break
Given this Lemma, one may therefore choose approximate identities
$ (E_n^{(k)})_{n \in \N}$ invariant with respect to $\theta$, and use these to construct ${\cal L}[E]$.
Henceforth we will assume that such a choice has been fixed, and we assume that approximate identities
$ (E_n^{(k)})_{n \in \N}$ have been chosen such that the constant vector $\psi_0:=1$ is in the range space of
each $E_n^{(k)}$ in the Schr\"odinger representation $\pi_0$.

Given this choice of approximate identities, we now have for ${\cal L}[E]$, that
the action $\alpha:\gauc \Lambda \to\aut{\mathfrak A}_{\Lambda}$ extends from
$\gauc \Lambda = G^{(\Lambda^0)}$ to all of $G^{\Lambda^0}$, which includes the constant maps, i.e
global gauge transformations. In particular on ${\mathfrak A}_{\Lambda} =
{\mathfrak F}_{\Lambda} \otimes {\cal L}[E]$,  we have a product action:
$\alpha_\gamma:=\alpha_\gamma^1\otimes\alpha_\gamma^2$, $\gamma\in G^{\Lambda^0}$ where as before
\[
\alpha_\gamma^1(a(f)):=a(\gamma\cdot f)\qquad \hbox{where}\qquad (\gamma\cdot f)(x):={\gamma(x)}f(x)\quad \hbox{for all}\quad x\in\Lambda^0,
\;f\in\ell^2(\Lambda^0,\Cn)
\]
since $f\mapsto \gamma\cdot f$ defines a unitary on $\ell^2(\Lambda^0,\Cn)$.
Moreover, by the invariance of $ (E_n^{(k)})_{n \in \N}$, the same formula
\[
 \alpha_\gamma^2\big(A_1\otimes\cdots\otimes A_k\otimes E[{\bf n}]_{k+1}\big):=
 \theta^1_\gamma(A_1)\otimes\cdots\otimes \theta^k_\gamma(A_k)\otimes E[{\bf n}]_{k+1}
 \]
 is valid, but now for all $\gamma\in G^{\Lambda^0}$. So global gauge transformations are given
 by $\alpha_\gamma$ where $\gamma(x)=g\in G$ for all $x\in\Lambda^0$.

\vspace{3mm}
\noindent {\bf Remarks:}
\begin{enumerate}
\item
Recall from Proposition~\ref{indLimitA} that
for a directed set $\cl S.$  of open, bounded convex subsets of $\R^3$ such that
$\bigcup\limits_{S\in\cl S.}S=\R^3,$ then
 \[  {\mathfrak A}_{\Lambda} =\ilim{\mathfrak A}_S=\ilim\big({\mathfrak F}_{S}\otimes{\cal L}_{S}[E]\big)\]
 where ${\mathfrak F}_{S}:=C^*\big(\mathop{\cup}\limits_{x\in\Lambda_S^0}{\mathfrak F}_x\big)$ and
 ${\cal L}_S[E]:=C^*\big(\mathop{\cup}\limits_{{\bf n}\in\N^\infty}\cl E._S[\bn]\big)$.
 With the choice of invariant approximate identities $ (E_n^{(k)})_{n \in \N}$ above,
 it is clear that the extended action $\alpha:G^{\Lambda^0} \to\aut{\mathfrak A}_{\Lambda}$ preserves
 each of the  ``local'' algebras ${\mathfrak A}_S={{\mathfrak F}_{S}\otimes{\cal L}_{S}[E]}$.
 Moreover, a ``local'' algebra ${\mathfrak A}_S$
 cannot tell the global gauge transformations apart from certain local gauge transformations.
  That is, given any global gauge transformation
 $\alpha_\gamma$ where $\gamma(x)=g\in G$ for all $x\in\Lambda^0$ and a ``local'' algebra ${\mathfrak A}_S$,
 then there is a $\gamma\s{\rm loc}.\in\gauc \Lambda$ such that
 $\alpha_\gamma\restriction{\mathfrak A}_S=\alpha_{\gamma\s{\rm loc}.}\restriction{\mathfrak A}_S$,
for example take $\gamma\s{\rm loc}.(x)=g=\gamma(x)$ if $x\in S$ and $\gamma\s{\rm loc}.(x)=e$
if $x\not\in S$. This is not true for the full algebra ${\mathfrak A}_{\Lambda} $ because
given a global gauge transformation $\alpha_\gamma$, we cannot find a $\gamma\s{\rm loc}.\in\gauc \Lambda$
which will work for all ${\mathfrak A}_S\subset{\mathfrak A}_{\Lambda}$.
\item
From Lemma~\ref{Lem.2.12B}, we obtain a very natural representation for  ${\cal L}[E]$
with the choice of approximate identity made here. For ${\mathfrak A}_{\Lambda} =
{\mathfrak F}_{\Lambda} \otimes {\cal L}[E]$ define a product representation
$\pi=\pi_{\rm Fock}\otimes\pi_\infty$ where $\pi_{\rm Fock}$ is the Fock representation of
${\mathfrak F}_{\Lambda}=\CAR {\al H.}.$, and $\pi_\infty$ is an infinite tensor product of
Schr\"odinger representations $\pi_0$ (one for each factor $\al L._\ell$ of ${\cal L}[E]$),
but where we choose the reference sequence to be just the sequence ${(\psi_0,\psi_0,\ldots)}$
where $\psi_0=1$ is the constant vector. This means we can consider the representation
space $\al H._\infty$ of $\pi_\infty$ to be spanned by elementary tensors of the type
\[
\varphi_1\otimes\cdots\otimes \varphi_k\otimes\psi_0\otimes\psi_0\otimes\cdots,\quad\varphi_i\in L^2(G).
\]
Then, if we consider the action of ${\cal L}[E]$ on it, we see
\begin{eqnarray*}
\pi_\infty\big(A_1\otimes\cdots\otimes A_k\otimes E[{\bf n}]_{k+1}\big)\big(\varphi_1\otimes\cdots \varphi_k\otimes\psi_0\otimes\psi_0\otimes\cdots\big)
\\[1mm]
=\pi_0(A_1)\varphi_1\otimes\cdots\otimes \pi_0(A_k)\varphi_k\otimes\psi_0\otimes\psi_0\otimes\cdots
\end{eqnarray*}
because $\pi_0(E_n^{(j)})\psi_0=\psi_0$ for all $n$ and $j$. Hence all of ${\cal L}[E]$  can be represented on $\al H._\infty$.
In fact, since each factor of the representation is covariant, and $\psi_0$ is an invariant vector, we also get that
$\pi$ is covariant w.r.t.  $\alpha:G^{\Lambda^0} \to\aut{\mathfrak A}_{\Lambda}$, and it has an invariant vector
$\Omega\otimes(\psi_0\otimes\psi_0\otimes\cdots)$ where $\Omega$ is the Fock vacuum vector. Thus the vector state of
this vector is a  $\alpha(G^{\Lambda^0})\hbox{--invariant}$ state on  ${\mathfrak A}_{\Lambda}$.
This is interesting as this means that we have an invariant state for the much larger group action
$\alpha:G^{\Lambda^0}\to\aut{\mathfrak A}_{\Lambda}$, not just for its restriction to the amenable group
$\gauc \Lambda$.

We claim that the representation $\pi=\pi_{\rm Fock}\otimes\pi_\infty$ is faithful. Since $\pi_{\rm Fock}$ is already known to be faithful,
we only have to show that $\pi_\infty$ is faithful (since the tensor norm for ${\mathfrak F}_{\Lambda} \otimes {\cal L}[E]$
is unique, using Theorem~4.9(iii), p208 in~\cite{Tak}). Recall that ${\cal L}[E]$ is the C*-algebra constructed from all
${\cal L}[{\bf n}]\subseteq M({\cal L}[{\bf 1}])$
in $M({\cal L}[{\bf 1}])$, hence we have a faithful embedding ${\cal L}[E]\subset M({\cal L}[{\bf 1}])$.
Now the restriction $\pi_\infty\restriction{\cal L}[{\bf 1}]$
is faithful  as  ${\cal L}[{\bf 1}]$ is simple and $\pi_\infty$ is nonzero on it.
But then the extension of $\pi_\infty$ to $M({\cal L}[{\bf 1}])$ is faithful, hence
$\pi_\infty$ is faithful on ${\cal L}[E]$.
\end{enumerate}

\subsection{Defining the full field algebra.}
\label{FFAlg}

There is physical information contained in the gauge action
 $\alpha:G^{\Lambda^0} \to\aut{\mathfrak A}_{\Lambda}$ as
 $\alpha(\gauc \Lambda)$ is the local gauge transformations
 and   $\alpha(G)$ is the global gauge transformations (identifying $G$
 with the constant maps in $G^{\Lambda^0}$).
It is therefore desirable to  extend the
field algebra ${\mathfrak A}_{\Lambda}$ to ensure that in physical representations, the generators of the
unitary implementers of $\alpha$ are
affiliated to our field algebra. Usually, one takes the crossed product, but in this context
e.g. the crossed product ``${{\mathfrak A}_{\Lambda}\rtimes_\alpha (\gauc \Lambda )}$'' cannot be defined because
$\gauc \Lambda$ is not locally compact. In fact for non-locally compact groups, it is
 a very hard question as to what  C*-algebra should play the role of the crossed product.
In such a situation,  the best one can do at the moment, is to endow the given group with the
discrete topology, which makes it locally compact, and then to use the crossed product w.r.t. this discrete group.
This has the disadvantage of having too many representations, in particular it allows those covariant representations
where the unitary implementers are not continuous w.r.t. the original group topology.
In the present context one may argue that as the gauge transformations will be factored out by a constraint procedure,
the topology of the gauge group is not physically relevant.

Concretely, our strategy is as follows.
Let $\gaue \Lambda $ denote our chosen group in $G^{\Lambda^0}$ of physically relevant transformations
(this should at least contain the local gauge transformations $\gauc \Lambda \subset G^{\Lambda^0}$).
Let $\gaued\Lambda$
denote $\gaue \Lambda $ equipped with
 the {\it discrete} topology. Then take the discrete crossed product
${\mathfrak A}_{\Lambda}\rtimes_\alpha (\gaued \Lambda)$. As it is convenient to have an identity
in our field algebra, we will take instead
\[
\al F._e:=({\mathfrak A}_{\Lambda}\oplus\C)\rtimes_\alpha (\gaued \Lambda)
\]
where ${\mathfrak A}_{\Lambda}\oplus\C$ denotes ${\mathfrak A}_{\Lambda}$ with an identity adjoined.
It is generated as a C*-algebra by a copy of ${\mathfrak A}_{\Lambda}$ as well as by unitaries
 $U_g$, $g\in\gaue \Lambda $ such that
 $U_gAU_g^*=\alpha_g(A)$ for all $A\in{\mathfrak A}_{\Lambda}$, and $U_gU_h=U_{gh}$. Algebraically
 \begin{eqnarray*}
\al F._e&=&({\mathfrak A}_{\Lambda}\oplus\C)\rtimes_\alpha (\gaued \Lambda)=C^*\left(
U\s\gaued \Lambda.\cup{\mathfrak A}_{\Lambda}\right)\quad\hbox{where}\quad
{\mathfrak A}_{\Lambda} :={\mathfrak F}_{\Lambda} \otimes {\cal L}[E]\\[1mm]
&=&\left[ U\s\gaued \Lambda.\cdot{\mathfrak A}_{\Lambda}\right]+[U\s\gaued \Lambda.]
\end{eqnarray*}
where we use the notation $[\cdot]$ for the closed linear space
generated by its argument.
The representations of ${\cal F}_e$ consist of {\it all} covariant
representations for $\alpha:\gaue \Lambda\to{\rm Aut}({\mathfrak A}_{\Lambda}\oplus\C)$, whether continuous or not.

The natural choice for our full field algebra, is $\al F._e$ where we take
$\gaue \Lambda $ to be the group generated in $G^{\Lambda^0}$ by $\gauc \Lambda$ and
$G$  (the constant maps in $G^{\Lambda^0}$), as this will include both local and global
gauge transformations. However, with our eye on the subsequent work below (enforcing constraints)
we will make the smaller choice where we take $\al F._e$ with
$\gaue \Lambda =\gauc \Lambda$.  The reason why we will not include unitaries corresponding to global
gauge transformations, is because locally these implement the same automorphisms as some local gauge transformations
(see remark~(1) at end of Subsect.~\ref{GlobalGT}). Thus, if we enforce local gauge invariance through constraints,
then the images of these unitaries will commute with all the local algebras, hence with the image of
${\mathfrak A}_{\Lambda}$, and hence will be of no physical relevance. Thus, to conclude, henceforth for
our full field algebra we will take
\[
\al F._e=({\mathfrak A}_{\Lambda}\oplus\C)\rtimes_\alpha (\gaud \Lambda)=C^*\left(
U\s\gaud \Lambda.\cup{\mathfrak A}_{\Lambda}\right)\,.
\]

\subsection{The local Gauss law.}
\label{GaussLaw}

We consider the local gauge transformations.
Given the action $\alpha:\gauc \Lambda \to\aut{\mathfrak A}_{\Lambda}$ defined above,
 an (abstract) Gauss law element will be a nonzero element in the range of the derived action
$d\alpha:\Lgau\Lambda \to {\rm Der}({\mathfrak A}_{\Lambda}^\infty)$ where ${\mathfrak A}_{\Lambda}^\infty$ is the
algebra of smooth elements of the action. Since $\alpha_\gamma:=\alpha_\gamma^1\otimes\alpha_\gamma^2$, it is of the
form
\[
d\alpha(\nu)=d\alpha^1(\nu)\otimes\un+\un\otimes d\alpha^2(\nu),\;\;
\nu\in \Lgau\Lambda,\qquad\hbox{on}\qquad
{\mathfrak F}_{\Lambda}^\infty \otimes {\cal L}[E]^\infty\subseteq {\mathfrak A}_{\Lambda}^\infty,
\]
i.e. it is a sum of a matter part and a radiation part.
The Gauss law condition, is simply the enforcement of of it as a constraint,
i.e. setting it to zero in an appropriate way. We will investigate this below.
Concrete Gauss law elements consist of implementers of the derivations $d\alpha(\nu)$ in
$\alpha\hbox{--covariant}$ representations by selfadjoint operators, and clearly  these will be
the generators of the unitaries implementing the one--parameter groups
$t\mapsto\alpha(\exp(t\nu))$.

To obtain an explicit form for  $d\alpha(\nu)$, recall that $\Lgau\Lambda$ consists of finite spans of
elements $\nu=Y\cdot\delta_x$ for $Y\in{\mathfrak g},\,x\in\Lambda^0,$ and these are
the generators of the one-parameter groups $t\mapsto\exp(tY\cdot \delta_x)\in\gauc \Lambda$.
Thus the matter part of the Gauss law, $d\alpha^1(Y\cdot\delta_x)\in {\rm Der}({\mathfrak F}_{\Lambda}^\infty )$, is
given by
\[
d\alpha^1(Y\cdot\delta_x)\big(a(f)\big)={d\over dt}a\left({\exp(tY\cdot \delta_x)}f\right)\Big|_{t=0}
=a\left(\delta_x\cdot {Y}f\right)\in{\mathfrak F}_x=\CAR V_x..
\]
In fact, as $\Cn$ is finite dimensional, this is defined for all $f\in{\mathfrak F}_{\Lambda}$, hence
${\mathfrak F}_{\Lambda}^\infty $ contains the dense *-algebra generated in ${\mathfrak F}_{\Lambda}$
by  the set $\{a(f)\,\mid\,f\in\ell^2(\Lambda^0,\Cn)\}$.

Next, we consider the radiation part of the Gauss law, hence $\alpha^2$. Recall that we have enumerated the links
$\Lambda^1=\{\ell_n=(x_n,y_n)\,\mid\,n\in\N\}$, and that for each link $\ell_k=(x_k,y_k)$ there is an action
$\theta^k:\gauc \Lambda \to\aut{\cal L}_k$ where $\theta^k_\gamma$ only depends on $\gamma(x_k)$ and $\gamma(y_k)$.
Thus $\alpha^2(\gamma)=\alpha^2\big(\exp(tY\cdot \delta_x)\big)$ will only affect the links which contain $x$.
Let $L(x):=\{k\in\N\,\mid\,\ell_k=(x,y_k)\;\mbox{or}\;\ell_k=(x_k,x)\}$. As $\Lambda$ is a cubic lattice, there are
at most 6 links connected to a vertex $x$ so $|L(x)|\leq 6$, hence
$L(x)=\{k_1,\,k_2,\ldots,k_j\}$ where $j\leq 6$.
  Then
\begin{eqnarray*}
 \alpha_\gamma^2\big(\mathop{\otimes}_{i=1}^{n}A_i\otimes E[{\bf n}]_{n}\big)&=& \\[1mm]
 &&\!\!\!\!\!\!\!\!\!\!\!\!\!\!\!\!\!\!\!\!\!\!\!\!\!\!\!\!\!\!\!\!\!\!\!\!\!\!\!\!\!\!\!\!\!\!\!\!\!\!\!\!\!
A_1\otimes\cdots A_{k_1-1}\otimes \theta^{k_1}_\gamma(A_{k_1})\otimes A_{k_1+1}\otimes\cdots
A_{k_j-1}\otimes \theta^{k_j}_\gamma(A_{k_j})\otimes A_{k_j+1}\otimes\cdots
A_{n}\otimes E[{\bf n}]_{n}
\end{eqnarray*}
for $n> k_j,$ and so
\[
d\alpha^2(Y\cdot \delta_x)=\sum_{k\in L(x)}d\theta^k(Y\cdot\delta_x)
=\sum_{k\in L(x)}\Big(d\sigma^k(Y\cdot\delta_x)+d\tau^k(Y\cdot\delta_x)\Big)
\]
since  $d\theta^k(\nu)=d\sigma^k(\nu)+d\tau^k(\nu)$ for $\nu\in \Lgau\Lambda$ on the span of
$(L^1(G)\cap C^\infty(G))\cdot C^\infty(G).$
 In particular, if $k\in L_1(x):=\{k\in L(x)\,\mid\,\ell_k=(x,y_k)\}$, then from (\ref{thetaform}) we
get for $A_k=\varphi\cdot f\in
{(L^1(G)\cap C^\infty(G))\cdot C^\infty(G)}$ that
\[
\theta^k_\gamma(A_k)(g)=\sigma^k_\gamma(\varphi)\cdot\tau^k_\gamma(f)\quad\hbox{where}\quad
\sigma^k_\gamma(\varphi)(g):=\varphi\big(e^{-tY}g\,e^{tY}\big)\quad\hbox{and}\quad
\]
$(\tau^k_\gamma f)(g):=f\big(e^{-tY}g\big)$ as $\gamma(x)=\exp(tY)$.
Then clearly
\begin{eqnarray*}
d\tau^k(Y\cdot \delta_x)(f)(g)&=&{d\over dt}f\big(e^{-tY}g\big)\big|_{t=0}=-(\wt{Y}f)(g)=-df(\wt{Y})(g)\qquad\hbox{and}\\[2mm]
d\sigma^k(Y\cdot \delta_x)(\varphi)(e^Z)&=&{d\over dt}\varphi\big(e^{-tY}e^Z\,e^{tY}\big)\big|_{t=0}
={d\over dt}\varphi\big(\exp(-t\,{\rm ad}_Y(Z))\big)\big|_{t=0}\\[2mm]
&=&{\rm ad}_Y(Z)(\varphi)(e^Z)\quad\forall\,Z\in{\mathfrak g}
\end{eqnarray*}
where $\wt{Y}$ is the right invariant vector field on $G$ generated by $t\to e^{tY}g$.
On the other hand, if   $k\in L_2(x):=L(x)\backslash L_1(x)=\{k\in L(x)\,\mid\,\ell_k=(x_k,x)\}$ then $\sigma^k_\gamma$ is the identity, so
\begin{eqnarray*}
\theta^k_\gamma(A_k)(g)&=&\varphi\cdot\tau^k_\gamma(f)\quad\hbox{where}\quad
(\tau^k_\gamma f)(g):=f\big(g\,e^{tY}\big) \\[2mm]
\hbox{hence}\qquad\qquad d\tau^k(Y\cdot \delta_x)(f)(g)&=&{d\over dt}f\big(g\,e^{tY}\big)\big|_{t=0}=(Yf)(g)=df(Y)(g)\,.
\end{eqnarray*}
So all components of the Gauss law elements have been made explicit
\[
d\alpha(Y\cdot \delta_x)=d\alpha^1(Y\cdot \delta_x)\otimes\un+\un\otimes \sum_{k\in L(x)}\Big(d\sigma^k(Y\cdot\delta_x)+d\tau^k(Y\cdot\delta_x)\Big)
\qquad\forall\,x\in\Lambda^0,Y\in{\mathfrak g}\,.
\]
For the algebra ${\mathfrak F}_{\Lambda}^\infty \otimes {\cal L}[E]^\infty\subseteq {\mathfrak A}_{\Lambda}^\infty$ on which this acts,
the first factor ${\mathfrak F}_{\Lambda}^\infty$ contains
$\hbox{*-alg}\{a(f)\mid f\in\ell^2(\Lambda^0,\Cn)\}$, and
the second factor ${\cal L}[E]^\infty$ contains the span of all elementary tensors
${A_1\otimes\cdots\otimes A_k\otimes E[{\bf n}]_{k+1}}\in {\cal L}[E]$ such that
$A_j\in {(L^1(G)\cap C^\infty(G))\cdot C^\infty(G)}\subset\cl L._{j}$ for all $j$.

 Note that the Gauss law elements will only be represented concretely in representations
 $\pi$ for which $t\to\pi(U_{\exp(t\nu)})$ is continuous for all $\nu\in\Lgau\Lambda.$
Moreover, in the case that $G$ is abelian, i.e. $G=\T$ (electromagnetism) we see that
$d\sigma^k=0$ for all $k$, which simplifies the last expression.

%
%

\section{Enforcement of local Gauss law Constraints.}
\label{GenConstraints}

Here we want to obtain the algebra of physical observables from our chosen field algebra
$\al F._e=({\mathfrak A}_{\Lambda}\oplus\C)\rtimes_\alpha (\gaud \Lambda)$
by enforcing the local Gauss law constraint, and by imposing gauge invariance in
an appropriate form. There is a wide range of methods in the literature for enforcing quantum constraints,
not all equivalent~\cite{Constr}.  For lattice QCD, so far constraint methods include
the method developed by Kijowski and Rudolph in \cite{KR}, the direct construction of explicit gauge invariant
quantities through either Wilson loops of Fermi bilinears connected with a flux line (cf.~\cite{KS}) or by
taking group averages over compact local gauge groups (cf.~\cite{Crz}), e.g. for finite lattices.

In this section we will consider two methods:
\begin{itemize}
\item{} The method developed by Kijowski and Rudolph in \cite{KR, JKR}, which is summarized below,
following Theorem~\ref{Teo.2.6}.
\item{} The {\bf T-procedure} developed by  Grundling and Hurst (reviewed in \cite{GrSrv}) is based on enforcing the
constraints as state conditions in the universal representation.
It is based on Dirac's method for enforcement of constraints, and it is
summarized below in Subsection~\ref{GL-Tproc}. This method is ideally suited to C*-algebras,
has produced correct results in explicit examples, will always include gauge invariant elements in the
algebra if there are any, and has managed to sidestep difficulties such as indefinite metric
representations or continuous spectrum problems.
\end{itemize}
For the case of a finite lattice, we will show below in Theorem~\ref{structurTradLocalObs}
that these two methods produce the same result.
We want to apply these methods to the system constructed  in Section~\ref{Gauge--Gauss}.


\subsection{Enforcing constraints by T-procedure - method.}
\label{GL-Tproc}

In this section we review the T-procedure for the enforcement of constraints,
and we show that the system defined in Section~\ref{Gauge--Gauss} satisfies its input assumptions.
A convenient review of the T-procedure is in \cite{GrSrv}.
The starting point is:
\begin{Definition}
A {\bf quantum system with constraints} is a
pair $(\al F.,\;\al C.)$ where the {\bf field algebra}
$\al F.$ is a unital {\rm C*}--algebra containing
the {\bf constraint set} $\al C.=\al C.^*.$ A
{\bf constraint condition} on $(\al F.,\,\al C.)$ consists of
the selection of the physical
state space by:
\[
  {\got S}_D:=\Big\{ \omega\in{\got S}({\al F.})\mid\pi_\omega(C)
              \Omega_\omega=0\quad {\forall}\, C\in {\al C.}\Big\}\,,
\]
where ${\got S}({\al F.})$ denotes the state space of $\al F.$,
and $(\pi_\omega,\al H._\omega,\Omega_\omega)$ denotes the
GNS--data of $\omega$. The elements of ${\got S}_D$ are called
{\bf Dirac states}.
The case of {\bf unitary constraints} means
that $\al C.=\al U.-\EINS$ for a set of unitaries
$\;\al U.\subset\al F._u$, and
for this we will
also use the notation $(\al F.,\,\al U.)$.
\end{Definition}
Thus in the GNS-representation of each Dirac state, the GNS cyclic vector
$\Omega_\omega$ satisfies the physical selection condition
$\pi_\omega(C)\psi=0$ for all $C\in {\al C.}$ for physical states $\psi$.
 The assumption is that all physical information is contained in the
pair $(\al F.,{\got S}_D)$.

In our case, of the system defined in Section~\ref{Gauge--Gauss}, we will take
the  field algebra defined above:
$\al F._e:=({\mathfrak A}_{\Lambda}\oplus\C)\rtimes_\alpha (\gaud \Lambda).$
In representations
 $\pi$ for which $t\to\pi(U_{\exp(t\nu)})$ is continuous for all $\nu\in\Lgau\Lambda,$
 the concrete Gauss law elements $\pi(d\alpha(\nu)))\in {\rm Der}\big(\pi(
 {\mathfrak F}_{\Lambda}^\infty \otimes {\cal L}[E]^\infty)\big)$
 are given by $\pi(d\alpha(\nu))(A)={i\big[B_\nu,A]}$ for $A\in{\pi(
 {\mathfrak F}_{\Lambda}^\infty \otimes {\cal L}[E]^\infty)}$ where
$\pi(U_{\exp(t\nu)})=\exp(itB_\nu)$. These are enforced as state constraints by selecting the physical subspace
by the condition
 $B_\nu\psi=0$. This condition is the same as $\pi(U_{\exp(t\nu)})\psi=\psi$ for all $t\in\R$.
 As $G$ is a compact connected Lie group, each element in $G$ is an exponential, hence
this also holds for any finite product of $G\hbox{'s}$ and hence for $\gauc \Lambda $.
 Thus the condition
 $B_\nu\psi=0$ for all $\nu$ is the same as $\pi(U_{g})\psi=\psi$ for all $g\in \gau \Lambda$.
 This justifies our choice for constraint set as $\al C.=U\s\gau \Lambda.-\EINS$, i.e. we have the
case of  unitary constraints with $\al U.=U\s\gau \Lambda.$.
Our system with unitary constraints is the pair $(\al F._e,\;U\s\gau \Lambda.)$.

For the general case of unitary constraints $(\al F.,\,\al U.)$, we have the following equivalent
characterizations of the Dirac states
(cf.~\cite[Theorem~2.19~(ii)]{Grundling85}):
\begin{eqnarray}
  \label{DiracU1}{\got S}_{{D}}&=&\Big\{ \omega\in{\got S}({\al F.})\mid
                   \omega(U)=1 \quad {\forall}\, U\in\al U.\Big\} \\[1mm]
  \label{DiracU2}              &=&\Big\{ \omega\in{\got S}({\al F.})\mid
                   \omega(FU)=\omega(F)=\omega(UF) \quad {\forall}\,
                   F\in\al F.,\; U\in\al U.\Big\}.\qquad
\end{eqnarray}
From these, we note that
${\got S}_{{D}}$ is already selected by any set $\al U._0\subset \al F._u$ which generates the same group as $\al U.$
in $\al F._u$.
In particular, in the context of our lattice model, a useful generating subset of $U\s\gau \Lambda.$ is
\[
\al U._0:=\{U_{\exp(t\nu)}\,\mid t\in\R,\;\nu=Y\cdot\delta_x\quad\hbox{for all}\quad
Y\in{\mathfrak g},\,x\in\Lambda^0\},
\]
and in fact the system we will analyze below is $(\al F._e,\,\al U._0)$.

Observe  that (\ref{DiracU2}) shows that $\alpha\s\gau \Lambda.$ leaves every Dirac state invariant,
i.e.~we have $\omega\circ\alpha_g=\omega$ for all
$\omega\in {\got S}_{{D}}$, $g\in \gau \Lambda$.
Since $\al F._e$ is a crossed product,
on the kinematic field algebra ${\mathfrak A}_{\Lambda}\subset\al F._e$ we also have the converse:
\begin{Proposition}
\label{GSisGD}
For the system above with unitary constraints $(\al F._e,\;U\s\gau\Lambda.)$, we have that
${\got S}_{{D}}\restriction {\mathfrak A}_{\Lambda}={\got S}^\gauc\restriction {\mathfrak A}_{\Lambda}$
where ${\got S}^\gauc=\Big\{ \omega\in{\got S}({\al F._e})\mid
                   \omega\circ\alpha_g=\omega \quad {\forall}\, g\in \gau \Lambda\Big\}$,
and where $\alpha_g$ was extended from ${\mathfrak A}_{\Lambda}$ to  ${\al F._e}$
by setting it to be  $\alpha_g={\rm Ad}(U_g)$.
\end{Proposition}
 \noindent {\bf Proof:}
We already know that ${\got S}_{{D}}\subseteq {\got S}^\gauc$, hence that
${\got S}_{{D}}\restriction {\mathfrak A}_{\Lambda}\subseteq{\got S}^\gauc\restriction {\mathfrak A}_{\Lambda}$.
We only need to prove the inclusion
${\got S}^\gauc\restriction {\mathfrak A}_{\Lambda}\subseteq{\got S}_{{D}}\restriction {\mathfrak A}_{\Lambda}$,
i.e. that any $\omega\in {\got S}^\gauc\restriction {\mathfrak A}_{\Lambda}$ has an extension to $\al F._e$
as a Dirac state.
First recall that
 \[
 \al F._e=({\mathfrak A}_{\Lambda}\oplus\C)\rtimes_\alpha (\gaud \Lambda)
=\big[U\s\gau \Lambda.\cdot({\mathfrak A}_{\Lambda}\oplus\C)\big]
\]
where we use the notation $[\cdot]$ for the closed linear space
generated by its argument.
Let $\omega\in {\got S}^\gauc({\mathfrak A}_{\Lambda})$, then by Corr.~2.3.17~\cite{Brat} we obtain a covariant
representation $(\pi_\omega,V^\omega)$ of the action $\alpha:\gau \Lambda\to{{\rm Aut}({\mathfrak A}_{\Lambda}\oplus\C)}$
such that $V^\omega_g\Omega_\omega=\Omega_\omega$ for all
$g\in \gau \Lambda$. By Prop.~7.6.4 and Theorem~7.6.6 in~\cite{Ped} we know that this covariant pair defines a
representation $\wt\pi:{({\mathfrak A}_{\Lambda}\oplus\C)\rtimes_\alpha \gaud\Lambda}\to\al B.(\al H._\omega)$
by $\wt\pi(A):=\pi_\omega(A)$ and $\wt\pi( U_g):=V^\omega_g$
for all $A\in{\mathfrak A}_{\Lambda}$, $g\in \gau \Lambda$.
It is obvious that this representation extends $\pi_\omega$, hence we can define an extension of
$\omega$ to $\al F._e$ by
$\wt\omega(F):={(\Omega_\omega,\wt\pi(F)\Omega_\omega)}$ for all $F\in \al F._e.$
Since $\wt\omega$ is a state, and as
$\wt\omega(U_g)={(\Omega_\omega,V_g^\omega\Omega_\omega)}=1$ it follows that $\wt\omega\in{\got S}_{{D}}$
on $\al F._e$, and this concludes the proof.
\hfill $\rule{2mm}{2mm}$
\vspace{3mm}\break
Thus on the kinematical field algebra ${\mathfrak A}_{\Lambda}$, the Dirac states and the $\gau \Lambda\hbox{--invariant}$
states are the same.

The choice of the Dirac states for a  constraint system $(\al F.,\;\al C.)$, determines a lot of structure.
First, let $N_\omega:={\{F\in\al F.\mid\omega(F^*F)=0\}}$ be the left kernel of
a state $\omega$ and let
$\al N.:=\cap\; \{N_{\omega}\mid\omega\in{\got S}_D \}$.
Then $\al N.=[\al F.\al C.]$
(where we use the notation $[\cdot]$ for the closed linear space
generated by its argument), as every closed
left ideal is the intersection of the left kernels which contains it
(cf.~3.13.5 in \cite{Ped}).
Thus $\al N.$ is the left ideal generated by $\al C.$.
Since $\al C.$ is selfadjoint and contained in
$\al N.$ we conclude $\al C.\subset {\rm C}^*(\al C.)\subset
\al N.\cap\al N.^*=[\al F.\al C.]\cap[\al CF.]$, where ${\rm C}^*(\cdot)$
denotes the C*--algebra in $\al F.$ generated by its argument.

By Theorem~5.2.2 in~\cite{Pa94}, we know that
if $\al A.$ is a Banach algebra with a bounded
left approximate identity and $T : \al A.\to \al B.(X)$ is a continuous representation
of $\al A.$ on the Banach space $X\,,$ then for each $y\in\overline{{\rm Span}(T(\al A.)X)}$ there are
elements $a\in\al A.$ and $x \in X$ with $y = T(a)x\,,$ i.e. $[T(\al A.)X]=T(\al A.)X$.
Thus, if $X =\al F.$ and $T : C^*(\al C.)\to \al B.(X)$ is defined by
$T(C)F:=CF$, then $\al N.^*=[\al CF.]=[C^*(\al C.)\al F.]=C^*(\al C.)\al F.$, hence $\al N.=\al F.C^*(\al C.)$.

\begin{Theorem}
\label{Teo.2.1}
Now for the Dirac states we have~\cite{GL}:
\begin{itemize}
\item[{\rm (i)}] ${\got S}_{{D}}\neq\emptyset\;$ iff
   $\;\EINS\not\in {\rm C}^*(\al C.)$ iff
$\;\EINS\not\in \al N.\cap\al N.^*=:\al D.$.
\item[{\rm (ii)}] $\omega\in {\got S}_D\;$ iff
   $\; \pi_{\omega}({\al D.})\Omega_{\omega}=0$.
\item[{\rm (iii)}] An extreme Dirac state is pure.
\end{itemize}
\end{Theorem}

We will call a constraint set $\al C.$ {\it first class} if
$\EINS\not\in {\rm C}^*(\al C.)$, and this is the nontriviality
condition which needs to be checked \cite[Section~3]{Grundling88a}.

For our system  $(\al F._e,\;U\s\gau\Lambda.)$, we automatically have ${\got S}_{{D}}\neq\emptyset$,
since $\al F._e$ always has the trivial Dirac state $\omega_0$ given by
$\omega_0\big(U\s\gauc \Lambda.{\mathfrak A}_{\Lambda}\big)=0$, $\omega_0(U\s\gauc \Lambda.)=1$
on $\al F._e$.
However, to verify that constraining will produce physically nontrivial results, we need to check
 via Proposition~\ref{GSisGD}  that there
are gauge invariant states on ${\mathfrak A}_{\Lambda}\subset\al F._e$,
as these will extend to Dirac states on $\al F._e$ for which $N_\omega\cap N^*_\omega$ will not
contain ${\mathfrak A}_{\Lambda}$. At the end of Subsection~\ref{GlobalGT} we constructed a representation
$\pi=\pi_{\rm Fock}\otimes\pi_\infty$ which was covariant and had a nonzero invariant vector.
The vector state of this invariant vector is therefore a gauge invariant state on ${\mathfrak A}_{\Lambda}$,
and shows that our constraint system is physically nontrivial.

We recall the rest of the T-procedure before we implement it for the present system.
Define
\[
 {\al O.} := \{ F\in {\al F.}\mid [F,\, D]:= FD-DF \in {\al D.}\quad
               {\forall}\, D\in{\al D.}\}.
\]
Then ${\al O.}$ is the C$^*$--algebraic analogue of Dirac's observables
(the weak commutant of the constraints) \cite{Dirac}.

\begin{Theorem}
\label{Teo.2.2}
With the preceding notation we have~\cite{GL}:
\begin{itemize}
\item[{\rm(i)}] $\al D.=\al N.\cap \al N.^*$ is the unique maximal
  {\rm C}$^*$--algebra in $\, \cap\; \{ {\rm Ker}\,\omega\mid \omega\in
  {\got S}_{{D}} \}$. Moreover $\al D.$ is a hereditary
  {\rm C}$^*$--subalgebra of $\al F.$, and $\al D.=[\al CFC.]$.
\item[{\rm(ii)}] ${\al O.} = {M}_{\al F.}({\al D.})
  :=\{ F\in{\cal F}\mid FD\in{\cal D}\ni DF\quad\forall\, D\in{\cal D}\}$,
  i.e.~it is the relative multiplier algebra of ${\al D.}$ in ${\al F.}$.
\item[{\rm(iii)}] $\al O.=\{F\in\al F.\mid\; [F,\,\al C.]\subset\al D.\}$.
\item[{\rm(iv)}] $\al D.=[\al OC.]=[\al CO.]=[\al COC.]$. (Thus  $\al D.=\al O.C^*(\al C.)=C^*(\al C.)\al O.$  by Theorem~5.2.2 in~\cite{Pa94} quoted above).
\item[{\rm(v)}] For the present case $\al C.=U\s\gau\Lambda.-\EINS$, we have
 $U\s\gau \Lambda.\subset\al O.$ and $\al O.={\{F\in\al F._e\mid\alpha_g(F)-F\in\al D.
\quad\forall\;g\in\gau \Lambda\}}$.
\end{itemize}
\end{Theorem}
 \noindent {\bf Proof:}
Only the last statement in (i) needs proof, as the rest is in~\cite{GL}.
Clearly $[\al CFC.]\subseteq \al N.\cap \al N.^*=\al D.$
is hereditary by Theorem~3.2.2 in \cite{Mur}. Since
\[
\al N.=[\al F.\al C.]=\{F\in\al F.\,\mid\,F^*F\in[\al CFC.]\}
\]
it follows from  Theorem~3.2.1 in \cite{Mur} that $[\al CFC.]=\al N.\cap \al N.^*=\al D.$.
\hfill $\rule{2mm}{2mm}$
\vspace{3mm}\break
Thus $\al D.$ is a closed two-sided ideal of $\al O.$ and it
is proper when ${\got S}_D\not=\emptyset$ (which is the case for our current example).
From (iii) above, we see that the traditional observables $\al C.'\subset\al O.$,
where $\al C.'$ denotes the relative commutant of $\al C.$ in $\al F._e$.
(In our case  $\al C.'$ is just the gauge invariant elements
of $\al F._e$.) Note also that two constraint sets $\al C._1,\;\al C._2$ which
select the same set of Dirac states ${\got S}_{{D}}$, will produce the same algebras
$\al D.$ and $\al O.$, but need not produce the same
traditional observables, i.e. $\al C._1'\not=\al C._2'$ is possible.
In examples, $\al O.$ is generally much harder to obtain explicitly than $\al C.'\subset\al O.$.
In our example, $U\s\gau \Lambda.$ and $\al U._0$ will
produce the same  $\al D.$ and $\al O.$.

Define the {\it maximal {\rm C}$^*$--algebra of physical observables} as
\[
 {\al R.}:={\al O.}/{\al D.}.
\]
This method of constructing $\al R.$  from $(\al F.,\,\al C.)$
is called the {\bf T--procedure}. We call the factoring map
$\xi:\al O.\to\al R.$ the {\bf constraining homomorphism}.
We require that after the T--procedure all physical
information is contained in the pair $({\al R.}\kern.4mm ,{\got S}
({\al R.}))$, where ${\got S}({\al R.})$ denotes the set of
states on $\al R.$.
The following result justifies the choice of $\al R.$ as the
algebra of physical observables (cf. Theorem 2.20 in \cite{Grundling85}):

\begin{Theorem}
\label{Teo.2.6}
 There exists a w$^*$--continuous isometric affine bijection
         between the Dirac states on ${\al O.}$ and the states on ${\al R.}$.
\end{Theorem}
An established alternative method for enforcing constraints (cf.~\cite{KR}) for lattice QCD, is to take the
traditional observables $\al C.'$ (gauge invariant observables) and then to factor out
by the ideal generated by the Gauss law (the state constraint $\al C.)$.
Since $\al C.$ need not be in $\al C.'$ (e.g. for nonabelian gauge theory), the term ``ideal generated by the Gauss law''
needs interpretation. The easiest interpretation of this ideal, is as the intersection of $\al C.'$ with the ideal
which $\al C.$ generates in  $C^*(\al C.'\cup\al C.)\subseteq\al O.$. By
Theorem~\ref{Teo.2.12} below, the ideal generated by $\al C.$ in  $C^*(\al C.'\cup\al C.)$
is just $C^*(\al C.'\cup\al C.)\cap\al D.$, hence
the ``ideal generated in $\al C.'$ by $\al C.$'' is
just $\al D.\cap \al C.'$. Thus the physical algebra obtained is
$\al C.'/(\al D.\cap \al C.')\subset {\al O.}/{\al D.}=\al R.$.
For particular field algebras, these algebras can coincide
(cf. \cite{GL} for the Weyl algebra with linear constraints), and below in Theorem~\ref{structurLocalObs}  we will show for
our model on a finite lattice, that they also do.

We can gain further understanding of the
algebras $\al D.,\;\al O.,\;\al R.$
through the hereditary property of $\al D..$
Denote by $\pi_u$ the universal representation of $\al F.$ on the
universal Hilbert space $\al H._u$ \cite[Section~3.7]{Ped}.
$\al F.''$ is the strong closure of $\pi_u(\al F.)$ and since $\pi_u$
is faithful we make the usual identification of $\al F.$
with a subalgebra of $\al F.''$, i.e.~generally omit explicit
indication of $\pi_u$. If $\omega\in{\got S}(\al F.)$, we will
use the same symbol for the unique normal
extension of $\omega$ from $\al F.$ to
$\al F.''$. Recall the definition from Pedersen~\cite{Ped}:
\begin{Definition}
For a C*-algebra $\al F.,$
a projection $P\in\al F.''$ is {\bf open} if $\al L.=\al F.\cap(\al F.''P)$
is a closed left ideal of $\al F..$
\end{Definition}
We then know from Theorem~3.10.7, Proposition~3.11.9 and Remark~3.11.10
in  Pedersen~\cite{Ped} that the open projections are in
bijection with hereditary C*-subalgebras of $\al F.$ by
$P\to P\al F.''P\cap\al F.\,.$
\begin{Theorem}
\label{Teo.2.7}
For a constrained system $(\al F.,\al C.)$ there is an open
 projection 
 $P\in\al F.''$ such that \cite{GL}:
\begin{itemize}
 \item[{\rm (i)}] $\al N.=\al F.''\,P\cap \al F.$,
 \item[{\rm (ii)}] $\al D.=P\,\al F.''\,P \cap \al F.$ and
 \item[{\rm (iii)}] ${\got S}_D=\{\omega\in{\got S}(\al F.)\mid\omega(P)=0\}$.
\end{itemize}
\end{Theorem}
\begin{Theorem}
\label{Teo.2.10}
Let $P$ be the open projection in Theorem~$\ref{Teo.2.7}$. Then \cite{GL}:
 \[
 \al O.=\{ A\in\al F. \mid PA(\EINS-P)=0=
                        (\EINS-P)AP \}=P'\cap \al F.
\]
\end{Theorem}
\noindent What these two last theorems mean, is that with respect to
the decomposition
\[
 \al H._u=P\,\al H._u\oplus (\EINS-P)\,\al H._u
\]
we may rewrite
\begin{eqnarray*}
 \al D.&=& \Big\{ F\in\al F.\;\Big|\; F=
          \left(\kern-1.5mm\begin{array}{cc}
              D \kern-1.6mm & 0 \\ 0 \kern-1.6mm & 0
          \end{array} \kern-1.5mm\right),\;
       D\in P\al F.''P \Big\}\;\;{\rm and}  \\
 \al O.&=& \Big\{ F\in\al F.\;\Big|\; F=
          \left(\kern-1.5mm\begin{array}{cc}
              A \kern-1.6mm & 0 \\ 0 \kern-1.6mm & B
          \end{array} \kern-1.5mm\right),\;
      A\in P\al F.''P,\; B\in(\EINS-P)\al F.''(\EINS-P) \Big\}\,.
\end{eqnarray*}
It is clear that in general $\al O.=P'\cap \al F.$ can be much greater than the
traditional observables $\al C.'\cap\al F.$.
To appreciate this difference, consider the example where
$\al F.=\al B.(\al H.)$ for a separable Hilbert space $\al H.,$
and let $\al C.=\al K.(\al H.)\equiv\;$compact operators.
Then $\al C.'=\C\EINS,$ but $\al O.=\al B.(\al H.).$

We can identify the
final algebra of physical observables $\al R.$ with a subalgebra of
$\al F.''\,:$

\begin{Theorem}
\label{Teo.2.11}
For $P$ as above we have:
\[
 \al R.\cong  
 (\EINS-P)\,(P'\cap\al F.) \subset \al F.''.
\]
\end{Theorem}
Notice that this just means that $\al R.$ is the restriction of $P'\cap\al F.$
to the subspace $(\EINS-P)\al H._u$ of the universal representation,
and that $(\EINS-P)\al H._u$ is the annihilator of $\al N.,$ hence of
$\al C.\,.$ Thus a simplified (equivalent) version of the T-procedure, is to select
the space $\{\psi\in\al H._u\,\mid\,\pi_u(\al C.)\psi=0\}=(\EINS-P)\al H._u$,
select the set of all elements of $\al F.$ which preserves this space together
with their adjoints (this is $P'\cap\al F.$), and restrict it to
$(\EINS-P)\al H._u$ to obtain $\al R.$. In other words, it is just the enforcement of the constraints
by state condition in the universal representation.

Regarding transformations of the system, consider the automorphisms of $\al F.$
which factor through to $\al R.$, i.e. those which preserve both $\al O.$ and $\al D.$.
Define
\[
\Upsilon:=\set\alpha\in\aut\al F.,{\al D.=\alpha(\al D.)}.,
\]
 then since
${\al O.=\al M._\al F.(\al D.)}$, an $\alpha\in\Upsilon$ also preserves $\al O.$ and so defines
canonically an automorphism $\alpha'$ on $\al R.$ when we factor out by $\al D.$.
Define the group homomorphism ${T:\Upsilon
\mapsto\aut\al R.}$ by ${T(\alpha)}=\alpha'$, then  ${\ker T}$ consists of all the transformations which become the identity
on the physical algebra $\al R.$, i.e. ``gauge transformations'' (in a different sense than encountered above).
In fact, in the case of our assumed constraint system $(\al F._e,\;U\s\gau \Lambda.)$
we obtain from Theorem~\ref{Teo.2.2}(v)and (ii)  that $\alpha\s\gau \Lambda.={\rm Ad}\,U\s\gau \Lambda.\subset\ker T,$ so we can
indeed claim that  $\al R.$ consists of gauge invariant observables,
though not necessarily obtained from the traditional gauge invariant observables
$U\s\gau \Lambda.'\subset\al F._e$.\medskip

 We now return for a more detailed analysis of our assumed constraint system $(\al F._e,\,\al U._0)$.
Our strategy in the rest of this paper, will be to first analyze the constraint systems for finite lattices,
and then to use these to analyze the full system.\medskip

 Recall from Proposition~\ref{indLimitA} that ${\mathfrak A}_{\Lambda}$ has an inductive limit structure over
any directed set $\cl S.$  of open, bounded convex subsets of $\R^3$ such that
$\bigcup\limits_{S\in\cl S.}S=\R^3,$ partially ordered by inclusion. In particular
\[  {\mathfrak A}_{\Lambda} =\ilim{\mathfrak A}_S=\ilim\big({\mathfrak F}_{S}\otimes{\cal L}_{S}[E]\big)\]
 where ${\mathfrak F}_{S}:=C^*\big(\mathop{\cup}\limits_{x\in\Lambda_S^0}{\mathfrak F}_x\big)$ and
 ${\cal L}_S[E]:=C^*\big(\mathop{\cup}\limits_{{\bf n}\in\N^\infty}\cl E._S[\bn]\big)$.
Here we will only be concerned with the particular case where $\cl S.$ is a linear increasing chain
$\cl S.=\{S_k\mid k\in\N\}$, $S_1\subset S_2\subset S_3\subset\cdots$.
Note that as each $S_i$ is open bounded and convex, it only contains finitely many lattice points.
We can equip
\[
\al U._0:=\{U_{\exp(t\nu)}\,\mid t\in\R,\;\nu=Y\cdot\delta_x\quad\hbox{for all}\quad
Y\in{\mathfrak g},\,x\in\Lambda^0\},
\]
with the same inductive limit structure as follows.
Let
\[
\al U._0^S:=\{U\in \al U._0 \,\mid [U,{\mathfrak A}_S]\not=0\}=
\{U_{\exp(tY\cdot\delta_x)}\,\mid t\in\R\backslash 0,\;
Y\in{\mathfrak g}\backslash 0,\,x\in S_e\}
\]
where the ``lattice envelope'' $S_e$ of $S\in\cl S.$ is
\[
S_e:=\{x\in\Lambda^0\,\mid\,\exists\; \ell=(x_\ell,y_\ell)\in\Lambda^1\;\;\hbox{such that}\;\;
\ell\cap S\not=\emptyset\;\;\hbox{and}\;\;
x_\ell=x\;\;\hbox{or\;}\;y_\ell=x \}\,.
\]
If  we denote $\al C._i:=\al U._0^{S_i}-\EINS,$ then $\al C._1\subset\al C._2\subset\ldots,$
and $\al C.=\al U._0-\EINS=\mathop{\cup}\limits_{i=1}^\infty
\al C._i$. The group generated by $\al U._0^{S_i}$ is denoted by $U\s\gauc S_i.$, so these sets of unitaries produce equivalent constraint sets.
We also obtain an inductive limit for
$\al F._e=\big[U\s\gaud \Lambda.\cdot({\mathfrak A}_{\Lambda}\oplus\C)\big]$ by
\[
\al F._e =\ilim\al F._S\quad\hbox{where}\quad \al F._S:=\big[ U\s\gauc S.\cdot({\mathfrak A}_S\oplus\C)\big]
=\big[ U\s\gauc S.\cdot({\mathfrak F}_{S}\otimes{\cal L}_{S}[E]\oplus\C)\big]\,.
\]
This suggests that we analyze the ``local constraint systems'' ${(\al F._{S_i},\al C._i)}$.
Below we will see that $i<j$ implies that $\al U._0^{S_i}=\al U._0^{S_j}\cap\al F._{S_i}$, hence the
set of constraint systems ${(\al F._{S_i},\al C._i)},$ $i\in\N$, is a
system of local quantum constraints in the sense of~\cite{GL} (Def.~3.3).
Such systems were studied in detail in~\cite{GL}, and in Section~\ref{GL-Tprfull} below, we will apply this analysis.
However, first we need to solve the constraint system for an individual
``local'' system ${(\al F._{S_i},\al C._i)}.$ To do so, we will solve the corresponding system for
a finite lattice in the next subsection.

We start with the constraint system
for the {\it finite} lattice, i.e. the system $(\al F._{S_i}^F,\,\al C._i)$ where
\[
\al F._{S_i}^F:=\big[ U\s\gauc S_i.\cdot(\CAR {\al H._{S_i}}.
\otimes{\cal L}^{(S_i)}\oplus\C)\big]
\subset M({\mathfrak A}_{\Lambda}\rtimes_\alpha (\gaud \Lambda))
\supset \al F._e
\]
with the same constraints  $\al C._i:=\al U._0^{S_i}-\EINS.$
Note that $\al F._{S_i}^F$ is not contained in $\al F._e$,
though $\al C._i\subset{\al F._e\cap\al F._{S_i}^F}$.
Moreover  $\al F._{S_i}^F$ only differs from  $\al F._{S_i}$
by the replacement of ${\cal L}^{S_i}$  by ${\cal L}_{S_i}[E]$.

\subsection{Enforcing the Gauss law constraint for finite lattices.}
\label{FinLat-Tproc}

In this subsection we will obtain a full analysis of the constraint data ${\big(\al D._i^F,\,\al O._i^F,\,\al R._i^F\big)}$
for the  the finite lattice system $(\al F._{S_i}^F,\,\al C._i)$  in $S_i$.
First observe that
\begin{eqnarray*}
\al F._{S_i}^F&=&\big[ U\s\gauc S_i.\cdot(\al A._{S_i}\oplus\C)\big]=\left(
\al A._{S_i}+\C\right)\rtimes_\alpha (\gaud S_i)\\[1mm]
&=&\left[ U\s\gauc S_i.\cdot\al A._{S_i}\right]+[U\s\gauc S_i.]\qquad\hbox{where}\qquad
\al A._{S_i}:={\mathfrak F}_{S_i}\otimes{\cal L}^{(S_i)}
\end{eqnarray*}
and $\gaud S_i:= \big\{\gamma\in\gauc \Lambda\,\mid\,{\rm supp}(\gamma)\subset (S_i)_e\big\}\cong\prod\limits_{x\in (S_i)_e}G$
with the discrete topology.
\begin{Lemma}
\label{Fideal}
With notation as above, we have that
$\left[ U\s\gauc S_i.\al A._{S_i}\right]$ is a closed two-sided ideal of $\al F._{S_i}^F$.
Moreover $\left[ U\s\gauc S_i.\al A._{S_i}\right]\cap[U\s\gauc S_i.]=\{0\}$, i.e.
$\al F._{S_i}^F\big/\left[ U\s\gauc S_i.\al A._{S_i}\right]\cong[U\s\gauc S_i.]$.
\end{Lemma}
 \noindent {\bf Proof:}
That $\left[ U\s\gauc S_i.\al A._{S_i}\right]$ is a closed two-sided ideal of $\al F._{S_i}^F$
is clear by construction, so we prove the second statement.
Consider a faithful representation $V:[U\s\gauc S_i.]\to\al B.(\al H.)$, and let
$\varphi:\al A._{S_i}+\C\to\al B.(\al H.)$ be the character $\varphi(A+\lambda\un)=\lambda\un$
for all $A\in \al A._{S_i}$, $\lambda\in\C$. Then the pair $(V,\varphi)$ defines a covariant representation,
i.e. $\varphi(\alpha_g(B))=V(U_g)\varphi(B)V(U_g)^*=\varphi(B)$ for all $B\in \al A._{S_i}+\C$
and $g\in \gauc S_i$. Thus it defines a representation $\pi$ of the crossed product
$\al F._{S_i}^F=\left(\al A._{S_i}+\C\right)\rtimes_\alpha (\gaud S_i)$ on $\al H.$ by
$\pi(B):=\varphi(B)$ for all $B\in \al A._{S_i}+\C$, and $\pi(U_g)=V(U_g)$ for all $g\in \gauc S_i$.
As $\al F._{S_i}^F=\left[ U\s\gauc S_i.\al A._{S_i}\right]+[U\s\gauc S_i.]$, it is obvious that
$\pi\big(\left[ U\s\gauc S_i.\cdot\al A._{S_i}\right]\big)=0$ and $\pi$ is
faithful on $[U\s\gauc S_i.]$, hence $\pi(\al F._{S_i}^F)\cong[U\s\gauc S_i.]$
and $\ker(\pi) = \left[ U\s\gauc S_i.\al A._{S_i}\right]$.
\hfill $\rule{2mm}{2mm}$
\vspace{3mm}\break
As $\al H._{S_i}$ is finite dimensional, ${\mathfrak F}_{S_i}=\CAR {\al H._{S_i}}.$ is just a
(full) matrix algebra, and as ${\cal L}^{(S_i)}={\bigotimes\{\cl L._{k}\,\mid\,\ell_k\cap S_i\not=\emptyset\}}$
is a finite tensor product of factors $\cl L._k\cong{\cal K}(\cl H.)$, it is isomorphic to ${\cal K}(\cl H.)$ and hence
$\CAR {\al H._{S_i}}.\otimes{\cal L}^{(S_i)}\cong{\cal K}(\cl H.)$. This has the following consequences:
\begin{itemize}
\item{} The algebra $\al A._{S_i}:=\CAR {\al H._{S_i}}.\otimes{\cal L}^{(S_i)}\cong{\cal K}(\cl H.)$ has (up to unitary equivalence) only one irreducible representation
$\pi:\al A._{S_i}\to\al B.(\al H._\pi)$. This representation
 $\pi$ is faithful, and $\pi\left(\al A._{S_i}\right)={\cal K}(\cl H._\pi)$.
 \item{} Note that the enforcement of the constraints  $\al C._i:=\al U._0^{S_i}-\EINS$ will put
 $[\al U._0^{S_i}]\subset\al F._{S_i}^F$ and hence $[U\s\gauc S_i.]$ equal to $\C\EINS$, hence the only
 nontrivial part of $\al F._{S_i}^F$ which needs to be analyzed w.r.t. constraints is the closed two-sided ideal
 $\left[ U\s\gauc S_i.\cdot\al A._{S_i}\right]= {\al A._{S_i} \rtimes_\alpha (\gaud S_i)}\subset\al F._{S_i}^F$.
\item{} For the action  $\alpha:\gauc S_i \to\aut\al A._{S_i}$, as $\gauc S_i\cong\prod\limits_{x\in (S_i)_e}G$
is compact, we know from \cite{Ros} that the invariance algebra is $\al A._{S_i}^\alpha=p\big(\al A._{S_i}
\rtimes_\alpha (\gauc S_i)\big)p$
for some projection $p\in M\big(\al A._{S_i}
\rtimes_\alpha (\gauc S_i)\big)$, where the equality is realised in the multiplier algebra $ M\big(\al A._{S_i}
\rtimes_\alpha (\gauc S_i)\big)$ using the imbedding of $\al A._{S_i}$ in it.
We will find a similar structure in Theorem~\ref{structurLocalObs} below.
\item{} All automorphisms of ${\cal K}(\cl H.)\cong\al A._{S_i}$ are inner, hence there are unitaries $W_g\in M(\al A._{S_i})$ implementing
$\alpha_g\in\aut\al A._{S_i}=\aut(\al A._{S_i}+\C),$ $g\in\gauc S_i$, and the unitaries are unique up to scalar multiples.
The map $g\mapsto W_g$ need not be a group homomorphism, and it is well known that the obstruction
for this to be the case, is a nontrivial $H^2(\gauc S_i,\T)$ (second Moore cohomology group), cf.~\cite{Raeb}.
Sufficient conditions for a trivial $H^2(\gauc S_i,\T)$ are in \cite{Vara}.
\end{itemize}
\begin{Proposition}
\label{A-irredRep}
Let $G$ be a connected compact Lie group. 
Then
\begin{itemize}
 \item[{\rm (i)}] there is a representation of $\al F.^F_{S_i}$ which is irreducible on the subalgebra
 $\al A._{S_i}\subset\al F.^F_{S_i}$.
Hence the  irreducible representation $\pi$ of  $\al A._{S_i}=\CAR {\al H._{S_i}}.\otimes{\cal L}^{(S_i)}$  is also covariant
for the action $\alpha:\gauc S_i \to\aut\al A._{S_i}$.
 \item[{\rm (ii)}] the action $\alpha:\gauc S_i \to\aut\al A._{S_i}$ is inner, i.e. there is a strictly continuous homomorphism
 $V:\gauc S_i \to UM(\al A._{S_i})$ which implements $\alpha$.
 \item[{\rm (iii)}] Let  $H$ denote either $\gauc S_i$ or $\gaud S_i$. Then we have an isomorphism \\
 $\varphi_H:\al A._{S_i}\rtimes_\alpha H\to\al A._{S_i}\otimes C^*(H)$. Explicitly it is obtained by
 defining $\varphi_H^{-1}(A\otimes f)\in C_c(H,\al A._{S_i})$ by
 $\varphi_H^{-1}(A\otimes f)(g)=AV_g^{-1}f(g)$ for $A\in \al A._{S_i}$, $f\in C_c(H)$ and $g\in H$.
 \item[{\rm (iv)}] We have that
\begin{eqnarray*}
\al F._{S_i}^F&=&\left(\al A._{S_i}+\C\right)\rtimes_\alpha (\gaud S_i)
=\al A._{S_i}\rtimes_\alpha (\gaud S_i)+ C^*(\gaud S_i)  \\[1mm]
\hbox{and that}&\quad&
\xi:\left[\al A._{S_i}\otimes C^*(\gaud S_i)+ \EINS\otimes C^*(\gaud S_i)\right]\to\al F._{S_i}^F
\end{eqnarray*}
is an isomorphism where $\xi(A\otimes f_1+\EINS\otimes f_2)(g)=AV_g^{-1}f_1(g)+V_g^{-1}f_2(g)$
for $A\in\al A._{S_i}$ and $f_i\in C_c(\gaud S_i)$.
\end{itemize}
\end{Proposition}
\noindent {\bf Proof:} (i) Recall that
\[
\alpha_\gamma:=\alpha_\gamma^1\otimes\alpha_\gamma^2\quad
\hbox{where}\quad
\alpha^1:\gauc S_i \to\aut\CAR {\al H._{S_i}}.\quad\hbox{and}\quad\alpha^2:\gauc S_i \to\aut{\cal L}^{(S_i)}\,.
\]
We show there is a product representation of irreducible covariant representations. For $\alpha^1:\gauc S_i \to\aut\CAR {\al H._{S_i}}.$,
we only need to take the Fock representation, which is irreducible and covariant. Regarding $\alpha^2:\gauc S_i \to\aut{\cal L}^{(S_i)}$,
recall that ${\cal L}^{(S_i)}={\bigotimes\limits_{\ell_k\in\Lambda^1_{S_i}}\cl L._{k}}$ where $\Lambda^1_{S_i}:=\{\ell\in
\Lambda^1\,\mid\,\ell\cap S_i\not=\emptyset\}$, and that $\alpha^2=\mathop{\otimes}\limits_{\ell_k\in\Lambda^1_{S_i}}\theta^k$.
However, by Lemma~\ref{Lem.2.12B}(i), we have for each $\cl L._k$ an irreducible representation $\pi_0$ which
is covariant w.r.t. ${\theta^k:G^2\to\aut\al L._k}$.
Thus, by taking the (finite) tensor
product of all of these covariant representations $\pi_0$ for $\theta^k$, $\ell_k\in\Lambda^1_{S_i}$, we obtain an
irreducible covariant representation for $\alpha^2:\gauc S_i \to\aut{\cal L}^{(S_i)}$, and tensoring it with the Fock
representation produces the desired irreducible covariant representation for $\alpha:\gauc S_i \to\aut\al A._{S_i}$.
As $\al F.^F_{S_i}$ is a crossed product, this defines then a representation of $\al F.^F_{S_i}$
which is irreducible on $\al A._{S_i}$. \\
(ii) The representation $\pi:\al F.^F_{S_i}\to\al B.(\al H._\pi)$ in (i)  is irreducible (hence faithful) on $\al A._{S_i}$.
As $\pi\left(\al A._{S_i}\right)={\cal K}(\cl H._\pi)$, which is an essential ideal in ${\cal B}(\cl H._\pi)=M({\cal K}(\cl H._\pi))$,
$\pi$ produces  a faithful homomorphism from ${\cal B}(\cl H._\pi)$ into $M(\al A._{S_i})$ by
Prop.~3.12.8 in~\cite{Ped}. Since $\pi$ is covariant we have the unitary implementers $U_g^\pi:=\pi(U_g)\in {\cal B}(\cl H._\pi)$,
which therefore define the unitaries $V_g\in UM(\al A._{S_i})$ which also implement $\alpha_g$. It is clear that
$g\to V_g$ is a homomorphism which is strictly continuous, since the strong operator topology and strict topology w.r.t. the compacts
coincide on unitaries. Thus the action $\alpha:\gauc S_i \to\aut\al A._{S_i}$ is inner. \\
(iii) By (ii), the action $\alpha$ is inner for $H$.
Thus by Lemma~2.68 and Remark~2.71 of~\cite{Wil},
the action $\alpha:H\to\aut\al A._{S_i}$ is exterior equivalent to the trivial action $\iota:H\to\aut\al A._{S_i}$,
hence the crossed products are isomorphic by $\psi_1:\al A._{S_i}\rtimes_\iota H\to \al A._{S_i}\rtimes_\alpha H$,
where the isomorphism is given by  $\psi_1(A\cdot f)(g):=Af(g)V_g^{-1}$ for $A\in \al A._{S_i}$ and $f\in C_c(H)$.    However, by
Lemma~2.73 of~\cite{Wil} we know that the crossed product $\al A._{S_i}\rtimes_\iota H$ is
isomorphic to the tensor product $\al A._{S_i}\otimes C^*(H)$ (using nuclearity of $\al A._{S_i}$ for the tensor norm).
Explicitly this isomorphism $\psi_0:\al A._{S_i}\otimes C^*(H)\to\al A._{S_i}\rtimes_\iota H$ is given by
$\psi_0(A\otimes f)=A\cdot f$. Thus we obtain the claimed isomorphism $\varphi_H^{-1}:=\psi_1\circ\psi_0,$ which explicitly
is given by $\varphi_H^{-1}(A\otimes f)(g)=\psi_1(A\cdot f)(g)=Af(g)V_g^{-1}$.\\
(iv) This follows from the previous part, keeping in mind that
$\al F._{S_i}^F=C^*\!\!\left(\al U._0^{S_i}\cup\al A._{S_i}\right)=C^*\!\!\left(U\s\gauc S_i.\cup\al A._{S_i}\right)=
\left[ U\s\gauc S_i.\cdot\al A._{S_i}+U\s\gauc S_i. \right],$ and that
$[U\s\gauc S_i.\cdot\al A._{S_i}]=\al A._{S_i}\rtimes_\alpha \gaud S_i$.
\hfill $\rule{2mm}{2mm}$
\vspace{3mm}\break
\noindent {\bf Remark:}
 Observe in part (iv) that if we extend $\xi$ to $M(\al A._{S_i})\otimes C^*(\gaud S_i)$, then  \\
$\xi(V_g\otimes\delta_g)
=\delta_g=U_g$ where $\delta_x(y)=1$ if $x=y$ and it is zero otherwise. So we may consider $U_g$ to be the product of
$V_g$ with some independent part which commutes with $\al A._{S_i}$. Note that $V_g^{-1}U_g$ commutes with
 all of $M(\al A._{S_i})$.

\begin{Theorem}
\label{structurLocalObs}
For the system $(\al F._{S_i}^F,\,\al C._i)$ above, we  have that
\begin{enumerate}
\item[(i)]
There is a projection $P_\alpha\in M(\al A._{S_i})$ such that
$\alpha_g(P_\alpha)=P_\alpha$ for all $g\in\gauc S_i$ and
\begin{eqnarray*}
{\al O._i^F\cap \al A._{S_i}}&=&P_\alpha\al A._{S_i}P_\alpha\oplus(\EINS-P_\alpha)\al A._{S_i}(\EINS-P_\alpha), \\[1mm]
\al D._i^F\cap \al A._{S_i}&=&(\EINS-P_\alpha)\al A._{S_i}(\EINS-P_\alpha)  \\[1mm]
\qquad\hbox{and}\qquad&&
\big(\al O._i^F\cap \al A._{S_i}\big)\Big/\big(\al D._i^F\cap \al A._{S_i}\big)\cong P_\alpha\al A._{S_i}P_\alpha.
\end{eqnarray*}
Moreover  $U_gP_\alpha=P_\alpha=P_\alpha U_g$ for all $g\in\gauc S_i$ and so $\al C._iP_\alpha=0$ where these relations hold
 in the algebra
$\big[ U\s\gauc S_i.\cdot(M(\al A._{S_i})\oplus\C)\big]=\left(
M(\al A._{S_i})+\C\right)\rtimes_\alpha (\gaud S_i)\supset \al F._{S_i}^F$.
\item[(ii)]  Let $\pi:\al F.^F_{S_i}\to\al B.(\al H._\pi)$ be a representation which is irreducible (hence faithful) on $\al A._{S_i}$. Then
$\pi(P_\alpha)$ is the projection onto $\al H._\pi^G:=\{\psi\in\al H._\pi\,\mid\, U_g^\pi\psi=\psi\;\forall\,g\in\gaud S_i\}$,
where $U_g^\pi:=\pi(U_g)$, and
\begin{eqnarray*}
\pi\big({\al O._i^F\cap \al A._{S_i}}\big)
&=&{\cal K}(\cl H._\pi^G)\oplus{\cal K}\big((\cl H._\pi^G)^\perp\big), \qquad
\pi\big({\al D._i^F\cap \al A._{S_i}}\big)
={\cal K}\big((\cl H._\pi^G)^\perp\big)  \\[1mm]
\qquad\hbox{and}\qquad&&
\wt\pi\big((\al O._i^F\cap \al A._{S_i})\big/(\al D._i^F\cap \al A._{S_i})\big)={\cal K}(\cl H._\pi^G)\,,
\end{eqnarray*}
where $\wt\pi$ is the restriction of $\pi\big({\al O._i^F\cap \al A._{S_i}}\big)$ to $\cl H._\pi^G$.
\item[(iii)] We have $\al H._\pi^G\not=\{0\}$.
\item[(iv)]   $P_\alpha=(\EINS-P_{S_i})P_J$ where $P_{S_i}\in (\al F.^F_{S_i})''$ is the open projection of Theorem~\ref{Teo.2.7}
for the constraint system $(\al F._{S_i}^F,\,\al C._i)$,
and $P_J\in(\al F._{S_i}^F)''$ is the central projection determined by the ideal $\left[ U\s\gauc S_i.\cdot\al A._{S_i}\right]$
of $\al F._{S_i}^F$.
\end{enumerate}
\end{Theorem}
\noindent {\bf Proof:}
(ii)
From Proposition~\ref{A-irredRep}, we have the representation $\pi:\al F.^F_{S_i}\to\al B.(\al H._\pi)$
which is irreducible and faithful on $\al A._{S_i}$.
As $\pi\left(\al A._{S_i}\right)={\cal K}(\cl H._\pi)$, which is an essential ideal in ${\cal B}(\cl H._\pi)=M({\cal K}(\cl H._\pi))$,
$\pi$ produces  a faithful homomorphism from ${\cal B}(\cl H._\pi)$ into $M(\al A._{S_i})$ by
Prop.~3.12.8 in~\cite{Ped}. Thus there exists a unique projection $P_\alpha\in M(\al A._{S_i})$ such that
$\tilde\pi(P_\alpha)$ is the projection $P^G$ onto $\al H._\pi^G$ where $\tilde\pi$ is the unique extension of
$\pi$ to  $M(\al A._{S_i})$ on $\cl H._\pi$. By definition we have that $\pi(U_g)\tilde\pi(P_\alpha)=\tilde\pi(P_\alpha)$
and so in the algebra
$\big[ U\s\gauc S_i.\cdot(M(\al A._{S_i})\oplus\C)\big]=\left(
M(\al A._{S_i})+\C\right)\rtimes_\alpha (\gaud S_i)\supset \al F._{S_i}^F$
we obtain $U_gP_\alpha=P_\alpha=P_\alpha U_g$ and so $\al C._iP_\alpha=0$.

 As $\pi$ is irreducible, it is cyclic, hence there is a projection
$P^\pi\in\pi_u(\al F.^F_{S_i})'$ on the  Hilbert space $\al H._u$ of the universal representation $\pi_u:\al F.^F_{S_i}\to\al B.(\al H._u)$ such that
$P^\pi\al H._u\to\al H._\pi$ and $P^\pi\circ\pi_u=\pi$.

Let $P_{S_i}\in (\al F.^F_{S_i})''$ be the open projection of Theorem~\ref{Teo.2.7} for the constraint system $(\al F._{S_i}^F,\,\al C._i)$,
specified by $\omega(P_{S_i})=0$  iff $\pi_\omega(\al C._i)\Omega_\omega=0$
for $\omega\in{\got S}(\al F.^F_{S_i})$.
Thus $\EINS-P_{S_i}$ projects onto the $U\s\gauc S_i.\hbox{--invariant}$ subspace in all representations, and so
 $P^G:=(\EINS-P_{S_i})P^\pi$ is the projection of $\al H._\pi$ onto
$\al H._\pi^G$. 

By Theorem~\ref{Teo.2.10}
we have that  ${\al O._i^F\cap \al A._{S_i}}=P_{S_i}'\cap \al A._{S_i}$, so
\begin{eqnarray*}
\pi\big({\al O._i^F\cap \al A._{S_i}}\big)&:=&P^\pi\big({P_{S_i}'\cap \al A._{S_i}}\big)\\[1mm]
&=&\big\{ \pi(A)\,\big|\,A\in\al A._{S_i}\;\;\hbox{such that}\;\;\big[\pi(A),\, P^\pi(\EINS-P_{S_i})\big]=0\big\}
\quad\hbox{as $P^\pi\in(\al F.^F_{S_i})'$}\\[1mm]
&=& (P^G)'\cap\pi(\al A._{S_i})\\[1mm]
=\Big\{\pi(A)\in\pi(\al A._{S_i})&\Big|& \pi(A)=
          \left(\kern-1.5mm\begin{array}{cc}
              C \kern-1.6mm & 0 \\ 0 \kern-1.6mm & D
          \end{array} \kern-1.5mm\right),\;
      C\in P^G\pi(\al A._{S_i})P^G,\; D\in(\EINS-P^G)\pi(\al A._{S_i})(\EINS-P^G) \Big\}\,,
\end{eqnarray*}
where the matrix decomposition corresponds to the decomposition $\al H._\pi=\al H._\pi^G\oplus(\al H._\pi^G)^\perp.$
Since $\pi\left(\al A._{S_i}\right)={\cal K}(\cl H._\pi)$ we have that $P^G\pi(\al A._{S_i})P^G\in\pi(\al A._{S_i})$, hence
\[
\pi\big({\al O._i^F\cap \al A._{S_i}}\big)=P^G\pi(\al A._{S_i})P^G\oplus(\EINS-P^G)\pi(\al A._{S_i})(\EINS-P^G)
={\cal K}(\cl H._\pi^G)\oplus{\cal K}\big((\cl H._\pi^G)^\perp\big) \,.
\]
Since $\al D._i^F\subseteq P_{S_i}\al O._i^F,$ we conclude that
\[
\pi\big({\al D._i^F\cap \al A._{S_i}}\big)=(\EINS-P^G)\pi(\al A._{S_i})(\EINS-P^G)
={\cal K}\big((\cl H._\pi^G)^\perp\big) \,,
\]
hence as $\pi$ is faithful, $\al O._i^F\cap \al A._{S_i}\cong{\cal K}(\cl H._\pi^G)\oplus{\cal K}\big((\cl H._\pi^G)^\perp\big)$
and $\al D._i^F\cap \al A._{S_i}\cong{\cal K}\big((\cl H._\pi^G)^\perp\big)$ and hence the physical algebra in $\al A._{S_i}$
is
\[
\big(\al O._i^F\cap \al A._{S_i}\big)\Big/\big(\al D._i^F\cap \al A._{S_i}\big)\cong{\cal K}(\cl H._\pi^G)\,,
\]
where the isomorphism is obtained from restricting $\pi\big({\al O._i^F\cap \al A._{S_i}}\big)$ to $\cl H._\pi^G$.

(i) Using the faithful homomorphism from ${\cal B}(\cl H._\pi)$ into $M(\al A._{S_i})$, we obtain the corresponding
statements in $\al A._{S_i}$ for the projection $P_\alpha\in M(\al A._{S_i})$ such that
$\tilde\pi(P_\alpha)=P^G$ from part (ii).

(iii) To see that $\al H._\pi^G\not=\{0\}$, recall from the proof of Proposition~\ref{A-irredRep}(i) that
we have a specific representation $\pi$ as in (ii), and it is
$\pi=\pi_1\otimes\pi_2$ where $\pi_1$ is the Fock representation on the CAR-part of $\al A._{S_i}=\CAR {\al H._{S_i}}.\otimes{\cal L}^{(S_i)}$,
and $\pi_2$ is the (finite) tensor product of  copies of the usual
regular representation of $C(G)\rtimes_\lambda G\cong\cl K.(\cl H.)\cong\al L._\ell$ on $L^2(G)$.
The implementing unitaries of $\alpha^1$ for $\pi_1$ act on the Fock space of the first factor by
second quantized unitaries, hence the vacuum vector is an invariant vector. The implementing unitaries
$W\s{(h,s)}.\in U(L^2(G))$
for the factor actions $\theta$ w.r.t. the  representation $\pi_0$ (cf. Lemma~\ref{Lem.2.12B})
are given by  $\left(W\s{(h,s)}.\psi\right)(g):=\psi(h^{-1}gs)$. It is clear that if $\psi$ is constant
then it is invariant w.r.t. $W$ (as $G$ is compact, the constant functions are in $L^2(G)$).
Thus, the tensor product of the Fock vacuum with copies of these constant vectors will produce a nonzero element in
$\al H._\pi^G$. Thus by (ii) the factor algebra
${(\al O._i^F\cap \al A._{S_i})\big/(\al D._i^F\cap \al A._{S_i})}$ is nontrivial, and as all
representations as in (ii) are faithful, we obtain from (ii) that ${\cal K}(\cl H._\pi^G)\not=\{0\}$ hence
 $\al H._\pi^G\not=\{0\}$ for all $\pi$ as in (ii).

(iv) All representations $\pi:\al F._{S_i}^F\to\al B.(\al H.)$ decompose uniquely into $\pi=\pi_1\oplus\pi_2$ where
$\pi_1(P_J)=\un$ and $\pi_2(P_J)=0$ as $\left[ U\s\gauc S_i.\cdot\al A._{S_i}\right]$ is an ideal
of $\al F._{S_i}^F$. It suffices to verify the claim in all representations of these two types, since then it
follows for the universal representation. If $\pi_1(P_J)=\un$, then $\pi_1$ is nondegenerate on $\left[ U\s\gauc S_i.\cdot\al A._{S_i}\right]\cong
\al A._{S_i}\rtimes_\alpha (\gaud S_i)$, hence on  $\al A._{S_i}$ by this property for crossed products.
In particular, for the representation in (ii) we get that
$\pi_1((\EINS-P_{S_i})P_J)=\pi_1(\EINS-P_{S_i})=P^G=\pi_1(P_\alpha)$.
 This  determines it in the multiplier of $\al A._{S_i}$,
hence for any representation which is nondegenerate on $\al A._{S_i}$, which includes all representations of $\al F._{S_i}^F$
such that $\pi(P_J)=\un$. On the other hand, for a representation $\pi_2$ with $\pi_2(P_J)=0$, we have that
$\pi_2((\EINS-P_{S_i})P_J)=0$. Moreover as $\al A._{S_i}\subset\left[ U\s\gauc S_i.\cdot\al A._{S_i}\right]$
we have (using  ideal structure):
\[
P_\alpha\in M(\al A._{S_i})\subset\al A._{S_i}''\subset\left[ U\s\gauc S_i.\cdot\al A._{S_i}\right]''
=P_J(\al F._{S_i}^F)''\subset(\al F._{S_i}^F)''
\]
and so it is clear that $\pi_2(P_J)=0$ implies that $\pi_2(P_\alpha)=0$. Hence we conclude that $P_\alpha=(\EINS-P_{S_i})P_J$.
\hfill $\rule{2mm}{2mm}$
\vspace{3mm}\break
\noindent {\bf Remarks:}
\begin{enumerate}
\item
Observe that as  ${\cal K}(\cl H._1)\cong{\cal K}(\cl H._2)$  iff   ${\rm dim}(\cl H._1)={\rm dim}(\cl H._2)$, we conclude
from (ii) that ${\rm dim}(\cl H._\pi^G)$ is the same for all representations  $\pi:\al F.^F_{S_i}\to\al B.(\al H._\pi)$
 which are irreducible  on $\al A._{S_i}$.
\item
 Note that as $\gauc S_i$ is compact and $V:\gauc S_i \to UM(\al A._{S_i})$ a strictly continuous homomorphism
(cf. Proposition~\ref{A-irredRep}), that  $P_\alpha$ is the group average of $V$ where the average is taken w.r.t. the
strict topology. Thus $P_\alpha$ is the projection onto the invariant subspace of $V$ in any representation.
\item
Observe that from Theorem~\ref{structurLocalObs}(i), $P_\alpha\al A._{S_i}P_\alpha$ is the gauge invariant
part of the observables ${\al O._i^F\cap \al A._{S_i}}$. Since $\alpha: \gauc S_i\to\aut\al A._{S_i}$ is
strongly continuous and $\gauc S_i$ is compact, we can obtain $P_\alpha\al A._{S_i}P_\alpha$ as a group average,
i.e.
\[
P_\alpha AP_\alpha=\int_{\gauc S_i}\alpha_g(A)\,d\mu(g)\quad\hbox{where $\mu$ is the normalized Haar measure of}\;\;\gauc S_i
\]
for $A\in\al A._{S_i}$.
This produces a well-known method for construction of gauge invariant observables for finite lattice systems.
 In fact, from Theorem~\ref{FLocalObs} below, we will see that for finite lattice systems,
$\al R._i^F=\al O._i^F\big/\al D._i^F\cong [P_\alpha\al A._{S_i}P_\alpha] +\C$, so all physically relevant
observables in  $\al F.^F_{S_i}$ can be obtained this way.
\end{enumerate}

We will now show that the
physical algebra obtained above, coincides with the physical algebra produced by the traditional constraint method
(cf. Theorem~4.1 in~\cite{JKR}):
\begin{Theorem}
\label{structurTradLocalObs}
For the system $(\al F._{S_i}^F,\,\al C._i)$ above, the set of traditional observables in
$\al A._{S_i}$, i.e. $\al C._i'\cap\al A._{S_i}$,  is the gauge invariant part $\al A.^G_{S_i}$ of $\al A._{S_i}$.
Let   $\pi:\al F.^F_{S_i}\to\al B.(\al H._\pi)$
be an irreducible representation which is irreducible on $\al A._{S_i}$, then
\[
\pi(\al A.^G_{S_i})=\pi\left(\al C._i'\cap\al A._{S_i}\right)={\cal K}(\cl H._\pi^G)\oplus
\Big({\cal K}\big((\cl H._\pi^G)^\perp\big)\cap \pi(\al C._i)'\Big)\,.
\]
The ``ideal generated in $\al A.^G_{S_i}$ by the Gauss law'' is taken to be
${\al D._i^F\cap \al A.^G_{S_i}}$, and its image w.r.t. $\pi$ is
\[
\pi\big({\al D._i^F\cap \al A.^G_{S_i}}\big)
={\cal K}\big((\cl H._\pi^G)^\perp\big)\cap \pi(\al C._i)' \,,
\]
hence the (traditional) algebra of physical observables is
\[
\al A.^G_{S_i}\big/({\al D._i^F\cap \al A.^G_{S_i}})\cong
{\cal K}(\cl H._\pi^G)
\]
where the isomorphism is obtained by restricting $\pi(\al A.^G_{S_i})$ to $\cl H._\pi^G$.
\end{Theorem}
\noindent {\bf Proof:}
As $\al A._{S_i}$ commutes with $\al U._0\backslash\al U._0^{S_i}$ we have
$\al A.^G_{S_i}=\al A.^{G_i}_{S_i}=\al A._{S_i}\cap(\al U._0^{S_i})'=\al A._{S_i}\cap\al C._i'
\subseteq{\al O._i^F\cap \al A._{S_i}}.$ Thus by Theorem~\ref{structurLocalObs}(ii):
\[
\pi(\al A.^G_{S_i})\subseteq\pi\big({\al O._i^F\cap \al A._{S_i}}\big)={\cal K}(\cl H._\pi^G)\oplus{\cal K}\big((\cl H._\pi^G)^\perp\big)
=\Big\{
          \left(\kern-1.5mm\begin{array}{cc}
              C \kern-1.6mm & 0 \\ 0 \kern-1.6mm & D
          \end{array} \kern-1.5mm\right)\,\Big|\,
      C\in {\cal K}(\cl H._\pi^G),\; D\in{\cal K}\big((\cl H._\pi^G)^\perp\big) \Big\}
\]
Note that for $g\in\gaud S_i$ we have the decomposition
\[
  U_g^\pi=\left(\begin{array}{cc} \EINS
& 0\\ 0 & V_g
\end{array}\right)\qquad\hbox{for}\qquad V_g=(\EINS-P^G)U_g^\pi\in U\big((\cl H._\pi^G)^\perp\big),
\]
hence
\begin{eqnarray*}
\pi(\al A.^G_{S_i})&\subseteq&\big(U_{\gaud S_i}^\pi\big)'\cap\Big\{
          \left(\kern-1.5mm\begin{array}{cc}
              C \kern-1.6mm & 0 \\ 0 \kern-1.6mm & D
          \end{array} \kern-1.5mm\right)\,\Big|\,
      C\in {\cal K}(\cl H._\pi^G),\; D\in{\cal K}\big((\cl H._\pi^G)^\perp\big) \Big\}  \\[1mm]
      &=&\Big\{
          \left(\kern-1.5mm\begin{array}{cc}
              C \kern-1.6mm & 0 \\ 0 \kern-1.6mm & D
          \end{array} \kern-1.5mm\right)\,\Big|\,
      C\in {\cal K}(\cl H._\pi^G),\; D\in{\cal K}\big((\cl H._\pi^G)^\perp\big)\cap \big(V\s\gaud S_i.\big)' \Big\} \\[1mm]
      &=& {\cal K}(\cl H._\pi^G)\oplus
\Big({\cal K}\big((\cl H._\pi^G)^\perp\big)\cap \pi(\al C._i)'\Big)\,.
\end{eqnarray*}
We prove that the inclusion is an equality. If $A+B\in {\cal K}(\cl H._\pi^G)\oplus
\Big({\cal K}\big((\cl H._\pi^G)^\perp\big)\cap\pi( \al C._i)'\Big)$, then as
$A\in {\cal K}(\cl H._\pi^G)\subset\pi\big({\al O._i^F\cap \al A._{S_i}}\big)\supset
{\cal K}\big((\cl H._\pi^G)^\perp\big)\cap \pi(\al C._i)'\ni B$, and $\pi$ is faithful
on $\al A._{S_i}$, there are unique $A_0,\,B_0\in{\al O._i^F\cap \al A._{S_i}}$ such that
$\pi(A_0)=A$ and $\pi(B_0)=B$. As $\pi$ is covariant and both $A$ and $B$ commute with all implementers
$U_g^\pi$, $g\in\gaud S_i$, we see that $A_0,\,B_0\in{\al O._i^F\cap \al A._{S_i}}$ are $G\hbox{--invariant}$,
using faithfulness of $\pi$ on $\al A._{S_i}$, hence $A_0+B_0\in\al A.^G_{S_i}$. Thus
\[
\pi(\al A.^G_{S_i})={\cal K}(\cl H._\pi^G)\oplus
\Big({\cal K}\big((\cl H._\pi^G)^\perp\big)\cap \pi(\al C._i)'\Big)\,.
\]
Then
\begin{eqnarray*}
\pi\big({\al D._i^F\cap \al A.^G_{S_i}}\big)&=&\pi\big({\al D._i^F\cap \al A._{S_i}}\big)\cap \pi(\al A.^G_{S_i})\\[1mm]
&=&{\cal K}\big((\cl H._\pi^G)^\perp\big)\cap\Big({\cal K}(\cl H._\pi^G)\oplus
\Big({\cal K}\big((\cl H._\pi^G)^\perp\big)\cap \pi(\al C._i)'\Big)\Big)  \\[1mm]
&=&{\cal K}\big((\cl H._\pi^G)^\perp\big)\cap \pi(\al C._i)' \,,
\end{eqnarray*}
and clearly $\pi(\al A.^G_{S_i})\restriction \cl H._\pi^G={\cal K}(\cl H._\pi^G)$, and the claim follows.
\hfill $\rule{2mm}{2mm}$
\vspace{3mm}\break
Now $\al F._{S_i}^F=C^*\!\left(\al U._0^{S_i}\cup\al A._{S_i} \right)$ hence to obtain the full algebras
$\al D._i^F$ and $\al O._i^F$  we need to consider the role of
$\al U._0^{S_i}$. By construction, $\al C._i\subset\al D._i^F\subset\al O._i^F\ni\EINS$,
hence $\al U._0^{S_i}=\al C._i+\EINS\subset\al O._i^F$.
\begin{Theorem}
\label{FLocalObs}
With notation as above, we have that
\begin{eqnarray*}
&\al O._i^F=\left[U\s\gauc S_i.\cdot(\al O._i^F\cap \al A._{S_i}+\C)\right]
=[P_\alpha\al A._{S_i}P_\alpha]+\left[(\EINS-P_\alpha)U\s\gauc S_i.\al A._{S_i}(\EINS-P_\alpha)\right]+\left[U\s\gauc S_i.\right]
\\[1mm]
&\hbox{and}\quad
 \al D._i^F=\al B._i+ [\al C._iU\s\gauc S_i.]\qquad\hbox{where}\\[1mm]
  &\al B._i:=[\al C._iU\s\gauc S_i.\big({\al D._i^F\cap \al A._{S_i}}\big)]=
  \left[\al C._iU\s\gauc S_i.\al A._{S_i}\al C._i\right]\qquad\qquad\qquad\qquad\\[1mm]
  &=\left[(\EINS-P_\alpha)U\s\gauc S_i.\al A._{S_i}(\EINS-P_\alpha)\right] =\al D._i^F\cap\left[ U\s\gauc S_i.\al A._{S_i}\right]
  \;.\qquad\\[1mm]
&\hbox{Moreover:}\qquad\qquad\al R._i^F=\al O._i^F\big/\al D._i^F\cong [P_\alpha\al A._{S_i}P_\alpha] +\C
\;\cong\;{\cal K}(\cl H._\pi^G)+\C\,,\qquad\qquad\qquad
\end{eqnarray*}
where the last isomorphism is concretely realized in the representation $\pi$ of Theorem~\ref{structurLocalObs}(ii).
\end{Theorem}
\noindent {\bf Proof:} From the fact that $\al U._0^{S_i}$ generates $U\s\gauc S_i.$ as a group, we have that
$C^*(\al U._0^{S_i})=[U\s\gauc S_i.]$ hence via the implementing relations we obtain:
\[
\al F._{S_i}^F=C^*\!\!\left(\al U._0^{S_i}\cup\al A._{S_i}\right)=\left(\al A._{S_i}+\C\right)\rtimes_\alpha (\gaud S_i)
=\left[ U\s\gauc S_i.\cdot\al A._{S_i}\right]+[U\s\gauc S_i.]\,.
\]
As $\al U._0^{S_i}\subset\al O._i^F$ it is obvious that
\[
    \al O._i^F\supseteq\left[U\s\gauc S_i.\cdot(\al O._i^F\cap \al A._{S_i}+\C)\right],
\]
so we show the converse inclusion.
By Theorem~\ref{Teo.2.10}
we have that  $\al O._i^F=P_{S_i}'\cap \al F._{S_i}^F$  where
$P_{S_i}\in (\al F.^F_{S_i})''$ is the open projection of Theorem~\ref{Teo.2.7} for the constraint system $(\al F._{S_i}^F,\,\al C._i)$,
specified by $\omega(P_{S_i})=0$  iff $\pi_\omega(\al C._i)\Omega_\omega=0$
for $\omega\in{\got S}(\al F.^F_{S_i})$. Let $B\in\al O._i^F\subset\left[ U\s\gauc S_i.\cdot\al A._{S_i}\right]+[U\s\gauc S_i.]=\al F._{S_i}^F$,
then since we already know that $[U\s\gauc S_i.]\subset\al O._i^F$, it suffices to assume that
$B\in\left[ U\s\gauc S_i.\cdot\al A._{S_i}\right]\cap\al O._i^F$, i.e.
\[
B=\lim_{n\to\infty}\sum_{j=1}^{N_n}U_j^{(n)}A_j^{(n)}\qquad\hbox{for}\qquad U_j^{(n)}\in U\s\gauc S_i.\quad\hbox{and}
\quad A_j^{(n)}\in \al A._{S_i}
\]
such that $[B,P_{S_i}]=0$, i.e. $ B=P_{S_i}BP_{S_i}+(\EINS-P_{S_i})B(\EINS-P_{S_i})$.
Recall from Theorem~\ref{structurLocalObs}(iv) that
 $P_\alpha=(\EINS-P_{S_i})P_J$ where
 $P_J\in(\al F._{S_i}^F)''$ is the central projection determined by the ideal $\left[ U\s\gauc S_i.\cdot\al A._{S_i}\right]$
of $\al F._{S_i}^F$. Thus, since $B$ is in this ideal, we get
\[
B=P_JBP_J=(\EINS-P_\alpha)B(\EINS-P_\alpha)+P_\alpha BP_\alpha
\]
and hence
\begin{equation}
\label{PalphDecomp}
B=\lim_{n\to\infty}\sum_{j=1}^{N_n}U_j^{(n)}\big((\EINS-P_\alpha)A_j^{(n)}(\EINS-P_\alpha)+P_\alpha A_j^{(n)}P_\alpha\big).
\end{equation}
However we have that $(\EINS-P_\alpha)A_j^{(n)}(\EINS-P_\alpha)+P_\alpha A_j^{(n)}P_\alpha\in \al O._i^F\cap \al A._{S_i}$
by Theorem~\ref{structurLocalObs}(i), and hence $B\in\left[U\s\gauc S_i.(\al O._i^F\cap \al A._{S_i})\right]$.
We conclude that
$\al O._i^F=\left[U\s\gauc S_i.\cdot(\al O._i^F\cap \al A._{S_i}+\C)\right],$ as claimed.

By Theorem~\ref{structurLocalObs}(i) we  have
a projection $P_\alpha\in M(\al A._{S_i})$ such that $U_gP_\alpha=P_\alpha=P_\alpha U_g$ for all $g\in\gauc S_i$ and
\begin{equation}
\label{OFinAi}
{\al O._i^F\cap \al A._{S_i}}= P_\alpha\al A._{S_i}P_\alpha\oplus(\EINS-P_\alpha)\al A._{S_i}(\EINS-P_\alpha)
\end{equation}
from which we obtain the second equality for $\al O._i^F$.

Next, recall from Theorem~\ref{Teo.2.2}(iv) that we have
\[
\al D._i^F=[\al C._i\al O._i^F]=[\al C._iU\s\gauc S_i.\big({\al O._i^F\cap \al A._{S_i}}\big)]+ [\al C._iU\s\gauc S_i.]\,.
\]
However $DP_{S_i}=D$ for all $D\in\al D._i^F\supset\al C._i$, hence
$[\al C._iU\s\gauc S_i.\big({\al O._i^F\cap \al A._{S_i}}\big)]=[\al C._iP_{S_i}U\s\gauc S_i.\big({\al O._i^F\cap \al A._{S_i}}\big)]
=[\al C._i(\un-P_\alpha)U\s\gauc S_i.\big({\al O._i^F\cap \al A._{S_i}}\big)]=[\al C._iU\s\gauc S_i.\big({\al D._i^F\cap \al A._{S_i}}\big)]$
by the decomposition in Equation~(\ref{PalphDecomp}) for elements of $\left[U\s\gauc S_i.(\al O._i^F\cap \al A._{S_i})\right].$
This establishes the first equality   $\al B._i=[\al C._iU\s\gauc S_i.\big({\al D._i^F\cap \al A._{S_i}}\big)]$
for $\al D._i^F.$

It is clear via Lemma~\ref{Fideal}, that $\al B._i=\al D._i^F\cap\left[ U\s\gauc S_i.\al A._{S_i}\right]$
since $\left[ U\s\gauc S_i.\al A._{S_i}\right]$ is a closed two-sided ideal of
$\al F._{S_i}^F
=\left[ U\s\gauc S_i.\al A._{S_i}\right]+[U\s\gauc S_i.]$
and $[\al C._iU\s\gauc S_i.]\subset[U\s\gauc S_i.]$.
For the second equality for $\al B._i$ use $\al C._iP_\alpha=0$ and
\begin{eqnarray*}
\al D._i^F&=&[\al C._i\al O._i^F\al C._i]=[\al C._i\Big([P_\alpha\al A._{S_i}P_\alpha]+\left[(\EINS-P_\alpha)U\s\gauc S_i.
\al A._{S_i}(\EINS-P_\alpha)\right]+\left[U\s\gauc S_i.\right]\Big)\al C._i]\\[1mm]
&=& \left[\al C._iU\s\gauc S_i.\al A._{S_i}\al C._i\right]+\left[U\s\gauc S_i.\al C._i\right]
\end{eqnarray*}
and so $\al B._i=\al D._i^F\cap\left[ U\s\gauc S_i.\al A._{S_i}\right]=\left[\al C._iU\s\gauc S_i.\al A._{S_i}\al C._i\right]$.
For the third equality recall from Theorem~\ref{structurLocalObs}
that $(\EINS-P_\alpha)\al A._{S_i}(\EINS-P_\alpha)=\al D._i^F\cap \al A._{S_i}\subset
\al D._i^F\cap\left[ U\s\gauc S_i.\al A._{S_i}\right]=\left[\al C._iU\s\gauc S_i.\al A._{S_i}\al C._i\right]$
hence
\[
\left[(\EINS-P_\alpha)U\s\gauc S_i.\al A._{S_i}(\EINS-P_\alpha)\right]\subseteq\left[\al C._iU\s\gauc S_i.\al A._{S_i}\al C._i\right]
\qquad\qquad\qquad\mathord{-}\!(*)
\]
To get equality, note that
${\al O._i^F\cap\left[ U\s\gauc S_i.\al A._{S_i}\right]}=
[P_\alpha\al A._{S_i}P_\alpha]+\left[(\EINS-P_\alpha)U\s\gauc S_i.\al A._{S_i}(\EINS-P_\alpha)\right]$ is a sum of two ideals with trivial intersection,
and that $\left[\al C._iU\s\gauc S_i.\al A._{S_i}\al C._i\right]=\al D._i^F\cap\left[ U\s\gauc S_i.\al A._{S_i}\right]\subset
\al O._i^F\cap\left[ U\s\gauc S_i.\al A._{S_i}\right]$ is also an ideal, as it is the intersection of two ideals.
By $(*)$ it suffices to show that
$\left[\al C._iU\s\gauc S_i.\al A._{S_i}\al C._i\right]\cap[P_\alpha\al A._{S_i}P_\alpha]=\{0\}$.
If $B\in [P_\alpha\al A._{S_i}P_\alpha]$ then $P_\alpha B=B$, however for any element $B\in\left[\al C._iU\s\gauc S_i.\al A._{S_i}\al C._i\right]$
we have $P_\alpha B=0$. It follows that $\left[\al C._iU\s\gauc S_i.\al A._{S_i}\al C._i\right]\cap[P_\alpha\al A._{S_i}P_\alpha]=\{0\}$
and hence
$\al B._i=\left[\al C._iU\s\gauc S_i.\al A._{S_i}\al C._i\right]=\left[(\EINS-P_\alpha)U\s\gauc S_i.\al A._{S_i}(\EINS-P_\alpha)\right].$

Finally, let $\xi:\al O._i^F\to\al O._i^F\big/\al D._i^F$ be the constraining homomorphism. Then $\xi(U\s\gauc S_i.)=\un$ hence
\[
\al R._i^F=\xi(\al O._i^F)=\xi\big( \al O._i^F\cap \al A._{S_i} \big)+\C=\big(\al O._i^F\cap \al A._{S_i}\big)\Big/\big(\al D._i^F\cap \al A._{S_i}\big)
+\C\cong P_\alpha\al A._{S_i}P_\alpha+\C
\]
by Theorem~\ref{structurLocalObs}(i), and using the faithful representation $\pi$ we also get from Theorem~\ref{structurLocalObs}(ii) that
$\al R._i^F\cong{\cal K}(\cl H._\pi^G)+\C$.
\hfill $\rule{2mm}{2mm}$
\vspace{3mm}\break
We have now fully specified the constraint data for the finite constraint systems
 $(\al F._{S_i}^F,\,\al C._i)$. For the physical algebra, we obtained the same result from two different methods.

\subsection{Solving the local constraint systems.}
\label{LocC-syst}

Our aim in this section is to use the results above for the finite lattice
constraint systems $(\al F._{S_i}^F,\,\al C._i)$ to solve the corresponding ``local''
constraint systems $(\al F._{S_i},\,\al C._i)$ in the infinite lattice. Recall that
\begin{eqnarray*}
\al F._{S_i}^F&=&\big[U\s\gauc S_i.\cdot(\al A._{S_i}\oplus\C)\big]=\left(\al A._{S_i}
+\C\right)\rtimes_\alpha (\gaud S_i)\\[1mm]
&=&\left[ U\s\gauc S_i.\cdot\al A._{S_i}\right]+[U\s\gauc S_i.]
\subset M({\mathfrak A}_{\Lambda}\rtimes_\alpha (\gaud \Lambda))
\supset \al F._e\qquad\hbox{and}  \\[1mm]
\al F._{S_i}&=&\big[U\s\gauc S_i.\cdot({\mathfrak A}_{S_i}\oplus\C)\big]
=\left[U\s\gauc S_i.\cdot{\mathfrak A}_{S_i}\right]+[U\s\gauc S_i.]\subset \al F._e
\end{eqnarray*}
where $\al A._{S_i}:={\mathfrak F}_{S_i}\otimes{\cal L}^{S_i}$ and
${\mathfrak A}_{S_i}:={\mathfrak F}_{S_i}\otimes{\cal L}_{S_i}[E]$
and we have
the same constraints  $\al C._i:=\al U._0^{S_i}-\EINS$
for both cases. As $\al F._{S_i}^F$  differs from  $\al F._{S_i}$ only
by the replacement of ${\cal L}^{S_i}$  by ${\cal L}_{S_i}[E]$, we examine the relation between these
algebras. Recall that
 \[
 {\cal L}_{S_i}[E]:=C^*\big(\bigcup_{{\bf n}\in\N^\infty}\cl E._{S_i}[\bn]\big)\subset {\cal L}[E],
 \]
 where $\cl E._{S_i}[\bn]$ denotes those elementary tensors in
$ \bigcup_{k \in \N} {\cal L}^{(k)} \otimes E[\bn]_{k+1}$ which can only differ from
$ E[\bn]_{1}=E_{n_{1}}^{(1)}\otimes E_{n_{2}}^{(2)}
\otimes\cdots$ in entries corresponding to links in $\Lambda_{S_i}^1.$

\begin{Lemma}
\label{LtensorE}
Let $E_{S_i}[\bn]\subset M( {\cal L}[E])$ consist of $ E[\bn]_{1}=E_{n_{1}}^{(1)}\otimes E_{n_{2}}^{(2)}
\otimes\cdots$ except for entries corresponding to links in $\Lambda_{S_i}^1$, where it is the identity.
Let $ {\cal T}_{S_i}[E] :=C^*\big(\bigcup\limits_{{\bf n}\in\N^\infty}E_{S_i}[\bn]\big)
=\big[ \bigcup\limits_{{\bf n}\in\N^\infty}E_{S_i}[\bn] \big]\subset M({\cal L}[E])$ denote the
``infinite tails'', then
\begin{eqnarray*}
 {\cal L}_{S_i}[E]&=&\big[{\cal L}^{S_i}\cdot {\cal T}_{S_i}[E]\big]\cong{\cal L}^{S_i}\otimes {\cal T}_{S_i}[E]\\[1mm]
\hbox{and}\qquad\qquad\qquad\al F._{S_i}&=&
\left[ U\s\gauc S_i.\cdot\al A._{S_i}\right]\otimes {\cal T}_{S_i}[E]+[U\s\gauc S_i.]\otimes\EINS\,.
\end{eqnarray*}
\end{Lemma}
\noindent {\bf Proof:}
Identify ${\cal L}^{S_i}$ with ${\cal L}^{S_i}\otimes\EINS\subset M({\cal L}[E]),$
by which we mean that for the elementary tensors, only the  entries corresponding to links in $\Lambda_{S_i}^1$
are not the identity. Observe that ${\cal L}_{S_i}[E]=\big[{\cal L}^{S_i}\cdot {\cal T}_{S_i}[E]\big]\,.$
In fact, as both  ${\cal L}^{S_i}\otimes\EINS$ and ${\cal T}_{S_i}[E]$ are generated by
elementary tensors with their only nontrivial parts in complementary factors, it is clear that
the algebraic span of the products of these generating tensors is an algebraic tensor product.
Recall from the line below equation~(\ref{4.4}) that the C*-norm is taken in $M({\cal L}[{\bf 1}])$,
and ${\cal L}[{\bf 1}]$ is a tensor product over the same index set as the elementary tensors above.
 It is well-known that for a tensor product
${\cal A}\tensor{\rm min}.{\cal B}$, its multiplier algebra ${M\big({\cal A}\tensor{\rm min}.{\cal B}\big)}$
in general strictly contains $M\big({\cal A}\big)\tensor{\rm min}.M\big({\cal B}\big)$ (cf.~\cite{APT} p286--287),
 and hence on this subalgebra the norm of the multiplier algebra is a cross norm. Thus the norm of
 $M({\cal L}[{\bf 1}])\supset {\cal L}^{S_i}\otimes\EINS\cup {\cal T}_{S_i}[E]$ is a cross norm,
 so if we take the closure of the algebraic tensor product mentioned above, we get that
 ${\cal L}_{S_i}[E]=\big[{\cal L}^{S_i}\cdot {\cal T}_{S_i}[E]\big]={\cal L}^{S_i}\otimes {\cal T}_{S_i}[E]\,.$

Note that ${\mathfrak A}_{S_i}={\mathfrak F}_{S_i}\otimes{\cal L}_{S_i}[E]=
{\mathfrak F}_{S_i}\otimes{\cal L}^{S_i}\otimes {\cal T}_{S_i}[E]=\al A._{S_i}\otimes {\cal T}_{S_i}[E]$.
Moreover $\left[ U\s\gauc S_i.\cdot{\mathfrak A}_{S_i}\right]={\mathfrak A}_{S_i}\rtimes_\alpha (\gaud S_i)
=\big(\al A._{S_i}\otimes {\cal T}_{S_i}[E]\big) \rtimes_\alpha (\gaud S_i)$, and the action $\alpha$ acts trivially on
$\EINS\otimes {\cal T}_{S_i}[E]$, hence it is the product action $\alpha=\alpha^F\otimes\iota$ where
$\alpha^F:\gaud S_i\to\Aut\al A._{S_i}$ is the restriction of $\alpha$ to the finite part
$\al A._{S_i}\otimes\EINS={\mathfrak F}_{S_i}\otimes{\cal L}^{S_i}\otimes\EINS$.
It follows from Lemma~2.75 in \cite{Wil}  that
$\big(\al A._{S_i}\otimes {\cal T}_{S_i}[E]\big) \rtimes_\alpha (\gaud S_i)
=\big(\al A._{S_i}\rtimes_{\alpha^F} (\gaud S_i)\big)\otimes {\cal T}_{S_i}[E]$,
using the fact that ${\cal T}_{S_i}[E]$ is commutative, hence nuclear.
Hence the implementing unitaries are of the form $U_g\otimes\EINS$, and so
\[
\left[U\s\gauc S_i.\cdot{\mathfrak A}_{S_i}\right]=
\big(\al A._{S_i}\rtimes_{\alpha^F} (\gaud S_i)\big)\otimes {\cal T}_{S_i}[E]
=\left[U\s\gauc S_i.\cdot\al A._{S_i}\right]\otimes {\cal T}_{S_i}[E]
\]
and this proves the last equality, using
$\al F._{S_i}=\left[ U\s\gauc S_i.\cdot{\mathfrak A}_{S_i}\right]+[U\s\gauc S_i.]$.
\hfill $\rule{2mm}{2mm}$
\vspace{3mm}\break
\noindent
\begin{Theorem}
\label{FiniteInField}
Given the constraint systems $(\al F._{S_i}^F,\,\al C._i)$ and  $(\al F._{S_i},\,\al C._i)$ above,
and their associated constraint data ${\big(\al D._i^F,\,\al O._i^F,\,\al R._i^F,\xi_i^F\big)}$ and
 ${\big(\al D._i,\,\al O._i,\,\al R._i,\xi_i\big)}$ respectively, then
 \[
 \al D._i
=\al B._i\otimes {\cal T}_{S_i}[E]+ [\al C._i U\s\gauc S_i.]\otimes\EINS
\]
where $\al B._i=[\al C._iU\s\gauc S_i.\big({\al D._i^F\cap \al A._{S_i}}\big)]=
  \left[\al C._iU\s\gauc S_i.\al A._{S_i}\al C._i\right]=\left[(\EINS-P_\alpha)U\s\gauc S_i.\al A._{S_i}(\EINS-P_\alpha)\right]$
  for a projection $P_\alpha\in M(\al A._{S_i})$ such that  $UP_\alpha=P_\alpha=P_\alpha U$ for all
$U\in U\s\gauc S_i.\supset\al U._0^{S_i}$ by Theorem~\ref{structurLocalObs}.
  Moreover
 \[
 \al O._i^F=\left[U\s\gauc S_i.(\al O._i^F\cap \al A._{S_i})\right]+\left[U\s\gauc S_i.\right]
 \quad\hbox{and}\quad
 \al O._i=\left[U\s\gauc S_i.(\al O._i^F\cap \al A._{S_i})\right]\otimes {\cal T}_{S_i}[E]+\left[U\s\gauc S_i.\right]
 \]
where
${\al O._i^F\cap \al A._{S_i}}=P_\alpha\al A._{S_i}P_\alpha\oplus(\EINS-P_\alpha)\al A._{S_i}(\EINS-P_\alpha)$.
Furthermore
\[
\al R._i=\xi_i^F\big({\al O._i^F\cap \al A._{S_i}}\big)\otimes  {\cal T}_{S_i}[E]+\C
\cong P_\alpha\al A._{S_i}P_\alpha\otimes  {\cal T}_{S_i}[E]+\C\cong
{\cal K}(\cl H._\pi^G)\otimes  {\cal T}_{S_i}[E]+\C
\]
where $\pi:\al F.^F_{S_i}\to\al B.(\al H._\pi)$ is any representation which is irreducible on $\al A._{S_i}$.
\end{Theorem}
\noindent {\bf Proof:}
First consider $\al F._{S_i}=
\left[U\s\gauc S_i.\cdot\al A._{S_i}\right]\otimes {\cal T}_{S_i}[E]+[U\s\gauc S_i.]\otimes\EINS$.
Note that $\left[U\s\gauc S_i.\cdot\al A._{S_i}\right]\otimes {\cal T}_{S_i}[E]$ is
a closed two--sided ideal of  $\al F._{S_i}$. By an analogous proof to Lemma~\ref{Fideal},
we see that $\left[U\s\gauc S_i.\cdot\al A._{S_i}\right]\otimes {\cal T}_{S_i}[E]\cap
[U\s\gauc S_i.]\otimes\EINS=\{0\}$, and hence decompositions in terms of these two spaces are unique.
By Theorem~\ref{Teo.2.2} we have $\al D._i=[\al C._i\al F._{S_i}\al C._i]$ hence
\begin{eqnarray*}
\al D._i&=&
\left[\al C._i U\s\gauc S_i.\al A._{S_i}\al C._i\right]\otimes {\cal T}_{S_i}[E]+[\al C._iU\s\gauc S_i.]\otimes\EINS \\[1mm]
&=&\al B._i\otimes {\cal T}_{S_i}[E]+ [\al C._i U\s\gauc S_i.]\otimes\EINS
\end{eqnarray*}
where $\al B._i=
  \left[\al C._iU\s\gauc S_i.\al A._{S_i}\al C._i\right]=[\al C._iU\s\gauc S_i.\big({\al D._i^F\cap \al A._{S_i}}\big)]
  =\left[(\EINS-P_\alpha)U\s\gauc S_i.\al A._{S_i}(\EINS-P_\alpha)\right]$
   by Theorem~\ref{FLocalObs}, which establishes the first claim.

The equality for $\al O._i^F$ follows directly from
$\al O._i^F=\left[U\s\gauc S_i.(\al O._i^F\cap \al A._{S_i}+\C)\right]$, obtained in
Theorem~\ref{FLocalObs}. Recall from Theorem~(\ref{OFinAi}) that
\[{\al O._i^F\cap \al A._{S_i}}= P_\alpha\al A._{S_i}P_\alpha\oplus(\EINS-P_\alpha)\al A._{S_i}(\EINS-P_\alpha).
\qquad\qquad\mathord{-}\!(*)\]
We now prove the stated equality for $\al O._i$.
Let $A\in \al F._{S_i}$, 
then by the decomposition for $\al F._{S_i}$
we may write $A=F+C$ where $F\in \left[U\s\gauc S_i.\cdot\al A._{S_i}\right]\otimes {\cal T}_{S_i}[E]$
and $C\in [U\s\gauc S_i.]\otimes\EINS$. As $C\in \al O._i$ already, we only have to consider $F$.
If $F\in\left[U\s\gauc S_i.\cdot(\al O._i^F\cap \al A._{S_i})\right]\otimes {\cal T}_{S_i}[E]$, then
\begin{eqnarray*}
&& F\al D._i=F\big(\al B._i\otimes {\cal T}_{S_i}[E]+ [\al C._i U\s\gauc S_i.]\otimes\EINS\big)
\subseteq\al B._i\otimes {\cal T}_{S_i}[E]\subset \al D._i\\[1mm]
\hbox{because} &&\left[U\s\gauc S_i.\cdot(\al O._i^F\cap \al A._{S_i})\right]\al B._i\subseteq
\left[U\s\gauc S_i.\cdot(\al O._i^F\cap \al A._{S_i})(\EINS-P_\alpha)U\s\gauc S_i.\al A._{S_i}(\EINS-P_\alpha)\right]\\[1mm]
&&\subseteq\left[(\EINS-P_\alpha)U\s\gauc S_i.\al A._{S_i}(\EINS-P_\alpha)\right]=\al B._i\qquad\hbox{(as $[P_\alpha, \al O._i^F]=0$)} \\[1mm]
\hbox{and}&&
\left[U\s\gauc S_i.\cdot(\al O._i^F\cap \al A._{S_i})\right][\al C._i U\s\gauc S_i.]
\subseteq\left[(\EINS-P_\alpha)U\s\gauc S_i.\al A._{S_i}(\EINS-P_\alpha)\right]=\al B._i
\end{eqnarray*}
using $\al C._iP_\alpha=0$, and the decomposition $(*)$ for ${\al O._i^F\cap \al A._{S_i}}$ stated above.
Likewise we also get that $\al D._iF\subseteq \al D._i$, and hence $F\in \al O._i$, so we have shown that
\[ \al O._i\supseteq\left[U\s\gauc S_i.(\al O._i^F\cap \al A._{S_i})\right]\otimes {\cal T}_{S_i}[E]+\left[U\s\gauc S_i.\right].\]
We prove the reverse inclusion.
Let $A\in \al O._i$, 
then as above
we may write $A=F+C$ where $F\in \left[U\s\gauc S_i.\cdot\al A._{S_i}\right]\otimes {\cal T}_{S_i}[E]$
and $C\in [U\s\gauc S_i.]\otimes\EINS$. As $C\in \al O._i$ we have  $F\in \al O._i$,
so we need to show that  $F\in \left[U\s\gauc S_i.(\al O._i^F\cap \al A._{S_i})\right]\otimes {\cal T}_{S_i}[E]$.
Let $\hat{P}_\alpha:=P_\alpha\otimes\EINS$, then  $\hat{P}_\alpha F \hat{P}_\alpha+(\EINS-\hat{P}_\alpha )F(\EINS- \hat{P}_\alpha)\in
\left[U\s\gauc S_i.(\al O._i^F\cap \al A._{S_i})\right]\otimes {\cal T}_{S_i}[E]$ by $(*)$, so it remains to show that
the remaining part of $F$:
 \[
 \tilde{F}:=\hat{P}_\alpha F (\EINS-\hat{P}_\alpha )+(\EINS- \hat{P}_\alpha) F\hat{P}_\alpha \in
\left[U\s\gauc S_i.(\al O._i^F\cap \al A._{S_i})\right]\otimes {\cal T}_{S_i}[E].
\]
Explicitly  $F\in \left[U\s\gauc S_i.\al A._{S_i}\right]\otimes {\cal T}_{S_i}[E]$ has the form
\begin{eqnarray*}
F&=&\lim_{n\to\infty}\sum_{j=1}^{N_n}U_j^{(n)}A_j^{(n)}\otimes T_j^{(n)}\qquad\hbox{for}\quad U_j^{(n)}\in U\s\gauc S_i.,\;\;
 A_j^{(n)}\in \al A._{S_i},\;\; T_j^{(n)}\in {\cal T}_{S_i}[E]\\[1mm]
\hbox{so}\quad\tilde{F}&:=&\hat{P}_\alpha F (\EINS-\hat{P}_\alpha )+(\EINS- \hat{P}_\alpha) F\hat{P}_\alpha\\[1mm]
&=&\lim\limits_{n\to\infty}\sum\limits_{j=1}^{N_n}
U_j^{(n)}\Big({P}_\alpha A_j^{(n)}(\EINS-{P}_\alpha )+(\EINS-{P}_\alpha )A_j^{(n)}{P}_\alpha \Big)\otimes T_j^{(n)}
\in \al O._i
\end{eqnarray*}
since the other part $\hat{P}_\alpha F \hat{P}_\alpha +(\EINS- \hat{P}_\alpha) F (\EINS-\hat{P}_\alpha )$ is
 in $\al O._i$. Thus $\tilde{F}$ is in the relative multiplier of $\al D._i$.
 Now $(\EINS-P_\alpha)\al A._{S_i}(\EINS-P_\alpha)\otimes {\cal T}_{S_i}[E]\subset\al D._i$, and so
 \[
 \tilde{F}\cdot(\EINS-P_\alpha)\al A._{S_i}(\EINS-P_\alpha)\otimes {\cal T}_{S_i}[E]
  =\hat{P}_\alpha F (\EINS-\hat{P}_\alpha )\cdot\Big((\EINS-P_\alpha)\al A._{S_i}(\EINS-P_\alpha)\otimes {\cal T}_{S_i}[E]
 \Big) \subset\al D._i,
  \]
and in fact it is in $\al D._i\cap\left[U\s\gauc S_i.\al A._{S_i}\right]\otimes {\cal T}_{S_i}[E]=\al B._i\otimes {\cal T}_{S_i}[E]$.
However $P_\alpha\in M(\al A._{S_i})$ by Theorem~\ref{structurLocalObs}(iv), so if $\{J_\lambda\mid\lambda\in\Lambda\}
\subset \al A._{S_i}$ is an approximate identity, then ${(\EINS-P_\alpha)J_\lambda(\EINS-P_\alpha)}\to(\EINS-P_\alpha)$
in the strict topology of $M( \al A._{S_i})$ hence in the strong operator topology of $ \al A._{S_i}''$.
Recall that  $ {\cal T}_{S_i}[E] :=C^*\big(\bigcup\limits_{{\bf n}\in\N^\infty}E_{S_i}[\bn]\big)
\subset M({\cal L}[E])$, and let $\{K_\gamma\mid\gamma\in\Gamma\}\subset{\cal T}_{S_i}[E]$ be an approximate identity of it.
Consider
 \begin{eqnarray*}
 \tilde{F}\cdot(\EINS-P_\alpha)J_\lambda(\EINS-P_\alpha)\otimes K_\gamma
  &=&\hat{P}_\alpha F (\EINS-\hat{P}_\alpha )\cdot\Big((\EINS-P_\alpha)J_\lambda(\EINS-P_\alpha)\otimes K_\gamma
 \Big) \\[1mm]
 &=&\lim_{n\to\infty}\sum\limits_{j=1}^{N_n}
U_j^{(n)}{P}_\alpha A_j^{(n)}(\EINS-{P}_\alpha )J_\lambda(\EINS-P_\alpha)\otimes T_j^{(n)}K_\gamma
\end{eqnarray*}
which is in $\al B._i\otimes {\cal T}_{S_i}[E]\subset\al D._i$.
 Construct the faithful representation
\[
\pi_1\otimes\pi_2:\left[U\s\gauc S_i. \al A._{S_i}\right]\otimes {\cal T}_{S_i}[E]\to
 \al B.(\al H._1\otimes\al H._2)
\]
where $\pi_1:\left[U\s\gauc S_i. \al A._{S_i}\right]\to
 \al B.(\al H._1)$ is the universal representation of $\left[U\s\gauc S_i. \al A._{S_i}\right]$ (which restricts on
$\al A._{S_i}$ to its universal representation on its essential subspace), and
$\pi_2:{\cal T}_{S_i}[E]\to
 \al B.(\al H._2)$ is the universal representation of ${\cal T}_{S_i}[E]$. Then
 \[
 \pi\Big(U_j^{(n)}{P}_\alpha A_j^{(n)}(\EINS-{P}_\alpha )J_\lambda(\EINS-P_\alpha)\otimes T_j^{(n)}K_\gamma\Big)
 \to \pi\Big(U_j^{(n)}{P}_\alpha A_j^{(n)}(\EINS-{P}_\alpha )\otimes T_j^{(n)}\Big)
 \]
as  $\lambda,\,\gamma\to \infty$ in strong operator topology. As the norm limit w.r.t. $n\,$ can be interchanged with the
strong operator limits w.r.t. $\lambda,\;\gamma$ (since the product is continuous w.r.t. strong operator topology), this implies that
 \begin{eqnarray*}
 \pi\Big(\tilde{F}\cdot(\EINS-P_\alpha)J_\lambda(\EINS-P_\alpha)\otimes K_\gamma\Big)
 &\to&  \pi\Big(\lim_{n\to\infty}\sum_{j=1}^{N_n}U_j^{(n)}{P}_\alpha A_j^{(n)}(\EINS-{P}_\alpha )\otimes T_j^{(n)}\Big)\\[1mm]
 &=&\pi\Big(  \hat{P}_\alpha F (\EINS-\hat{P}_\alpha )\Big)\in\pi(\al B._i\otimes {\cal T}_{S_i}[E])^{\mathord{-}\rm s.op}
\end{eqnarray*}
as  $\lambda,\,\gamma\to \infty$ in strong operator topology. As $\al B._i
  =\left[(\EINS-P_\alpha)U\s\gauc S_i.\al A._{S_i}(\EINS-P_\alpha)\right]$,
we have $(\EINS-\hat{P}_\alpha )B=B$ for $B\in\al B._i\otimes {\cal T}_{S_i}[E]$,
and hence $\pi(\EINS-\hat{P}_\alpha )\tilde{B}=\tilde{B}$ for all $\tilde{B}\in
\pi(\al B._i\otimes {\cal T}_{S_i}[E])^{\mathord{-}\rm s.op}$. Thus
\[
\pi\Big(  \hat{P}_\alpha F (\EINS-\hat{P}_\alpha )\Big)=\pi(\EINS-\hat{P}_\alpha ) \pi\big(  \hat{P}_\alpha F (\EINS-\hat{P}_\alpha )\big)=0
\]
and as $\pi$ is faithful, $ \hat{P}_\alpha F (\EINS-\hat{P}_\alpha )=0$. Likewise we get that
$ (\EINS-\hat{P}_\alpha ) F \hat{P}_\alpha=0$, and hence
$F=\hat{P}_\alpha F \hat{P}_\alpha+(\EINS-\hat{P}_\alpha )F(\EINS- \hat{P}_\alpha)=0\in
\left[U\s\gauc S_i.(\al O._i^F\cap \al A._{S_i})\right]\otimes {\cal T}_{S_i}[E]$ and so
\[ \al O._i=\left[U\s\gauc S_i.(\al O._i^F\cap \al A._{S_i})\right]\otimes {\cal T}_{S_i}[E]+\left[U\s\gauc S_i.\right].\]

To obtain the claimed equality for $\al R._i$, we consider the factor map $\xi_i:\al O._i\to\al R._i$. Since
$\left[U\s\gauc S_i.(\al O._i^F\cap \al A._{S_i})\right]\otimes {\cal T}_{S_i}[E]\cap\left[U\s\gauc S_i.\right]=\{0\}$,
we can analyze $\xi_i\big(\left[U\s\gauc S_i.\right]\big)$ and
${\xi_i\big(\left[U\s\gauc S_i.(\al O._i^F\cap \al A._{S_i})\right]\otimes {\cal T}_{S_i}[E]\big)}$
independently. By construction, as  $\al D._i=[\al C._i\al O._i\al C._i]$ by Theorem~\ref{Teo.2.2},
thus factoring $\al O._i$ by $\al D._i$, is a homomorphism which puts $\al C._i=\al U._0^{S_i}-\EINS$ to zero, hence
$\xi_i(\al U._0^{S_i})=\EINS$ and as $U\s\gauc S_i.$ is generated as a group by $\al U._0^{S_i}$,
we have $\xi_i(U\s\gauc S_i.)=\EINS$ and hence $\xi_i\big(\left[U\s\gauc S_i.\right]\big)=\C\EINS$.

Next, recall that as ${\cal T}_{S_i}[E]$ is commutative (hence nuclear), the tensor norm of
${\left[U\s\gauc S_i.(\al O._i^F\cap \al A._{S_i})\right]\otimes {\cal T}_{S_i}[E]}$ is unique.
Thus by II.9.6.6 in~\cite{Bla1} we have that
$\ker(\check\xi_i^F\otimes\iota)=\ker(\check\xi_i^F)\otimes{\cal T}_{S_i}[E]$
where $\iota$ is the identity map of ${\cal T}_{S_i}[E]$ and
where $\check\xi_i^F:\left[U\s\gauc S_i.(\al O._i^F\cap \al A._{S_i})\right]\to\al R._i^F$ is the restriction of
$\xi_i^F$ to ${\left[U\s\gauc S_i.(\al O._i^F\cap \al A._{S_i})\right]}=\al O._i^F\cap\left[ U\s\gauc S_i.\al A._{S_i}\right]$.
Now $\ker(\check\xi_i^F)=\al D._i^F\cap\left[ U\s\gauc S_i.\al A._{S_i}\right]=\al B._i$ by Theorem~\ref{FLocalObs},
hence
$\ker(\check\xi_i^F\otimes\iota)=\al B._i\otimes{\cal T}_{S_i}[E]=\al D._i\cap\left[ U\s\gauc S_i.\al A._{S_i}\right]\otimes{\cal T}_{S_i}[E]$.
As this is precisely the kernel of
$\xi_i$ restricted to ${\left[U\s\gauc S_i.(\al O._i^F\cap \al A._{S_i})\right]\otimes {\cal T}_{S_i}[E]}$, we conclude that
$\xi_i$ coincides with $\check\xi_i^F\otimes\iota$ on ${\left[U\s\gauc S_i.(\al O._i^F\cap \al A._{S_i})\right]\otimes {\cal T}_{S_i}[E]}$,
thus
\[
\xi_i\big(\left[U\s\gauc S_i.(\al O._i^F\cap \al A._{S_i})\right]\otimes {\cal T}_{S_i}[E]\big)
=\xi_i^F\big(\al O._i^F\cap \al A._{S_i}\big)\otimes {\cal T}_{S_i}[E]
\]
using $\check\xi_i^F\big(\left[U\s\gauc S_i.(\al O._i^F\cap \al A._{S_i})\right]\big)=\xi_i^F\big(\al O._i^F\cap \al A._{S_i}\big)$
by $\xi_i(U\s\gauc S_i.)=\EINS$. Combining this with the previous paragraph, we obtain
\[
\al R._i=\xi_i(\al O._i)=\xi_i\big(\left[U\s\gauc S_i.(\al O._i^F\cap \al A._{S_i})\right]\otimes {\cal T}_{S_i}[E]
+\left[U\s\gauc S_i.\right]\big)=
\xi_i^F\big({\al O._i^F\cap \al A._{S_i}}\big)\otimes  {\cal T}_{S_i}[E]+\C
\]
as claimed. The remaining equalities are obtained from Theorem~\ref{FLocalObs}.
\hfill $\rule{2mm}{2mm}$
\vspace{3mm}\break

\subsection{Solving the full system of constraints.}
\label{GL-Tprfull}

Our aim in this section is to solve the constraint problem for the full system
$(\al F._e,\al C.)$. As remarked above, the
set of local constraint systems ${(\al F._{S_i},\al C._i)},$ $i\in\N$
which it comprises of, is a
system of local quantum constraints in the sense of~\cite{GL} (Def.~3.3).
Such systems were studied in detail in~\cite{GL}, and now we will recall and apply
the relevant parts of that analysis.
\def\B{\gamma}
\begin{Definition}
\label{loc.cons}
A system of {\bf local quantum constraints} consists of
the following:
\begin{itemize}
\item[{\rm (1)}]
A directed set $\widetilde\Gamma$ of {\rm C*}--algebras with a common
 identity $\EINS$, partially ordered by inclusion, defining an inductive limit
 {\rm C*}--algebra $\al F._0$ (over $\widetilde\Gamma$).
We will call the elements of $\widetilde\Gamma$ the {\bf local
field algebras} and $\al F._0$ the \b quasi--local algebra..
There is  directed index set $\Gamma$ together with a
surjection $\al F.\colon\ \Gamma\to \widetilde\Gamma$ which is
 order preserving,
 i.e.~if $\B_1\leq\B_2$, then $\al F.(\B_1)\subseteq\al F.(\B_2)$.
\item[{\rm (2)}] (Local Constraints) There is a map $\al U.$ from $\Gamma$ to
the set of first class subsets of the unitaries in the
 local field algebras such that
\begin{itemize}
\item[] $\;\al U.(\B)\subset\al F.(\B)_u\;$ for all $\;\B\in\Gamma$, and
\item[] if $\B_1\leq\B_2$, then 
$\al U.(\B_1)=\al U.(\B_2)\cap
 \al F.(\B_1)$.
\end{itemize}
\end{itemize}
\end{Definition}
\noindent This definition was adapted from Definition~3.3 in~\cite{GL}, where $\Gamma$ and $\al F.$ had additional structure,
which we will not need. We first verify
that our current system ${(\al F._{S_i},\al U._0^{S_i})},$ $i\in\N$ satisfies  these conditions.
\begin{Proposition}
\label{LocalQC}
Let $\Gamma=\N$ with its usual order, and let
\[
\al F.(i):=\al F._{S_i}
=\left[U\s\gauc S_i.\cdot{\mathfrak A}_{S_i}\right]+[U\s\gauc S_i.]\subset \al F._e
=\ilim\al F._{S_i}
\]
where we use the notation established above.
Define $\al U.(i):=\al U._0^{S_i}$, then the system $\{{(\al F.(i),\al U.(i))}\mid i\in\N\}
=\{{(\al F._{S_i},\al U._0^{S_i})}\mid i\in\N\}$
is a system of local quantum constraints.
\end{Proposition}
 \noindent {\bf Proof:}
It is clear that (1) is satisfied.
Regarding (2), it is obvious that if $i<j$, then
$\al U._0^{S_i}\subseteq\al U._0^{S_j}\cap\al F._{S_i} $,
so it suffices to show that no $U\in\al U._0^{S_j}\backslash\al U._0^{S_i}$ is
in $\al F._{S_i}$.
Recall that
 \begin{eqnarray*}
 \al U._0^S&:=&\{U\in \al U._0 \,\mid [U,{\mathfrak A}_S]\not=0\}=
\{U_{\exp(tY\cdot\delta_x)}\,\mid t\in\R\backslash 0,\;
Y\in{\mathfrak g}\backslash 0,\,x\in S_e\}\qquad\hbox{where} \\[1mm]
\al U._0&:=&\{U_{\exp(t\nu)}\,\mid t\in\R,\;\nu=Y\cdot\delta_x\quad\hbox{for all}\quad
Y\in{\mathfrak g},\,x\in\Lambda^0\}\qquad\hbox{and}\\[1mm]
S_e&:=&\{x\in\Lambda^0\,\mid\,\exists\; \ell=(x_\ell,y_\ell)\in\Lambda^1\;\;\hbox{such that}\;\;
\ell\cap S\not=\emptyset\;\;\hbox{and}\;\;
x_\ell=x\;\;\hbox{or\;}\;y_\ell=x \}\,.
 \end{eqnarray*}
 Now $ \al F._S=\left[U\s\gauc S.\cdot{\mathfrak A}_{S}\right]+[U\s\gauc S.]$
   and we have uniqueness for decompositions in terms
of these two spaces.
If $i<j$, then
$\left[U\s\gauc S_i.\cdot{\mathfrak A}_{S_i}\right]\subseteq
\left[U\s\gauc S_j.\cdot{\mathfrak A}_{S_j}\right]$
and $[U\s\gauc S_i.]\subseteq[U\s\gauc S_j.] $
hence $\al U._0^{S_j}\cap\al F._{S_i}\subset[U\s\gauc S_i.]\cap[U\s\gauc S_j.]= [U\s\gauc S_i.]$.
Now $[U\s\gauc S_j.]=C^*(\gaud S_j)$ so as $\gauc S_j=(\gauc S_i)\times H$ where
$H:={\{\gamma\in\gauc S_j\mid\supp(\gamma)\cap(S_i)_e=\emptyset\}}$,
we have $C^*(\gaud S_j)=C^*(\gaud S_i)\tensor{\rm max}.C^*(H_d)$.
Thus $[U\s\gauc S_i.]=C^*(\gaud S_i)\tensor{\rm max}.\C\EINS$.

Moreover $U\in\al U._0^{S_j}\backslash\al U._0^{S_i}$ is of the form
$U=U_{\exp(tY\cdot\delta_x)}$ for $ t\in\R\backslash 0,$
$Y\in{\mathfrak g}\backslash 0$ and $x\in (S_j)_e\backslash(S_i)_e$
so $\exp(tY\cdot\delta_x)\in H$ hence $U\in \C\EINS\tensor{\rm max}.C^*(H_d)$.
As $U$ implements a nontrivial automorphism, it cannot be a multiple of the identity, hence
$U\not\in C^*(\gaud S_i)\tensor{\rm max}.\C\EINS=\al U._0^{S_j}\cap\al F._{S_i}$.
Thus   $\al U._0^{S_j}\cap\al F._{S_i} =\al U._0^{S_i}$
and so (2) is satisfied.
 \hfill $\rule{2mm}{2mm}$
\vspace{3mm}\break
 Given a system of local quantum constraints, $\B\to(\al F.(\B),\,
\al U.(\B))$, we can apply the T--procedure to each  system
$(\al F.(\B),\, \al U.(\B))$, to obtain the ``local'' objects:
\begin{eqnarray*}
{\got S}_D^\B &:=& \Big\{ \omega\in{\got S}({\al F.}(\B))\mid
                \omega(U)=1\quad {\forall}\, U\in \al U.(\B)\Big\}
         \kern2mm=\kern2mm {\got S}_D(\al F.(\B))\,, \\[2mm]
\al D.(\B) &:=& [\al F.(\B)\,\al C.(\B)]\cap[\al C.(\B)\,\al F.(\B)]\,,\\[2mm]
\al O.(\B) &:=& \{ F\in \al F.(\B)\mid  FD-DF \in \al D.(\B) \quad
                  {\forall}\, D\in\al D.(\B) \}\kern2mm = \kern2mm
                  M_{\al F.(\B)}(\al D.(\B))\,, \\[2mm]
\al R.(\B) &:=& \al O.(\B) / \al D.(\B)\qquad\hbox{and constraint homomorphism}\qquad
\xi_\B:\al O.(\B)\to\al R.(\B)\,.
\end{eqnarray*}
In the case of our system $\{{(\al F._{S_i},\al U._0^{S_i})}\mid i\in\N\}$, this corresponds to
the constraint data  ${\big({\got S}_D^i ,\al D._i,\,\al O._i,\,\al R._i,\xi_i\big)}$
analyzed in Theorem~\ref{FiniteInField}. We need to determine
what the inclusions in Definition~\ref{loc.cons}
imply for the associated
objects ${\big({\got S}_D^{\B},\,\al D.(\B),\,\al O.(\B),
\al R.(\B),\,\xi_\B\big)}$.
\begin{Proposition}
\label{Teo.4.2}
Let $\Gamma\ni\B\to (\al F.(\B),\al U.(\B))$ be a system of local
quantum constraints.
 Let   $\B_1\leq\B_2\;$ imply that
$\;\al O.(\B_1)\subseteq\al O.(\B_2)\;$ and $\;\al D.(\B_1)
 ={\al D.(\B_2)\cap\al O.(\B_1)}$.
 Then the constraint homomorphism $\xi_{\B_2}:\al O.(\B_2)\to\al R.(\B_2)$
 coincides on $\al O.(\B_1)$ with $\xi_{\B_1}$, and hence it defines
a unital {*--monomorphism} $\iota_{12}\colon\ \al R.(\B_1)
\into\al R.(\B_2)$.
 In this case, the net $\B\to\al R.(\B)$ has an inductive limit, which we
 denote by $\al R._0:=\mathop{{\rm lim}}\limits_{\longrightarrow} \al R.(\B)$.
 Now we may consistently write  $\;\al R.(\B_1)\subset\al R.(\B_2)$ if $\B_1\leq\B_2$.
\end{Proposition}
 \noindent {\bf Proof:}
Let   $\B_1\leq\B_2\;$
 and $\;\al O.(\B_1)\subseteq\al O.(\B_2)\;$ and $\;\al D.(\B_1)
 ={\al D.(\B_2)\cap\al O.(\B_1)}$.
{}From
\[
\al R.(\B_2) =\al O.(\B_2) / \al D.(\B_2) = \Big(\al O.(\B_1) / \al D.(\B_2) \Big) \cup
             \Big( (\al O.(\B_2)\setminus\al O.(\B_1)) / \al D.(\B_2)\Big)\,,
\]
 it is enough to show that $\al O.(\B_1) / \al D.(\B_2)\cong \al R.(\B_1)=
\al O.(\B_1) / \al D.(\B_1)$. Now, in $\al O.(\B_1)$ a
 $\al D.(\B_2)\hbox{--equivalence}$ class consists of $A,B\in\al O.(\B_1)$ such that
 $A-B\in\al D.(\B_2)$ and therefore $A-B\in\al D.(\B_2)\cap\al O.(\B_1)=\al D.(\B_1)$.
 This implies $\al O.(\B_1) / \al D.(\B_2)\cong \al O.(\B_1) / \al D.(\B_1)=\al R.(\B_1)$.
Moreover, since $\EINS\in\al O.(\B_1)\subset\al O.(\B_2)$,
and the $\al D.(\B_1)$--equivalence class of $\EINS$ is contained in
the $\al D.(\B_2)$--equivalence class of $\EINS$, it follows that
the identity maps to the identity.
We obtain
for $\B_1\leq\B_2$ a unital monomorphism $\iota_{12}\colon\ \al R.(\B_1)
\into\al R.(\B_2)$. Next
we have to verify that these monomorphisms  satisfy
Takeda's criterion: $\iota_{13}=\iota_{23}\circ\iota_{12}$
(cf.~\cite{Takeda55}), which will ensure the existence of
the inductive limit $\al R._0$, and in which case we can
write simply inclusion $\al R.(\B_1)\subset\al R.(\B_2)$
for $\iota_{12}$. Recall that $\iota_{12}
(A+\al D.(\B_1))=A+\al D.(\B_2)$ for $A\in\al O.(\B_1)$.
Let $\B_1\leq\B_2\leq\B_3$, then by assumption
$\al O.(\B_1) \subset\al O.(\B_2)\subset\al O.(\B_3)$, and so
for $A\in\al O.(\B_1)$,
$\iota_{23}\big(\iota_{12}(A+\al D.(\B_1))\big)=\iota_{23}(
A+\al D.(\B_2))=A+\al D.(\B_3)=\iota_{13}(A+\al D.(\B_1))$.
This establishes Takeda's criterion.
 \hfill $\rule{2mm}{2mm}$
\vspace{3mm}\break
The pair of conditions $\;\al O.(\B_1)\subseteq\al O.(\B_2)\;$ and $\;\al D.(\B_1)
 ={\al D.(\B_2)\cap\al O.(\B_1)}$ were analyzed in Subsections~3.1 and 3.2 of~\cite{GL}
 where together they were given the name of reduction isotony.
\begin{Theorem}
With notation established above, the system of local quantum constraints
$\{{(\al F._{S_i},\al U._0^{S_i})}\mid i\in\N\}$
satisfies
\begin{enumerate}
\item[(i)]
$\;\al O._i\subseteq\al O._j\;$ and $\;\al D._i
 ={\al D._j\cap\al O._i}$ if $i\leq j$. Thus  $\;\al R._i\subseteq\al R._j\,,$
and there is an inductive limit, which we
 denote by $\al R._0:=\mathop{{\rm lim}}\limits_{\longrightarrow} \al R._i$.
 \item[(ii)] $\;\al O._i\subseteq\al O.\;$ and $\;\al D._i
 ={\al D.\cap\al O._i}$ where ${\big({\got S}_D ,\al D.,\,\al O.,\,\al R.,\xi\big)}$
 is the constraint data  for the full system $\al F._e=\big[U\s\gaud \Lambda.\cdot({\mathfrak A}_{\Lambda}\oplus\C)\big]
=\ilim\al F._{S_i}$ with constraints $\al C.=\al U._0-\EINS=\mathop{\cup}\limits_{i=1}^\infty
\al C._i$. Thus $\xi$ coincides with $\xi_i$ on $\;\al O._i$, and hence defines
a unital {*--monomorphism} $\iota_{i}\colon\ \al R._i
\into\al R.$ which is compatible with the containments $\;\al R._i\subseteq\al R._j\,,$ hence we
denote it by $\;\al R._i\subseteq\al R..$ Thus $\al R._0=\mathop{{\rm lim}}\limits_{\longrightarrow} \al R._i\subseteq\al R.$.
 \item[(iii)]
  $\;\al D._i={\al D.\cap\al F._{S_i}}$, $\al O._i=\al O.\cap\al F._{S_i}$
  and $\;\al R._0=\xi\big(C^*(\al O.\cap\mathop{\cup}\limits_{i\in\N}\al F._{S_i})\big)$.
 \end{enumerate}
\end{Theorem}
\noindent {\bf Proof:}
Let $i\leq j$ and recall
\[
\al F._{S_i}
=\left[U\s\gauc S_i.\cdot{\mathfrak A}_{S_i}\right]+[U\s\gauc S_i.]
\quad\hbox{and}\quad
\al U._0^{S_i}=\{U\in \al U._0 \,\mid [U,{\mathfrak A}_{S_i}]\not=0\}
\]
where
${\mathfrak A}_{S_i}:={\mathfrak F}_{S_i}\otimes{\cal L}_{S_i}[E]$.
Thus for  $U\in\al U._0\backslash\al U._0^{S_i}$ we have $ [U,{\mathfrak A}_{S_i}]=0$ and
 $ [U,U\s\gauc S_i.]=0$, and so  $ [U,\al F._{S_i}]=0$. Now from Lemma~3.3 in \cite{GL} we have that
 \begin{eqnarray*}
 \al O._i\subseteq\al O._j\qquad&\hbox{iff}&\qquad
 \al O._i\subseteq\big\{F\in\al F._{S_i}\mid
 UFU^{-1}-F\in\al D._j\;\;\forall\; U\in\al U._0^{S_j}\backslash\al U._0^{S_i}\big\} \\[1mm]
  \al O._i\subseteq\al O.\qquad&\hbox{iff}&\qquad
 \al O._i\subseteq\big\{F\in\al F._{S_i}\mid
 UFU^{-1}-F\in\al D.\;\;\forall\; U\in\al U._0\backslash\al U._0^{S_i}\big\}
 \end{eqnarray*}
and by the previous lines we have that $UFU^{-1}-F=0\in\al D._j\cap\al D.$ for all $F\in\al F._{S_i}\supseteq\al O._i$
and $ U\in\al U._0\backslash\al U._0^{S_i}$. So these requirements are always satisfied, hence
$\al O._i\subseteq\al O._j$ and $\al O._i\subseteq\al O.$ as claimed.

Next, to show that $\;\al D._i={\al D._j\cap\al O._i}$ note from the definition of $\al D.$ that
 $\al D._i\subseteq \al D._j$, hence $\;\al D._i\subseteq{\al D._j\cap\al O._i}$.
 Since $\al O._i\subseteq\al O._j$ we may regard these as new field algebras, so
using Lemma~3.2 in \cite{GL} we will have that  $\;\al D._i={\al D._j\cap\al O._i}$
if we can prove that every Dirac state on $\al O._i$ (w.r.t. constraints $\al C._i$)
extends to a Dirac state on $\al O._j$ (w.r.t. constraints $\al C._j$).
In fact, it is enough to prove that every Dirac state on $\al O._i$ (w.r.t. constraints $\al C._i$)
extends to a Dirac state on $\al E.:=C^*(\al O._i\cup U\s\gauc S_j.)\subseteq\al O._j$ (w.r.t. constraints $\al C._j$)
because any further extension of such a state by the Hahn--Banach theorem to $\al O._j$ remains a Dirac state
w.r.t.  $\al C._j$.

Now (as observed in the proof of Proposition~\ref{LocalQC}) we have that
 $\gauc S_j={(\gauc S_i)\times H}$ where
$H:={\{\gamma\in\gauc S_j\mid\supp(\gamma)\cap(S_i)_e=\emptyset\}}$, and $\alpha_H$ acts trivially on
$\al F._{S_i}$, hence on $\al O._i$. Since $U\s\gauc S_i.\subset\al O._i$, we have
$\al E.=C^*(\al O._i\cup U_H)$ and recalling that the
unitaries $U\s\gauc S_i.$ were the implementing unitaries of $\alpha\s\gauc S_i.$ in the original crossed product,
this means that $\al E.=\al O._i\rtimes_\iota H_d$ where $\iota:H\to\aut\al O._i$ is the trivial action.
Thus by Lemma 2.73 in \cite{Wil}, we obtain that
 $\al E.=\al O._i\rtimes_\iota H_d=\al O._i\tensor{\rm max}. C^*( H_d)$.
 Now by Theorem~4.9 in~\cite{Tak}, given any two states $\omega_1$ on $\al O._i$ and  $\omega_2$
 on $C^*( H_d)$, we can define a state $\omega_1\otimes\omega_2$ on
 $\al E.=\al O._i\tensor{\rm max}. C^*( H_d)$ by $\omega_1\otimes\omega_2(A\otimes B)=\omega_1(A)\omega_2(B)$
 for all $A\in \al O._i$ and $B\in C^*( H_d)$. In particular, as $\al C._j$ is first-class, we can choose
 a state $\omega_2$ on $C^*( H_d)=C^*(U_H)$  such that $\omega_2(U_h)=1$ for all $h\in H$.
 Thus if $\omega_1$ is a Dirac state on $\al O._i$ (w.r.t. constraints $\al C._i$), then
 $\omega_1\otimes\omega_2$ is a Dirac state on $\al E.$ (w.r.t. constraints $\al C._j$)
 which extends $\omega_1$. This concludes the proof that $\;\al D._i={\al D._j\cap\al O._i}$,
 and thus (i) is proven.

For (ii), the same argument with suitable replacements proves that $\;\al D._i={\al D.\cap\al O._i}$.
Moreover, if we replace $\al O._i$ by $\al F._{S_i}$, this argument also proves that
 $\;\al D._i={\al D.\cap\al F._{S_i}}$.
To see that $\iota_{i}\colon\ \al R._i
\into\al R.$  is compatible with the containments $\;\al R._i\subseteq\al R._j\,,$
i.e. with the monomorphism $\iota_{ij}\colon\ \al R._i
\into\al R._j$ obtained from (i), recall that $\iota_{ij}
(A+\al D._i)=A+\al D._j$ for $A\in\al O._i$.
Then by assumption
$\al O._i \subset\al O._j\subset\al O.$, and so
for $A\in\al O._i$,
$\iota_{j}\big(\iota_{ij}(A+\al D._i)\big)=\iota_{j}(
A+\al D._j)=A+\al D.=\iota_{i}(A+\al D._i)$. Thus $\iota_{j}\circ\iota_{ij}=\iota_{i}$, and
this also proves that the set of monomorphisms ${\{\iota_i\mid i\in\N\}}$ defines a monomorphism of
$\al R._0=\mathop{{\rm lim}}\limits_{\longrightarrow} \al R._i$ into $\al R.$ by
the universal property of inductive limits (Theorem~L.2.1. in~\cite{WO93}).

(iii) We already have above that $\;\al D._i={\al D.\cap\al F._{S_i}}$, so we
prove that $\al O._i=\al O.\cap\al F._{S_i}$. As $\;\al O._i\subseteq\al O.\;$  by (ii), we have
$\al O._i\subseteq\al O.\cap\al F._{S_i}$. Conversely, if $A\in\al O.\cap\al F._{S_i}$, then
$A\al D._i=A({\al D.\cap\al O._i})\subseteq\al D.\cap\al F._{S_i}=\al D._i$. Likewise
$\al D._iA\subseteq\al D._i$ hence $A\in\al O._i$. Thus $\al O._i=\al O.\cap\al F._{S_i}$.
Since $\xi:\al O.\to\al R.$ takes each $\al O._i=\al O.\cap\al F._{S_i}$ to $\al R._i\subset\al R._0$
and it is a homomorphism, it takes $C^*(\al O.\cap\mathop{\cup}\limits_{i\in\N}\al F._{S_i})
=C^*(\mathop{\cup}\limits_{i\in\N}\al O._i)$ to $\al R._0$. Since all $\al R._i\subset
\xi\big(C^*(\al O.\cap\mathop{\cup}\limits_{i\in\N}\al F._{S_i})\big)$, and these generate $\al R._0$
it is clear that we have the claimed equality.
 \hfill $\rule{2mm}{2mm}$
\vspace{3mm}\break
Thus we have shown that the local physical observable algebras $\al R._i$ (obtained in
 Theorem~\ref{FiniteInField}), combine into an inductive limit $\al R._0$, and produce a
 large part of the full observable algebra $\al R.$.
If $\al R.\not=\al R._0$, then the extra elements must be obtained from
$\al O.\big\backslash C^*(\al O.\cap\mathop{\cup}\limits_{i\in\N}\al F._{S_i})$,
i.e. these do not come from ``local'' observables, so we may regard the elements
of $\al R.\backslash\al R._0$ as global observables.
 It is not clear if there are any.
 A natural representation for $\al R.$ is obtained from the
 representation we constructed
at the end of Subsection~\ref{GlobalGT},
$\pi=\pi_{\rm Fock}\otimes\pi_\infty$ which was covariant and had a nonzero invariant vector.
Restriction of $\pi(\al O.)$ to the gauge invariant subspace of $\al H._\pi$ produces a representation
of $\al R.$.

We would like to understand the inclusions  $\;\al R._i\subseteq\al R._j\,,$
in terms of the concrete characterization in Theorem~\ref{FiniteInField}:
\[
\al R._i=\xi_i(\al O._i)=\xi_i^F\big({\al O._i^F\cap \al A._{S_i}}\big)\otimes  {\cal T}_{S_i}[E]+\C
\cong {\cal K}(\cl H._\pi^G)\otimes  {\cal T}_{S_i}[E]+\C
\]
where $\pi:\al F.^F_{S_i}\to\al B.(\al H._\pi)$ is any representation which is irreducible on $\al A._{S_i}$.
Since for $i<j$ we have $\al R._i=\xi_i(\al O._i)=\xi_j(\al O._i)\subset\xi_j(\al O._j)=\al R._j$, we need to
consider the inclusion $\;\al O._i\subseteq\al O._j\,.$ By Theorem~\ref{FiniteInField}
 \[
 \al O._i=\left[U\s\gauc S_i.(\al O._i^F\cap \al A._{S_i})\right]\otimes {\cal T}_{S_i}[E]+\left[U\s\gauc S_i.\right]
 \quad\hbox{where}\quad
  \al O._i^F=\left[U\s\gauc S_i.(\al O._i^F\cap \al A._{S_i})\right]+\left[U\s\gauc S_i.\right]
 \]
and  $\al A._{S_i}:={\mathfrak F}_{S_i}\otimes{\cal L}^{S_i}$. As $\left[U\s\gauc S_i.\right]\subseteq\left[U\s\gauc S_j.\right]$
and $\xi_j\big(\left[U\s\gauc S_j.\right]\big)=\C$, we only need to examine the inclusion
${\left[U\s\gauc S_i.(\al O._i^F\cap \al A._{S_i})\right]\otimes {\cal T}_{S_i}[E]}
\subseteq{\left[U\s\gauc S_j.(\al O._j^F\cap \al A._{S_j})\right]\otimes {\cal T}_{S_j}[E]}$.
\begin{itemize}
\item{} We show how  ${\cal L}_{S_i}[E]\subseteq{\cal L}_{S_j}[E]$.
 Now ${\cal L}_{S_i}[E]\cong{\cal L}^{S_i}\otimes {\cal T}_{S_i}[E]$,  and so it is generated by elements of the type
 $L\otimes E_{S_i}[\bn]$ where $L\in{\cal L}^{S_i}$ and $E_{S_i}[\bn]$ is
as in Lemma~\ref{LtensorE}. Then $E_{S_i}[\bn]=E_{ij}\otimes E_{S_j}[\bn]$ where
$E_{ij}$ is the finite tensor product consisting of those entries of
 $ E_{n_{1}}^{(1)}\otimes E_{n_{2}}^{(2)}
\otimes\cdots$  corresponding to links in $\Lambda_{S_j}^1\backslash\Lambda_{S_i}^1$.
Now ${\cal L}^{S_i}\otimes E_{ij}\subset{\cal L}^{S_j}$ since $(E_n^{(k)})_{n \in \N}\subset{\cal L}_k$,
hence $L\otimes E_{S_i}[\bn]=L\otimes E_{ij}\otimes E_{S_j}[\bn]\in {\cal L}^{S_j}\otimes {\cal T}_{S_j}[E]$.
Thus we have identified
${\cal L}^{S_i}\otimes {\cal T}_{S_i}[E]\subseteq{\cal L}^{S_j}\otimes {\cal T}_{S_j}[E] $
and hence ${\cal L}_{S_i}[E]\subseteq{\cal L}_{S_j}[E]$.
 \item{} Since ${\mathfrak F}_{S_i}\subseteq{\mathfrak F}_{S_j}$ it follows that
 $\al A._{S_i}\otimes E_{ij}={\mathfrak F}_{S_i}\otimes{\cal L}^{S_i}\otimes E_{ij}\subseteq{\mathfrak F}_{S_j}\otimes{\cal L}^{S_j}=\al A._{S_j}$.
 We claim that $(\al O._i^F\cap \al A._{S_i})\otimes E_{ij}\subseteq(\al O._j^F\cap \al A._{S_j})$, and hence
$(\al O._i^F\cap \al A._{S_i})\otimes {\cal T}_{S_i}[E] \subseteq(\al O._j^F\cap \al A._{S_j})\otimes {\cal T}_{S_j}[E]$ .
That $(\al O._i^F\cap \al A._{S_i})\otimes E_{ij}\subseteq \al A._{S_j}$ is obvious. To see that it is in $\al O._j^F$
note that ${\big[\al U._0^{S_j}\backslash\al U._0^{S_i} ,(\al O._i^F\cap \al A._{S_i})\otimes E_{ij}\big]}=0$, and that
\[
{\big[\al U._0^{S_i},(\al O._i^F\cap \al A._{S_i})\otimes E_{ij}\big]}\subseteq
\al B._i^F\otimes E_{ij}\subseteq\al D._j^F
\]
 where  via Theorem~\ref{FLocalObs}
we have  $\al B._i=\al D._i^F\cap\left[ U\s\gauc S_i.\al A._{S_i}\right]=
  \left[\al C._iU\s\gauc S_i.\al A._{S_i}\al C._i\right]$, and the last inclusion follows from
 $\al B._i^F\otimes E_{ij}=\left[\al C._iU\s\gauc S_i.\al A._{S_i}\al C._i\right]\otimes E_{ij}
 \subseteq \left[\al C._i\big(U\s\gauc S_i.\al A._{S_i}\otimes E_{ij}\big)\al C._i\right] \subseteq\al D._j^F$.
 This inclusion
 \[
 (\al O._i^F\cap \al A._{S_i})\otimes {\cal T}_{S_i}[E] \subseteq(\al O._j^F\cap \al A._{S_j})\otimes {\cal T}_{S_j}[E]
 \]
 fully specifies the inclusion  $\;\al R._i\subseteq\al R._j\,$ because
$ \al R._i=\xi_i\big((\al O._i^F\cap \al A._{S_i})\otimes  {\cal T}_{S_i}[E]\big)+\C$.
\end{itemize}

\noindent
We conclude that we have concretely characterized the algebra of local physical observables
 $\al R._0=\mathop{{\rm lim}}\limits_{\longrightarrow} \al R._i\subseteq\al R.$, but that the existence and
 nature of the global physical observables $\al R.\backslash\al R._0$ remain an open question.


\subsection{Physical observables.}
\label{ObsPh}


We consider types of gauge invariant observables for lattice QCD which appeared in the literature (cf.~\cite{KS}).
As we know from Theorem~\ref{Teo.2.2}(iii), any gauge invariant element of ${\cal F}_e$ will be in $\al O.$ and hence will produce an element of our physical algebra
 $\al R.$. Moreover, if a gauge invariant $A\in\al F._e$ is constructed from a finite lattice,
i.e. it is in some $\al F._{S_i}$, then we obtain a gauge invariant element in
$\al O.\cap\mathop{\cup}\limits_{i\in\N}\al F._{S_i}
=\mathop{\cup}\limits_{i\in\N}\al O._i$ hence in $\al R._0$.
We consider three types.
\begin{itemize}
\item[(1)] As remarked above Theorem~\ref{structurTradLocalObs},
given $A\in\al A._{S_i}$
we only need to apply the well-known method of taking the group average over
$\gauc S_i$   to obtain a gauge invariant observable
\[
P_\alpha AP_\alpha=\int_{\gauc S_i}\alpha_g(A)\,d\mu(g)\quad\hbox{where $\mu$ is the normalized Haar measure of}\;\;\gauc S_i.
\]
Moreover from Theorem~\ref{FLocalObs}, for finite lattice systems, as
$\al R._i^F=\al O._i^F\big/\al D._i^F\cong [P_\alpha\al A._{S_i}P_\alpha] +\C$, all physically relevant
observables can be obtained this way. For the local algebra $\al F._{S_i}=
\left[ U\s\gauc S_i.\cdot\al A._{S_i}\right]\otimes {\cal T}_{S_i}[E]+[U\s\gauc S_i.]\otimes\EINS$,
the analogous statement holds for an $A\in\al A._{S_i}\otimes {\cal T}_{S_i}[E]$, and by
Theorem~\ref{FiniteInField} we have $\al R._i
\cong P_\alpha\al A._{S_i}P_\alpha\otimes  {\cal T}_{S_i}[E]+\C$ so again, all of $\al R._i$
can be obtained this way.  Thus  all of $\al R._0=C^*\big(\mathop{\cup}\limits_{i\in\N}\al R._i\big)$ can be obtained by local averages over the
gauge group.
In the (Euclidean) functional integral approach of lattice QCD, this idea is useful (cf.~\cite{OS}).
Here the functional integral measure is the (possibly infinite) product measure of the normalized Haar measure
of $G$, one for each link of the lattice. For a local observable, i.e.
a function depending on only finitely many link variables (in $G$), its expectation value is
its functional integral, which reduces to the integral w.r.t. the finite
product measure of the Haar measures of the links involved.

\item[(2)] We start with gauge invariant variables of pure gauge type, and consider the well-known Wilson loops cf.~\cite{Wilson}.
To construct a Wilson loop, we choose an oriented loop $L=\{\ell_1,\ell_2,\ldots,
 \ell_m\}\subset\Lambda^1$, $\ell_j=(x_j,y_j)$, such that  $y_j=x_{j+1}$ for $j=1,\ldots,{m-1}$
 and $y_m=x_1$. Let $G_k=G$ be the configuration space of $\ell_k$. Denoting the components of a gauge potential $\Phi$
 by $\Phi_{ij}(\ell_k)\in C(G_k)$ as in equation \eqref{Def-QuConn}, the matrix components of the quantum connection
at $\ell_k$ are given by $T_{\Phi_{ij}(\ell_k)}$.
To construct the gauge invariant observable associated with the loop, define (summing over repeated indices):
 \begin{eqnarray*}
 W(L)&:=&\Phi_{i_1i_2}(\ell_1)(g_1)\,\Phi_{i_2i_3}(\ell_3)(g_2)\cdots \Phi_{i_{m-1}i_1}(\ell_m)(g_m)\\[1mm]
 &=&(e_{i_1},g_1g_2\cdots g_me_{i_1})= {\rm Tr}(g_1g_2\cdots g_m)\,.
  \end{eqnarray*}
This defines a gauge invariant element $W(L)\in C(G_1)\otimes\cdots\otimes C(G_m)={C(G_1\times\cdots G_m)}$.
Wilson loops of particular importance, are those where the paths are plaquettes,
i.e. $L={(\ell_1,\ell_2,\ell_3,\ell_4)}\in\Lambda^2$ as such $W(L)$ occur in the lattice Hamiltonian.

As remarked in Subsection~\ref{GFA}, $C(G_j)\subset M(\cl L._{\ell_j})$ where $\cl L._{\ell_j}=
{C(G_j)\rtimes_\lambda G_j}$, it is not actually contained in $\cl L._{\ell_j}$.
Let $S_i$ contain the loop, then we embed $C(G_1)\otimes\cdots\otimes C(G_m)$ (hence $W(L)$)
 in $M({\mathfrak A}_{S_i})=M({\mathfrak F}_{S_i}\otimes{\cal L}_{S_i}[E])=
 M\big({\mathfrak F}_{S_i}\otimes{\cal L}^{(S_i)}\otimes {\cal T}_{S_i}[E]\big)$
 by setting it equal to $\un$ in
 those factors of the tensor product not corresponding to some $\cl L._{\ell_j}$.
 Recalling that $\al R._i
\cong P_\alpha\al A._{S_i}P_\alpha\otimes  {\cal T}_{S_i}[E]+\C$ where
$\al A._{S_i}:={\mathfrak F}_{S_i}\otimes{\cal L}^{(S_i)}$
and that $P_\alpha M(\al A._{S_i})P_\alpha$ is the gauge invariant part of $ M(\al A._{S_i})$,
we obtain that $W(L)\in P_\alpha M(\al A._{S_i})P_\alpha\otimes  {\cal T}_{S_i}[E]
\subseteq M\big(P_\alpha\al A._{S_i}P_\alpha\otimes  {\cal T}_{S_i}[E]\big)$
so $W(L)$ is in the multiplier algebra of a subalgebra of $\al R._0$.

\item[(3)]  Another method of constructing gauge invariant observables, is
  by Fermi bilinears connected with a Wilson line (cf.~\cite{KS}).
  Consider a path  $C=\{\ell_1,\ell_2,\ldots,
 \ell_m\}\subset\Lambda^1$, $\ell_j=(x_j,y_j)$, such that  $y_j=x_{j+1}$ for $j=1,\ldots,m-1$.
 We take notation as above, so
 $G_k=G$ is the configuration space of $\ell_k,$ and
$\Phi_{ij}(\ell_k)(g_k):=(e_i,g_ke_j)$, $g_k\in G_k$.
To construct a gauge invariant observable associated with the path, consider (with summation convention):
 \begin{eqnarray*}
 Q(C)&:=& \psi^*_{i_1}(x_1)\,\Phi_{i_1i_2}(\ell_1)\,\Phi_{i_2i_3}(\ell_3)\cdots \Phi_{i_{m-1}i_m}(\ell_m)\, \psi_{ i_m}(y_m)\\[1mm]
&\in& {\mathfrak F}_{S_i}\otimes C(G_1)\otimes\cdots\otimes C(G_m)
  \end{eqnarray*}
where $S_i$ contains the path and we assume $\Cn=\C^k$
(otherwise $\Cn=\C^k\times{\bf W}$ and there are more indices). Then $ Q(C)$ is gauge invariant.
As above,
 we embed ${\mathfrak F}_{S_i}\otimes C(G_1)\otimes\cdots\otimes C(G_m)$
 in $M({\mathfrak A}_{S_i})=M({\mathfrak F}_{S_i}\otimes{\cal L}_{S_i}[E])=
 M\big({\mathfrak F}_{S_i}\otimes{\cal L}^{(S_i)}\otimes {\cal T}_{S_i}[E]\big)$
 by setting it equal to $\un$ in
 those factors of the tensor product not corresponding to ${\mathfrak F}_{S_i}$ or some $\cl L._{\ell_j}$.
 We obtain $ Q(C)\in M({\mathfrak A}_{S_i})$. As it is gauge invariant, it is in fact in the multiplier of the gauge invariant
 part of ${\mathfrak A}_{S_i}$ which is $P_\alpha\al A._{S_i}P_\alpha\otimes  {\cal T}_{S_i}[E]$ and so
 $ Q(C)$ is in the multiplier algebra of a subalgebra of $\al R._0$ as well. \\
 In  this way mesons (bilinears in a
quark and an antiquark) and baryons (trilinears in quarks) are constructed in lattice QCD.
\end{itemize}

In the representation $\pi=\pi_{\rm Fock}\otimes\pi_\infty$ it is also possible to build unbounded observables
as gauge invariant operators.
For example we can  build gauge invariant combinations of the gluonic and the colour electric field
generators, and in the finite lattice context,
such operators were analyzed in \cite{KR1,JKR}.
As an example of such a gauge invariant operator, in the context of a finite lattice,
we state the Hamiltonian, where we disregard terms by which
$H$ has to be supplemented in order to avoid the
doubling problem. It is
\begin{eqnarray}
\label{Hamiltonian}
  H & = & \tfrac{a}{2} \sum_{\ell \in \Lambda^1}
  E_{ij}(\ell) E_{ji}(\ell)
  + \tfrac{1}{2 g^2 a}\sum_{p \in \Lambda^2}
 ( W (p) + W(p)^*) \nonumber \\
  & + & i\tfrac{a}{2} \sum_{\ell \in \Lambda^1}
  \bar\psi_n(x_\ell)  \big[\underline\gamma\cdot(y_\ell-x_\ell)\big]_{ni}
  \Phi_{ij} (\ell)\psi_j (y_\ell)  + h.c.
  \nonumber \\
  & + & ma^3 \sum_{x \in \Lambda^0} \bar \psi_i (x)  \psi_i(x)\,,
\end{eqnarray}
where $a$ is the assumed lattice spacing; $W (p)$ is the Wilson loop operator for the
plaquette $p= (\ell_1, \ell_2 ,\ell_3, \ell_4)$; the vector
$y_\ell-x_\ell$ for a link $\ell=(x_\ell,y_\ell)$  is the vector of length $a$ pointing from $x_\ell$ to $y_\ell$
 and h.c. means the Hermitean conjugate. As usual, $ \bar \psi_i (x)=\psi_j(x)^*(\gamma_0)_{ji}$.
 We have omitted the bispinor and flavour indices.
  Note that the summands occurring in \eqref{Hamiltonian} are all gauge invariant and hence
 observables, some unbounded.

\section{Conclusion.}


We have extended the finite QCD lattice model in~\cite{KR, KR1} to an infinite lattice.
We defined both local and global gauge transformations on it, and we identified
the Gauss law constraint. Using the T-procedure and the local structure of the
constraints we solved the constraint system, and identified
the algebra of local physical observables.

There are three directions in which this model needs to be developed in future work.
First, the open question of the existence and
 nature of the global physical observables $\al R.\backslash\al R._0$ needs to be settled.
 Second,
 we need to analyze boundary effects, i.e do colour charge analysis and connect to the
 results for the finite lattices in \cite{KR,KR1}.
Third, and more ambitiously, we need to define and analyze the dynamics of the system,
and obtain suitable ground states. This is already in the context of finite lattices, a very
hard task. We refer to \cite{Huebsch,RS} for the discussion of an exactly solvable model of the above type.

\appendix

\section{More on subsystems of constraints.}

Assume that $\al C.\subset\al A.\subset\al F.$ where $\al C.$ is
a first--class constraint set, and $\al A.,\;\al F.$ are unital
C*--algebras.
Now there are
two constrained systems to consider;-  $(\al A.,\,\al C.)$
and $(\al F.,\,\al C.).$ The first one produces the algebras
$\al D.\subset\al O.\subseteq\al A.,$ and the second produces
$\al D.\s{\al F.}.\subset\al O.\s{\al F.}.\subseteq\al F.,$
where as usual,
\begin{eqnarray*}
\al N.&=& [\al AC.]=\al A.\cdot C^*(\al C.),\qquad \al D.=\al N.\cap\al N.^*,\qquad
\al O.=M\s{\al A.}.(\al D.)
\qquad\hbox{and}  \\
\al N.\s{\al F.}.&=&[\al FC.]=\al F.\cdot C^*(\al C.),\qquad
\al D.\s{\al F.}.=\al N.\s{\al F.}.\cap
\al N.^*\s{\al F.}.,\qquad
\al O.\s{\al F.}.=M\s{\al F.}.(\al D.\s{\al F.}.)
\end{eqnarray*}
with constraining homomorphisms $\xi:\al O.\to\al R.=\al O./\al D.$ and
$\xi\s{\al F.}.:\al O.\s{\al F.}.\to\al R.\s{\al F.}.=\al O.\s{\al F.}./\al D.\s{\al F.}.$.
Then we have (cf. Theorem~3.2 of \cite{Grundling88b}):
\begin{Theorem}
\label{Teo.2.12}
Given as above the constraint systems $\al C.\subset\al A.\subset\al F.$
then
\[ \al N.\s{\al F.}.\cap\al A.=\al N.,\qquad
   \al D.\s{\al F.}.\cap\al A.=\al D.,\qquad
\qquad\hbox{and}\qquad
    \al O.\s{\al F.}.\cap\al A.=\al O.\;.
\]
Hence $\al R.=\al O./\al D.=(\al O.\s{\al F.}.\cap\al A.)\big/
(\al D.\s{\al F.}.\cap\al A.)\,,$ thus
$\;\xi\s{\al F.}.\restriction \al O.=\xi$.
\end{Theorem}
Thus we can always enlarge our given algebra to a larger more convenient one,
then we only need to intersect our constraint algebras $\al D.,$ $\al O.,$ with the original algebra
to obtain our required constraint algebras.



\begin{thebibliography}{99}




\bibitem{APT} Akemann, C.A., Pedersen, G.K., Tomiyama, J.:
Multipliers of C*-algebras. J. Funct. Anal. {\bf 13}, 277--301 (1973)



\bibitem{Bla1}
Blackadar, B.: Operator Algebras. Springer 2006

\bibitem{Bla2}
Blackadar, B.: Infinite tensor products of $C^*$-algebras,
Pac. J. Math. {\bf 77} (1977), 313--334


\bibitem{Brat}
Bratteli,~O., Robinson,~D.~W.:  Operator Algebras and Quantum
Statistical Mechanics 1,
Springer 1987 New York Inc.

 \bibitem{CR} Carey, A.L., Ruijsenaars, S.N.M.:
 On fermion gauge groups, current algebras and Kac--Moody algebras.
 Acta Applic. Math. {\bf 10}, 1--86 (1987)


\bibitem{Constr}
Costello, P.: The mathematics of the BRST-constraint method.
ArXiv:0905.3570 \\
M. Henneaux, C. Teitelboim: Quantization of Gauge Systems.
Princeton University Press, Princeton 1992\\
Landsman, N.P.: Rieffel induction as generalised quantum Marsden--Weinstein
reduction. J. Geom. Phys. \textbf{15}, 285--319 (1995)\\
Giulini, D., Marolf, D.: On the generality of refined algebraic
quantization. Class. Quant. Grav. {\bf 16}, 2479-2488 (1999)\\
Klauder, J., Ann. Physics \textbf{254}, 419--453 (1997)\\
Faddeev, L., Jackiw, R.: Hamiltonian reduction of unconstrained and constrained systems.
Phys. Rev. Lett. \textbf{60}, 1692 (1988)


\bibitem{Crz}
Creutz, Michael: Quarks, gluons and lattices.
Cambridge University Press 1983




\bibitem{Dirac}
Dirac, P.A.M.: Lectures on Quantum Mechanics. Belfer Graduate School of
  Science: Yeshiva University 1964







\bibitem{Gl03}
Gl\"ockner, H., Direct limit Lie groups and manifolds,
J. Math.\ Kyoto Univ.  {\bf 43} (2003), 1--26


\bibitem{GrNe}
Grundling, H., Neeb, K-H.:
Full regularity for a C*-algebra of the Canonical Commutation Relations,
 Rev. Math. Phys. {\bf 21} (2009), 587--613

\bibitem{GrSrv}
Grundling, H.: Quantum constraints. Rep. Math. Phys. {\bf 57}, 97-120 (2006)

\bibitem{Grundling85}
Grundling, H., Hurst, C.A.: Algebraic quantization of systems with a
  gauge degeneracy. Commun. Math. Phys. \textbf{98}, 369--390 (1985)


\bibitem{Grundling88a}
Grundling, H., Hurst, C.A.:
 The quantum theory of second class constraints: Kinematics.
  Commun. Math. Phys. \textbf{119}, 75--93 (1988) [Erratum: ibid. {\bf
  122}, 527--529 (1989)]

\bibitem{Grundling88b}
Grundling, H.: Systems with outer constraints. Gupta--Bleuler
  electromagnetism as an algebraic field theory. Commun. Math. Phys.
  \textbf{114}, 69--91 (1988

\bibitem{GL}
Grundling, H., Lledo, F. Local Quantum Constraints.
Rev. Math. Phys. \textbf{12}, 1159--1218 (2000)

\bibitem{Haag92}
Haag, R.: Local Quantum Physics. Berlin: Springer Verlag 1992

\bibitem{Han}
Hannabuss, K.: Some C*-algebras associated to quantum gauge theories.
	ArXiv:1008.0496v2

\bibitem{Huebsch}
Huebschmann,~J., Rudolph,~G. and Schmidt,~M.: A lattice gauge model for quantum mechanics on a stratified space.
Commun.~Math.~Phys. 286, 459-494 (2009)



\bibitem{Is}
Isham, C.J.: Modern differential geometry for physicists (2nd ed.).
Singapore: World Scientific 1999.


\bibitem{JKR}
Jarvis,~P.~D., Kijowski,~J. and  Rudolph,~G. : On the Structure of
the Observable Algebra of QCD on the Lattice. J.~Phys.~A: Math.~Gen. 38 (2005) 5359-5377

\bibitem{KR83} Kadison, R.~V., and Ringrose, J.~R., Fundamentals of the Theory of Operator
Algebras II, New York, Academic Press 1983

\bibitem{KR}
Kijowski,~J., Rudolph,~G.:
On the Gauss law and global charge for quantum chromodynamics.
J.~Math.~Phys. {\bf 43} (2002) 1796-1808

\bibitem{KR1}
Kijowski,~J., Rudolph,~G.:
Charge superselection sectors for QCD on the lattice,
J.~Math.~Physics Vol.~46, 032303 (2005)

\bibitem{KRT}
Kijowski,~J., Rudolph, G., Thielman,~A.:
Algebra of Observables and Charge
Superselection Sectors for QED on the Lattice.
Commun. Math. Phys. {\bf 188}, 535-564
(1997)


\bibitem{KRS}
 Kijowski,~J., Rudolph, G.,  Sliwa,~C.:
On the Structure of the Observable Algebra
for QED on the Lattice.
Lett. Math. Phys. {\bf 43}, 299-308 (1998)

\bibitem{KS}
 Kogut,~J.,  Susskind,~L.:
 Hamiltonian formulation of Wilson's lattice gauge theories.
 Phys. Rev. D {\bf 11}, 395--408 (1975)

\bibitem{K}
Kogut,~J.: Three Lectures on Lattice Gauge Theory. CLNS-347 (1976), Lecture Series Presented at the International Summer School, McGill
University, June 21-26, 1976

 \bibitem{La1}
 Langmann, E.: Fermion current algebras and Schwinger terms in
 (3+1)--dimensions. Commun. Math. Phys. {\bf 162}, 1--32 (1994)



 \bibitem{Mic1}
Mickelsson, J.: Current algebra representations for 3+1 dimensional
Dirac--Yang--Mills theory. Commun. Math. Phys. {\bf 117}, 261 (1988)



\bibitem{Mur}
Murphy, G.~J., {\rm C}$^*$--Algebras and Operator Theory, Boston,
  Academic Press, 1990
  
\bibitem{WN}  Napiorkowski, K., Woronowicz, S.: Operator theory in C*-framework, Reports on Mathematical Physics 
{\bf 31}, 353-371 (1992)

 \bibitem{OS}
Osterwalder, K., Seiler, E.: Gauge Field Theories on a Lattice. Ann. Phys. {\bf 110}, 440--471 (1978)



\bibitem{Pa94} Palmer, T. W., Banach Algebras and the General Theory of
$C^*$-algebras. Volume I; Algebras and Banach Algebras, Cambridge
Univ. Press, 1994

\bibitem{Ped}
G.~K.~Pedersen,  $C^*$-Algebras and their Automorphism Groups.
Academic Press 1989, London


\bibitem{Raeb}
Raeburn, I.: Dynamical systems and Operator Algebras.
Proceedings of the Centre for Mathematics and its Applications, Volume 36, p109, 1999.
National Symposium on Functional Analysis, Optimization and Applications, 1998 at
The University of Newcastle (the electronic MS is at www.math.dartmouth.edu/archive/m123f00/public\_html/DynSys5US.pdf)

\bibitem{Rief}
Rieffel, M.A.: On the uniqueness of the Heisenberg commutation relations,
Duke Mathematical Journal 39 (1972), 745--752

\bibitem{Ros}
Rosenberg, J.: Appendix to O. Bratteli's paper on ``Crossed
products of UHF algebras". Duke Math. J. {\bf 46} (1979), 25--26


\bibitem{RSV}
Rudolph,~G., Schmidt,~M. and Volobuev,~I.P.: Classification of Gauge Orbit Types
for SUn-Gauge Theories. J.~Math.~Phys.~Anal.~Geom. 5, (2002), 201-241 \\
Rudolph,~G., Schmidt,~M. and Volobuev,~I.P.: Partial Ordering of Gauge Orbit Types for SUn-Gauge Theories'', J.~Geom.~Phys. 42 (2002) 106-138


\bibitem{RSV1}
Rudolph,~G., Schmidt,~M. and Volobuev,~I.P.: On the Gauge Orbit Space Stratification: A Review, J.~Phys.~A: Math.~Gen. 35 (2002) R1-R50


\bibitem{RS}
Rudolph,~G., Schmidt,~M.: On the algebra of quantum observables for a certain gauge model. J.~Math.~Phys. 50, 052102 (2009)



\bibitem{Seiler}
Seiler,~E.:  Gauge Theories as a Problem of Constructive
Quantum Field Theory and Statistical Mechanics, Lecture Notes in
Phys., vol. 159, Springer (1982) \\
Seiler,~E.: ``Constructive
Quantum Field Theory: Fermions'', in Gauge Theories:
Fundamental Interactions and Rigorous Results, eds. P.~Dita,
V.~Georgescu, R.~Purice










\bibitem{Takeda55}
Takeda, Z., Inductive limit and infinite direct product of operator
  algebras. Tohoku Math. J. \textbf{7}, 67--86 (1955)


\bibitem{Tak}
Takesaki, M.: Theory of operator algebras I, New York,
Springer--Verlag, 1979.

\bibitem{Tak3}
Takesaki, M.: Theory of Operator Algebras III, Springer-Verlag, Berlin, 2003

\bibitem{Vara}
Varadarajan, V.S. Geometry of Quantum Theory, second edition, Springer-Verlag, New York, 1985.


\bibitem{WO93} Wegge-Olsen, N.\ E., K--theory and $C^*$-algebras,
Oxford Science Publications, 1993


\bibitem{Wil}
Williams, D.P.: Crossed products of C*-algebras. Providence, American Mathematical Society, 2007

\bibitem{Wilson}
Wilson,~K.G.: Confinement of quarks. Phys. Rev. D10, 2445 (1974)


\bibitem{unb}
 Woronowicz, S.L.: C*-algebras generated by unbounded elements.
 Rev. Math. Phys. {\bf 7}, 481--521 (1995)

































\end{thebibliography}
\end{document}